%% file: 0_main_Arxiv2026.tex
\documentclass[opre,nonblindrev]{informs3} 
\pdfoutput=1
\OneAndAHalfSpacedXI

\usepackage[sort&compress]{natbib}
 \bibpunct[, ]{(}{)}{,}{a}{}{,}%
 \def\bibfont{\small}%
\TheoremsNumberedThrough     
\ECRepeatTheorems

\EquationsNumberedThrough    

\MANUSCRIPTNO{ }

\usepackage{booktabs}
\usepackage{algorithm}
\usepackage{algorithmicx}
\usepackage[noend]{algpseudocode}
\usepackage{xcolor}
\usepackage{bbm}
\usepackage{subcaption}
\usepackage{cleveref}
\usepackage{placeins}
\usepackage{tikz}
\usepackage{tikz-3dplot}
\usetikzlibrary{positioning}
\usepackage{accents}

\def\RR{{\mathbb R}}
\def\R{{\mathcal R}}
\def\NN{{\mathbb N}}

\def\P{{\mathcal P}}
\def\EE{{\mathbb E}}

\def\PP{{\mathbb P}}

\def\G{{\mathcal G}}
\def\N{{\mathcal N}}

\def\S{{\mathcal S}}

\def\A{{\mathcal A}}

\def\D{{\mathcal D}}

\DeclareMathOperator{\alg}{ALG}

\newcommand{\revised}[1]{\textcolor{black}{#1}}

\allowdisplaybreaks

\renewcommand{\theARTICLETOP}{}

\allowdisplaybreaks

\renewcommand{\theARTICLETOP}{}

\begin{document}


\RUNAUTHOR{Torrico et al.}

\RUNTITLE{Equitable Congestion Pricing}

\TITLE{Equitable Congestion Pricing under the Markovian Traffic Model: An Application to Bogot\'a}

\ARTICLEAUTHORS{%
\AUTHOR{Alfredo Torrico}
\AFF{University of Southern California, \EMAIL{atorrico@usc.edu}}
\AUTHOR{Natthawut Boonsiriphatthanajaroen}
\AFF{Cornell University, \EMAIL{nb463@cornell.edu}} 
\AUTHOR{Nikhil Garg, Andrea Lodi}
\AFF{Jacobs Institute, Cornell Tech, \EMAIL{ngarg@cornell.edu,andrea.lodi@cornell.edu}}
\AUTHOR{Hugo Mainguy}
\AFF{University of Maryland, \EMAIL{hmainguy@umd.edu}}
} 

\ABSTRACT{%
{Congestion pricing} is used to raise revenues and reduce traffic and pollution. However, people have heterogeneous spatial demand patterns and willingness (or ability) to pay tolls, and so pricing may have substantial equity implications. We develop a data-driven approach to design congestion pricing given policymakers' equity and efficiency objectives. First, algorithmically, we extend the Markovian traffic equilibrium setting introduced by Baillon \& Cominetti (2008) to model heterogeneous populations and incorporate prices and outside options such as public transit. 
In this setting, we show that a unique equilibrium exists.
 Second, via a detailed case study, we empirically evaluate various pricing schemes using data collected by an industry partner in the city of Bogot\'a, one of the most congested cities in the world. We find that pricing personalized to each economic stratum can be substantially more efficient and equitable than uniform pricing; however, non-personalized but area-based pricing can recover much of the gap. 
}


\maketitle

%



\input{1_Introduction}
\input{1-1_Contributions}

\input{1-2_Related_Work}

\input{2_Model}

\input{3_Algorithm}

\input{3-4_Price_Opt}
\input{4_Data}
\input{5_Results}

\input{7_Conclusions}

%
%
%

\ACKNOWLEDGMENT{The authors would like to thank Emma Frejinger for preliminary discussions on the model, Roberto Cominetti for the helpful discussions on Theorem~\ref{thm:existence_uniqueness} and the code of their original method,  Frederic Charlier and Laura Iguavita from ClearRoad for providing data from their pilot in Bogota.}



\bibliographystyle{informs2014trsc} 
\bibliography{congestion_bibliography}


\newpage
\renewcommand{\thechapter}{A\arabic{chapter}}
\begin{APPENDICES}
\input{999_Appendix_A}
\input{999_Appendix_B}
\input{999_Appendix_C}

\input{999_Appendix_D}

\input{999_Appendix_E}

\end{APPENDICES}


\end{document}

%% file: 1_Introduction.tex

\section{Introduction}\label{sec:introduction}

Given increasing congestion, pollution, and budget concerns, cities are turning to \textit{congestion pricing} to charge drivers to use the roadways. Modern approaches propose to leverage technology that allow, for example, roadway, time-of-day, or per-person specific prices without needing to deploy substantial on-the-road infrastructure -- using either mobile phone applications or sensors on cars \citep{clearroad,cramton2018set}. The promise of these technologies is to enable data-driven prices, much like advances in algorithmic pricing have transformed ride-hailing platforms.

However, making pricing decisions in a data-driven manner is challenging. First, there are several potential desiderata when making such decisions: reducing congestion and raising revenue, for example. Furthermore, the city may desire the policy to have \textit{equitable} impacts, for a suitable notion of equity: For instance, that prices are evenly distributed spatially, or that people of different wealth levels are still able to access the roadways at equitable rates. Second, the impact of a given policy, for any fixed objective, is difficult to predict: It is unclear how any individual will react to a given set of prices -- will they react by paying the fee and continuing to drive, by reducing the number of trips they take, or by changing the timing or the route they take? Further complicating this calculation is that behavior and ultimate objective values are \textit{equilibria} values: Each potential driver's behavior depends on each option's travel time, which in turn depends on the behavior of other drivers -- and behavior is likely to differ by, e.g., wealth. 

In this work, we demonstrate how a government can make such decisions in a data-driven manner, overcoming some of these challenges. Our contributions are as follows:

%% file: 1-1_Contributions.tex

\paragraph{Algorithmic and Methodological Contributions.} We consider the Markovian traffic equilibrium introduced by \citet{baillon2008markovian}. Travelers make stochastic arc-level decisions, capturing variability in travelers' perception of route costs. To study the impact across different socioeconomic groups, we extend this model to the case with multiple agent types, and introduce two new aspects to make the model more suitable to the study of congestion pricing: (1) Users can choose an outside option at the beginning of the trip, e.g., using public transit, and (2) the route cost is composed of both a time component and a price component. In this setting, we show that a flow equilibrium exists and is unique.
This approach provides a tractable, extensible model to study congestion pricing and other design questions. In particular, it captures both stochastic behavior and heterogeneous cost-time valuations, where users can also choose an alternative outside option. Given a set of prices, we algorithmically find the equilibrium flow -- and thus the fraction of trips completed per agent type, revenue collected, travel speed, and other metrics.

\paragraph{Empirical Applied Contributions.}
We use our methodological framework to study the design of congestion pricing in Bogot\'a, Colombia,  currently one of the most congested cities in the world. We empirically evaluate different pricing schemes in the equilibrium model and analyze their impact with respect to revenue, equity, and efficiency -- based on travel times, costs paid, fraction of trips completed, etc. We calibrate our model using demand matrices (origin and destination pairs for members of different socioeconomic groups, called \textit{strata} in Bogot\'a) that were calculated using data collected by our industry partner. We ask: What are the implications of different pricing schemes? We especially focus on equity impacts on different strata. 

We evaluate the following schemes: (i) \textit{uniform} pricing, in which all {primary} roads are priced the same per distance; (ii) \textit{personalized} pricing, in which these prices may differ by socioeconomic group; (iii) \textit{area} pricing, in which prices may differ by geography but are the same for all primary roads and strata within a geographic area.\footnote{\revised{We remark that we did not include a cordon pricing policy in our analysis, since preliminary results showed significant underperformance relative to other pricing policies. More importantly, evidence from our industry partner's data show that most demand do not concentrate in an specific local zone (as in NYC) and the majority of drivers use major roads that run across the city.}} These schemes represent various approaches a government may pursue: (i) uniform pricing is the simplest to implement; on the other hand, (ii) personalized pricing may be more ``equitable'' to those with lower ability to pay tolls but would require more sophisticated tolling infrastructure that is able to personalize -- and more generally be technically or politically intractable; as we show, (iii) area pricing can adequately interpolate (i) and (ii) as it is similar to uniform pricing in terms of implementation and can achieve the equity levels of per-stratum pricing.

Our primary empirical findings are twofold. First, uniform pricing is highly inequitable, in terms of the proportion of trips started, average speed, and usage of primary roads by each strata. Second, area pricing can recover much of the equity benefits of personalized pricing while increasing overall revenue. This follows because where people live and, more generally, the origin-destination matrix varies substantially by stratum. Therefore, area pricing can be a proxy for personalized pricing. In particular, area pricing -- even when is coarsely implemented -- can achieve higher welfare for each social group, higher total welfare, and higher revenue than uniform pricing, and is generally comparable to the welfare levels of personalized pricing.

%% file: 1-2_Related_Work.tex

\subsection*{Related Literature}\label{sec:related_literature}

\paragraph{Network Equilibrium and Efficiency.}
Traffic equilibrium models (also known as \emph{selfish routing or congestion games}) have been extensively studied by different communities. For brevity, we only present the work that is most relevant in our context. The notions of network equilibria can be traced back to \citet{knight1924some} and later formalized by
\citet{wardrop1952road}. \citet{beckmann1956studies} presented the first mathematical formulation of traffic equilibrium, and, since then, different models have been studied including equilibrium analysis for non-atomic \citep{pigou1912wealth,beckmann1956studies,daganzo1982unconstrained,fukushima1984dual,braess2005paradox} and atomic agents \citep{monderer1996potential,rosenthal1973class}; for more details, we refer to \citep{florian1995network,florian1999network,Nisan_Roughgarden_Tardos_Vazirani_2007}. Stochastic equilibrium models have also been widely studied in the literature, see, e.g., ~\citep{dial1971probabilistic,daganzo1977stochastic,baillon2008markovian}. For more details see \citep{baillon2008markovian,florian1999network} and the references therein.

From a modeling perspective, our work is based on the Markovian model introduced by \citet{baillon2008markovian} where agents make random choices at each node in their route according to the cost of the remaining path. The authors show that a traffic equilibrium can be found by solving a strictly convex program. We emphasize that this model, and thus our work, captures only the spatial dynamics and not inter-temporal dynamics; for the latter, we refer to~\citep{florian1999network}. In our work, one could rerun our analyses using data from different inter- or intra-day data. 

Network performance (efficiency) in selfish routing models has also been analyzed, in particular, the notion of \emph{price of anarchy} that measures the gap between the equilibrium cost and the social optimum cost, see, e.g.,~\citep{johari2004efficiency,roughgarden2002howbad,roughgarden2004bounding,correa2004selfish, roughgarden2005selfish,cominetti2021price}.

\paragraph{Congestion Pricing.}
One way to ameliorate the inefficiencies in selfish routing is by affecting agents' behavior via network pricing. The initial ideas on \emph{marginal congestion pricing} were informally discussed by \citet{pigou1912wealth} who asserts that, on each edge, a user should pay the price (tax or toll) that is equivalent to the amount of delay that their presence provokes on other users. Decades later, \citet{beckmann1956studies} have shown that under the assumption of homogeneous users, there exists a set of edge prices for which the equilibrium attains the minimum social cost, i.e., inefficiencies disappear. Later, \citet{coletolls} have shown an analogous result for the heterogeneous case and, moreover, have proved that the optimal prices can be computed efficiently. The design of congestion pricing schemes can take mainly two forms: first-best and second-best. The former refers to the postulates on marginal edge pricing of \citet{pigou1912wealth}. Other works along this line are~\citep{arnott1994economics,smith1979marginal}. The second-best refers to the schemes that restrict the set of edges that can be priced. Most of this literature focuses on solving a mathematical programming formulation with equilibrium constraints, see, e.g.,~\citep{larsson1998side,labbe1998bilevel,lim2002transportation,patriksson2002mathematical,brotcorne2001bilevel,ferrari2002road}; for more details on pricing schemes, we refer to the survey by~\citet{florian1999network}. The hardness of computing optimal second-best prices is studied, e.g., in~\citep{bonifaci2011efficiency,harks2015computing,hoefer2008taxing}. Several works have considered heterogeneous congestion pricing schemes, e.g., ~\citep{feng2023collaborative,lazar2020optimal,mehr2019pricing}. In particular, \citet{brown2016study} studied the case of affine cost functions and users with different willingness to pay. Most of the literature has focused either on deterministic equilibrium models or homogeneous pricing schemes. Our work, instead, studies congestion pricing schemes in a stochastic equilibrium model with heterogeneous users.

\paragraph{Equity in Congestion Pricing.}
It has long been established that congestion pricing has a direct impact on social equity, see, e.g., \citep{levinson2010equity,wachs2005then,eliasson2016congestion,eliasson2006equity,gemici2018wealth}. Therefore, the literature has focused on ameliorating its effects by studying different strategies. One of those strategies involves congestion pricing and \emph{revenue refunding} schemes in which the revenue collected by the tolls is redistributed among the population -- for example, in terms of infrastructure investments~\citep{goodwin1990make,small1992using}. From a technical standpoint, these schemes have been analyzed for single-bottleneck models~\citep{arnott1994economics,bernstein1993congestion}, parallel networks~\citep{adler2001direct} and single origin-destination networks~\citep{eliasson2001road}. The second strategy corresponds to the design of \emph{Pareto-improving} schemes~\citep{lawphongpanich2007pareto,lawphongpanich2010solving,song2009nonnegative}, aiming to minimize the total congestion while ensuring that no user is worse off as compared to no-pricing. \citet{guo2010pareto} and \citet{jalota2021efficiency}
have investigated schemes that are both revenue refunding and Pareto-improving. Other equitable strategies such as tradable credit schemes have been studied by \citet{wu2012design}.

Most related is the recent and complementary work of \citet{maheshwari2024congestion}, who also developed a computational framework to analyze the efficiency and equity of various congestion pricing schemes in terms of how they affect different populations that are both spatially distributed and have different willingness (or ability) to pay tolls. While sharing the same high-level motivation, our work differs in the notion of equilibria, algorithmic framework, data, and application. In particular, \citet{maheshwari2024congestion} pre-fix a subset of feasible routes between each origin-destination pair and do not model outside options. Moreover, their solution does not induce unique route selections. Their approach is more computationally efficient (a linear program to obtain prices, unlike a grid search over prices as we require) but does not allow calculation of the same metrics such as travel time or cost paid by the group.\footnote{We thank the first author of \citet{maheshwari2024congestion}  for this discussion.} By applying their framework to data from the San Francisco Bay Area in the United States, they find that their schemes improve both efficiency and equity compared to the status quo, and that schemes personalized to different populations can be more equitable. \revised{For this, they consider a \emph{significantly smaller} network (17 nodes, 22 edges) in which only major roads between larger areas are considered. Finally, they obtain similar findings showing that spatial pricing is competitive with heterogeneous pricing, although their approach is not as granular as ours, which may overestimate certain user's behavior.}

\paragraph{Equity and Pricing in Operations.} Finally, our work broadly connects to a growing interest in analyzing and considering the fairness or equity implications of canonical operational questions, especially in government, spanning both theoretical and empirical work (e.g., \cite{bertsimas2011price,bertsimas2012efficiency,barre24,singh2022fair,rahmattalabi2022learning,jo2023fairness,liu2023quantifying,liusla24,manshadi2021fair}). 
One line of work, in particular, considers fair personalized pricing schemes in non-traffic settings \citep{kallus2021fairness,cohen2021dynamic,cohen2022price}. To this literature, our work contributes an empirical example in which specific design choices -- e.g., spatial rather than uniform pricing -- yield substantially different equity outcomes, along with an analysis of revenue-equity tradeoffs. We further note that a large literature in operations has recently considered algorithmic spatio-temporal pricing in ride-hailing marketplaces, including its potentially heterogeneous effects on riders and drivers in equilibria \citep{ma2022spatio,garg2022driver,freund2021pricing,castillo2023benefits,hu2022surge,yan2020dynamic,bimpikis2019spatial,lobel2021revenue,castro2023autonomous}.

%% file: 2_Model.tex

\section{Model}\label{sec:model}
At a high level, our model is as follows: As in the \textit{Markovian Traffic Equilibrium} model, non-atomic agents have demand for each origin-destination pair and (random) perceived route costs. The cost is an arc-based function of the travel time, price, and sensitivity parameters. In this traffic model, agents' decisions are made at each node during their trip, which leads to a Markov chain with arc-based transition probabilities. Agents may also choose to take the outside option at the beginning of their trip, which is dependent on the observed flows. Given the above, the agents' choices induce a traffic equilibrium that directly impacts the travel time and cost of each trip.
In our formulation, agents belong to different \textit{types} (socioeconomic groups), each with different demand distributions, outside option valuations, and sensitivities to travel time and price. 

\paragraph{Notation.} We write in bold to indicate a vector/matrix and in italic to indicate scalar values, for example, $\mathbf{z} = (z_i)_{i\in[n]}$. To denote a random variable, we put a tilde over the corresponding letter and the letter without it represents its expected value, for example, $\EE[\tilde{z}]=z$. For a parameter or variable that relates to the outside option, we write a circle accent, for example, $\mathring{z}$. We provide a summary of the most relevant notation in Tables~\ref{tab:notation} and~\ref{tab:notation2}.

\subsection{Background: The Markovian Traffic Equilibrium Model}\label{sec:background}
In the following, we briefly describe the original \emph{Markovian traffic equilibrium} model introduced by \citet{baillon2008markovian}. Consider a traffic network represented by a strongly connected digraph $\G=(\N,\A)$, where $\N$ is the set of nodes and $\A$ the set of arcs, and a subset of nodes $\D\subseteq \N$ that denotes the destinations.
In this network, there is a set of non-atomic agents who each make routing decisions between origin-destination (OD) pairs. We emphasize that the original model in \citep{baillon2008markovian} assumes that all agents are of the same type.
Let $g_{i,d}\geq 0$ be the aggregated demand for trips between the OD pair $(i,d)$.
The overall goal of a traffic equilibrium model is to capture how these demands flow throughout the network. An equilibrium (if any) highly depends on how agents perceive \emph{route costs} that may be determined by multiple factors such as travel time, distance and monetary costs. In a deterministic model, every agent has the same perceived cost for a given route, while in a stochastic model, perceptions can vary across agents. In particular, in the Markovian traffic equilibrium model, agents' behavior is assumed to be an arc-based recursive decision process where, at each node along the route, the agent chooses the arc (from a set of available options) with the lowest \textit{random cost to-go}, i.e., the cost of the remaining route towards the destination.

Formally, let $f_a$ be the expected flow through arc $a\in\A$. Then, the cost of arc $a\in\A$ perceived by each non-atomic user is the perturbation of a deterministic cost, specifically, $c_a+\tilde{\varepsilon}_{a}$, where $\tilde{\varepsilon}_{a}$ is a continuous random variable with zero  mean and $c_a = \ell_a(f_a)$ is the expected cost of arc $a$ that is a function of the expected flow $f_a$.\footnote{Function $\ell_a$ is also known as the latency of arc $a$.} Note that $c_a$ captures the expected cost perception of arc $a$ (which is the same across all users) and $\tilde{\varepsilon}_a$ is the variability in perception. We assume that the cost function $\ell_a:\RR\to(0,\infty)$ is a strictly increasing continuous function.
With the above, the cost of a route $r\in\R_{i,d}$ is given by $\sum_{a\in r}(c_a+\tilde{\varepsilon}_a)$, where $\R_{i,d}$ is the set of all possible simple paths for the OD pair $(i,d)$. Since the cost of a route is a random variable, then the optimal cost $\tilde{\tau}_{i,d} := \min\left\{\sum_{a\in r}(c_a+\tilde{\varepsilon}_a): \ r\in\R_{i,d}\right\}$ and the cost of an arc $a=(i_a,j_a)$ with respect to destination $d$ (cost to-go) defined by $\tilde{z}_{a,d} = c_a +\tilde{\varepsilon}_a+ \tilde{\tau}_{j_a,d}$ are also random variables. Since the random cost of each arc is the sum of two terms, expectation plus variability, then we can equivalently write the random optimal cost and cost to-go as  
$\tilde{\tau}_{i,d} = \tau_{i,d} + \tilde{\varepsilon}_{i,d}$ and $\tilde{z}_{a,d} = z_{a,d} + \tilde{\varepsilon}_{a,d}$ where $\tilde{\varepsilon}_{i,d}$ and $\tilde{\varepsilon}_{a,d}$ are continuous random variables with mean 0, respectively.\footnote{Note that, with a slight abuse of notation, we write random variabilities indexed over arcs $\tilde{\varepsilon}_{a}$ for all arcs $a$, over OD pairs $\tilde{\varepsilon}_{i,d}$ for all OD pairs $(i,d)$ and over pairs (arc,destination) $\tilde{\varepsilon}_{a,d}$ for all arcs $a$ and destination $d$. } 

The sequential decision process of a non-atomic agent that is traveling between the OD pair $(i,d)$ is as follows: Upon reaching a node $i$, they observe the random cost to-go $\tilde{z}_{a,d}$ of every outgoing arc $a\in \A_i^+=\{a\in \A: \ a=(i,j)\}$ and choose the one with the lowest value. Formally, the agents' decisions are captured by the following component-wise non-decreasing, concave and smooth functions, which are also known as choice models: For every pair of nodes $i\in\N$ and $d\in\D$, consider the function
\[
\varphi_{i,d}(\mathbf{z}_d) := \EE\left[\min_{a\in \A_i^+}\left\{z_{a,d}+\tilde{\varepsilon}_{a,d}\right\}\right],
\]
where $\mathbf{z}_d$ is the vector whose components are $z_{a,d}$ for all $a\in\A$. Note that $\varphi_{i,d}$ defines the expected optimal decision of the users at node $i$, i.e., they choose the minimum cost to-go. Given these choice functions, for each destination $d\in\D$, the sequential process can be captured by a Markov chain over the network $\G$ whose transition probabilities are defined as: For each node $i\neq d$ and arc $a=(i,j)\in\A_i^+$,
\[
P_{i,j}^{d} := \PP\left(\tilde{z}_{a,d}\leq \tilde{z}_{e,d}, \ \forall e\in \A_{i}^+\right) = \frac{\partial \varphi_{i,d}}{\partial z_{a,d}}(\mathbf{z}_d),
\]
and zero otherwise. In words, at node $i$ the probability that the agent transitions to node $j$ is equal to the probability that the cost to-go of arc $a$ has the lowest random value. To ease the exposition, we do not make explicit the dependence of the transition probabilities on $\mathbf{z}_d$, i.e., $P_{i,j}^d = P_{i,j}^d(\mathbf{z}_d)$. For more technical details on the derivation of the second equality above, we refer the reader to \citet{baillon2008markovian}.
Moreover, in this Markov chain, each destination $d\in\D$ is assumed to be an absorbing state, i.e., we have $P_{d,d}^{d} = 1$ and $P_{d,j}^{d} = 0$ for all $j\neq d$. Consequently, these transition probabilities determine how the expected flow that enters a node splits among outgoing arcs. Formally, let $x_{i,d}$ be the expected flow that enters node $i$ towards destination $d$. Then, for each node $i\in\N$ and outgoing arc $a=(i,j)\in \A_i^+$, we write the expected flow that \emph{traverses arc $a$ towards destination $d$} as 
\begin{equation}\label{eq:flow_split}
v_{a,d} := x_{i,d}\cdot P^d_{i,j}.
\end{equation}
Intuitively, for a given realization of the perception variabilities of the non-atomic users, $v_{a,d}$ corresponds to the fraction of users that reached node $i$ and chose to continue their route to $d$ via the outgoing arc $a=(i,j)$. From this, the flow conservation constraint for each node $i\in\N$ and destination $d\in\D$ corresponds to 
\begin{equation}\label{eq:flow_conservation}
\revised{x_{i,d} = g_{i,d} + \sum_{a=(i',i)\in \A_i^-}x_{i',d}\cdot P_{i',i}^d= g_{i,d} + \sum_{a\in \A_i^-}v_{a,d},}
\end{equation}
where $\A_i^-$ is the set of incoming arcs. In words, \revised{Equation~\eqref{eq:flow_conservation} states that the expected flow that enters node $i$ towards destination $d$ equals the demand to $d$ at that node plus the flow from incoming arcs.} 
Finally, the total flow that traverses $a\in\A$ \revised{satisfies} 
\begin{equation}\label{eq:arc_flow}
    f_a = \sum_{d\in\D}v_{a,d}.
\end{equation}
\revised{Note that \eqref{eq:arc_flow} can be interpreted as a fixed-point equation since each value $v_{a,d}$ depends on the flow vector $\mathbf{f}\in\RR_+^{|\A|}$ via the transition probabilities $P_{i,j}^d = P_{i,j}^d(\mathbf{z}_d)$ and the expected costs to-go $z_{a,d}$, which itself depends on the arc costs $c_a=\ell_a(f_a)$.}

Our goal now is to relate the expected cost to-go of an arc $a$ with the expected optimal cost of the remaining route after passing through $a$. For this, recall that we wrote the random cost to-go of an arc $a$ towards destination $d$ in two different ways: $\tilde{z}_{a,d}=\tilde{c}_a+\tilde{\tau}_{j_a,d}$ and $\tilde{z}_{a,d}=z_{a,d}+\tilde{\varepsilon}_{a,d}$. Then, due to the Bellman's principle, we can recursively express the random optimal cost  $\tilde{\tau}_{i,d}$ for any node $i\in\N$ along the route as the minimum cost to-go (among all outgoing arcs from $i$), i.e.,
$\tilde{\tau}_{i,d} = \min_{a\in \A_i^+}\left\{\tilde{z}_{a,d}\right\}$. Moreover, the expected value satisfies
\[
  \tau_{i,d} = \EE[\tilde{\tau}_{i,d}]= \EE\left[\min_{a\in \A_i^+}\left\{\tilde{z}_{a,d}\right\}\right] =
  \EE\left[\min_{a\in \A_i^+}\left\{z_{a,d}+\tilde{\varepsilon}_{a,d}\right\}\right]=
  \varphi_{i,d}(\mathbf{z}_d),
\]
where the second equality is because $\tilde{z}_{a,d}=z_{a,d}+\tilde{\varepsilon}_{a,d}$. On the other hand, we also have
that 
\[
z_{a,d} = \EE[\tilde{z}_{a,d}] = \EE[\tilde{c}_a+\tilde{\tau}_{j_a,d}]= c_a + \tau_{j_a,d},
\] 
where we used that: (i) \revised{$\EE[\tilde{\varepsilon}_{a,d}]=0$}, (ii) $\tilde{z}_{a,d}=\tilde{c}_a+\tilde{\tau}_{j_a,d}$ and (iii) \revised{$\tilde{c}_a$ and $\tilde{\tau}_{j_a,d}$ also have variability $\tilde{\varepsilon}_{a}$ and $\tilde{\varepsilon}_{j_a,d}$ with zero mean}. 
Consequently, using both calculations above, we can derive the equation
\begin{equation*}
z_{a,d} = c_a + \varphi_{j_a,d}(\mathbf{z}_d)  \qquad \text{for all} \ a\in\A, \; d\in\D,
\end{equation*}
which, equivalently, defines the fixed-point equation over $\{\tau_i\}_{i\in\N}$
\begin{equation}\label{eq:fixed_point_opt2}
\tau_{i,d} = \varphi_{i,d}\left(\big(c_a+\tau_{j_a,d}\big)_{a\in \A_i^+}\right), \qquad \text{for all} \ i\in\N.  
\end{equation}
We are now ready to define the notion of equilibrium in this stochastic model. 
\begin{definition}[\citet{baillon2008markovian}]
A vector $\mathbf{f}\in\RR^{\A}$ is a \emph{Markovian traffic equilibrium (MTE)} if and only if $\mathbf{f}$ satisfies \eqref{eq:arc_flow}, where the values $v_{a,d}$ satisfy the flow conservation constraints \eqref{eq:flow_conservation} with $\mathbf{z}_d$ solving the fixed-point equation \eqref{eq:fixed_point_opt2} for $c_a = \ell_a(f_a)$.
\end{definition}
\vspace{0.5em}
\citet{baillon2008markovian} show that, under certain conditions, there exists a unique MTE that can be obtained by solving a smooth strictly convex program. We conclude this section by summarizing the most relevant notation in Table~\ref{tab:notation}, which will help the reader to follow the next section.

\begin{table}[htpb]
 \centerline{\begin{minipage}[t]{1.0\textwidth}
\caption{Summary of the most relevant notation in Sections~\ref{sec:background}. 
}
\label{tab:notation}
\centering
\begin{tabular}{llll}
\toprule
$\N$ & Set of nodes & $\tau_{i,d}$& Expected optimal cost of OD pair $(i,d)$ \\
$\A$ & Set of directed arcs & $P_{i,j}^d$& Transition probability of arc $a=(i,j)$ to dest. $d$\\
$\D$ & Set of destinations & $x_{i,d}$& Expected flow entering $i$ towards dest. $d$\\
$\S$ & Set of possible strata & $v_{a,d}$& Expected flow through arc $a$ towards dest. $d$\\
$g_{i,d}$ &  Demand between OD pair $(i,d)$ &$f_{a}$& Total expected flow through arc $a$\\
$c_a $ & Expected cost of arc $a$ & $\varphi_{i,d}$ & Choice model function for OD pair $(i,d)$  \\
$\ell_a$ & Time cost function of arc $a$ & & \\
$z_{a,d}$ & Expected cost to-go of arc $a$ to dest. $d$ & &\\
\bottomrule
\end{tabular}
\end{minipage}}
\end{table}

\subsection{Our Model: Multiple Agent Types and Monetary Costs}\label{sec:our_model}
One limitation of the model presented in the previous section is that it assumes that every non-atomic user has the same expected cost per arc, $c_a$. Similarly, the cost perception variability $\tilde{\varepsilon}_a$ does not vary among different agents and just depends on each arc. While in principle this may be true when \emph{time} is the only aspect considered in the arc's cost, the cost perception can significantly vary among different users when other components such as monetary costs are also included. In particular, congestion pricing can affect the users' decision-making process since individuals differ in their willingness to pay. Intuitively, low-income populations might be less able to pay tolls, while the high-income population could easily afford price increases without drastically changing their behavior. In this section, we introduce a Markovian traffic equilibrium model with two additions: (i) a monetary component is included in each arc cost that allows us to study different pricing schemes; (ii) multiple agent types (socioeconomic groups that we call \emph{stratum}) who differentiate in their willingness-to-pay, modeling different users' behaviors. In the next section, we extend the model to allow for an outside option in which agents can decide not to take the trip via car at all.

Consider a discrete set $\S$ of strata (i.e., agent types). For each stratum $s\in\S$, we denote by $\D^s$ the set of possible destinations and $g_{i,d}^s$ the demand of stratum for the OD pair $(i,d)$ with $d\in \D^s$. \revised{The monetary cost associated to each arc $a\in\A$ is determined by its price $p_a^s\in\RR_+$ that may be dependent on each stratum $s\in\S$.}
On the other hand, as in the previous section, let $\ell_a$ be the \emph{time cost} function of arc $a\in\A$, which we assume is not dependent on the strata. We denote by $t_a$ the expected time cost of that arc, which is flow-dependent via function $\ell_a$, i.e., $t_a = \ell_a(f_a)$ for a flow $f_a\geq0$ going through arc $a$.
\revised{Finally, let $\theta^s\geq0$ the willingness-to-pay of stratum $s$, i.e., the amount of money that $s$ is willing to pay for using a road. For simplicity, we assume that this parameter is not dependent on each arc
or on each OD pair, however, our equilibrium existence results follow in the more general case.} 
Given this, the total expected cost of stratum $s\in\S$ for using arc $a\in\A$ with price $p_a^s\geq0$ is
\begin{equation}\label{eq:expected_arc_cost_strata}
\revised{c_a^s:= t_a + \frac{1}{\theta^s}\cdot p_a^s = \ell_a(f_a) + \frac{1}{\theta^s}\cdot p_a^s.}
\end{equation}
We can now analogously define the notation that we introduce in the previous section but to our stratum-dependent model. For each stratum $s\in\S$, we denote by $z_{a,d}^s$ and $\tau_{i,d}^s$ the expected cost to-go of arc $a\in\A$ towards destination $d\in\D^s$ and the expected optimal cost  from node $i\in\N$ to destination $d\in\D^s$, respectively. Therefore, the choice model of stratum $s\in\S$ is similarly defined as: For $i\in\N$ and $d\in\D^s$,
\[
\varphi_{i,d}^s(\mathbf{z}^s_d):= \EE\left[\min_{a\in \A_i^+}\left\{z_{a,d}^s+\tilde{\varepsilon}^s_{a,d}\right\}\right],
\]
where $\tilde{\varepsilon}^s_{a,d}$ is the random variability of the cost to-go of arc $a$ towards destination $d$ perceived by stratum $s$. 
Therefore, analogous to the previous section, we can conclude that \(\tau_{i,d}^s = \varphi_{i,d}^s(\mathbf{z}^s_d)\) for each $s\in\S$, $i\in\N$ and $d\in\D^s$. The fixed-point equation can be equivalently derived as we did for~\eqref{eq:fixed_point_opt2}, which in the stratum-dependent context corresponds to
\begin{align}\label{eq:fixed_point_opt2_general}
\tau^s_{i,d} &= \varphi^s_{i,d}\left(\big(c^s_a+\tau^s_{j_a,d}\big)_{a\in \A_i^+}\right) \notag \\
&= \varphi^s_{i,d}\left(\bigg(t_a+\frac{1}{\theta_s}\cdot p_a^s+\tau^s_{j_a,d}\bigg)_{a\in \A_i^+}\right), \qquad \text{for all} \ s\in\S,\; i\in\N, \; d\in\D^s.  
\end{align}
Then, as in \eqref{eq:flow_conservation}, we can obtain the following equations for the model with multiple strata:
\begin{subequations}\label{eq:functional_flow_conservation_general}
\begin{align}
    v^s_{a,d} &= x^s_{i,d}\cdot P^{s,d}_{i,j}, \hspace{4.5em} \text{for all} \ i\in\N, \; a=(i,j)\in\A_i^+,\;  s\in\S, \; d\in\D^s,\\
  x^s_{i,d} &= g^s_{i,d} + \sum_{e\in \A_i^-}v^s_{e,d}, \qquad \text{for all} \ s\in\S, \; i\in\N, \; d\in\D^s,
\end{align} 
\end{subequations}
where $v^s_{a,d}$, $x^s_{i,d}$ and $P^{s,d}_{i,j}$ have analogous definitions as the ones in the previous section, while they are now dependent on the stratum $s$. As before, to avoid extra notation in the transition probabilities, we write $P^{s,d}_{i,j} = P^{s,d}_{i,j}(\mathbf{z}_d^s)$.  Finally, the total expected flow that traverses arc $a$ is the sum of the flow produced by all strata, i.e.,
\begin{equation}\label{eq:arc_flow_general}
    f_a := \sum_{s\in\S}\sum_{d\in\D^s}v^s_{a,d}.
\end{equation}
We are now ready to define an equilibrium flow for our setting.
\begin{definition}\label{def:equilibrium_general}
A vector $\mathbf{f}\in\RR^{\A}$ is a \emph{MTE flow under multiple strata and monetary costs} if and only if $\mathbf{f}$ satisfies \eqref{eq:arc_flow_general}, where the values $v^s_{a,d}$ satisfy the flow conservation constraints \eqref{eq:functional_flow_conservation_general} with $z_{a,d}^s = t_a + \frac{1}{\theta_s}\cdot p_a^s+\tau^s_{j_a,d}$ where $\tau_{\cdot,d}^s$ solves \eqref{eq:fixed_point_opt2_general} for $t_a = \ell_a(f_a)$.
\end{definition}

\subsubsection{MTE: Existence and Uniqueness.}\label{sec:existence_uniqueness}
In this section, we show that, for the general model with multiple strata and monetary costs, a Markovian traffic equilibrium flow exists and it is unique. Before proving our main result, we denote by
\begin{equation}\label{eq:good_set}
{T}:= \left\{\mathbf{t}\in\RR^{\A}: \ \exists \ {\tau} \text{ such that } {\tau}_{i,d}^{s}<\varphi_{i,d}^{s}\Bigg(\bigg(t_a+\frac{1}{\theta_s}\cdot p_a^s+{\tau}_{j_a,d}^{s}\bigg)_{a\in\A_i^+}\Bigg), \  \forall\; i\neq d, \ s\in\S\right\},  
\end{equation}
which, in words, corresponds to the set of expected travel times for which one can guarantee that users reach their destination in finite time with probability one. To prove existence and uniqueness in our context, one can show that the system of equations~\eqref{eq:functional_flow_conservation_general} has a unique solution when $\mathbf{t}\in{T}$ and, consequently, these equations define unique implicit functions $\mathbf{x}^s_{\cdot,d}=\mathbf{x}^s_{\cdot,d}(\mathbf{t})$, $\mathbf{v}^s_{\cdot,d}=\mathbf{v}^s_{\cdot,d}(\mathbf{t})$, $\tau^s_{\cdot,d}=\tau^s_{\cdot,d}(\mathbf{t})$ for each stratum $s\in\S$ and destination $d\in\D^s$. Moreover, these functions are concave, smooth and component-wise nondecreasing. We include the proof of these results in Appendix~\ref{app:existence_uniqueness}. 

By using the implicit functions above, one can redefine the notion of equilibrium in our setting as $\mathbf{f}\in\RR_+^{\A}$ being a MTE if and only if $f_a = \sum_{s\in\S}\sum_{d\in\D^s}v_{a,d}^s(\mathbf{t})$ with $\mathbf{t}$ satisfying $t_a = \ell_a(f_a)$ for all $a\in\A$. Or equivalently, $\mathbf{t}$ satisfying $\ell^{-1}_a(t_a)=\sum_{s\in\S}\sum_{d\in\D^s}v_{a,d}^s(\mathbf{t})$. Given this, our main result is as follows.
\begin{theorem}\label{thm:existence_uniqueness}
Assume $\mathbf{t}^0\in T$ where $t_a^0 = \ell_a(0)$ for all $a\in\A$. Then, there exists a \emph{unique MTE} given by $f_a^\star = \ell_a^{-1}(t^\star_a)$, where $\mathbf{t}^\star$ is the unique solution of the smooth strictly convex program
\begin{equation}\label{eq:convex_program}
\min_{\mathbf{t}\in T}\Phi(\mathbf{t}):= \sum_{a\in\A}\int_{t_a^0}^{t_a}\ell_a^{-1}(y)\;\mathrm{d}y-\sum_{s\in\S}\sum_{d\in\D^s}\sum_{i\neq d}g_{i,d}^s\tau_{i,d}^s(\mathbf{t})    
\end{equation}
\end{theorem}
\proof{Proof.}
First, recall that for each stratum $s\in\S$ and destination $d\in\D^s$ functions $\tau^s_{\cdot,d}(\mathbf{t})$ are concave, smooth and component-wise nondecreasing. Thanks to Lemma~\ref{lemma:convex_coercive} in Appendix~\ref{app:existence_uniqueness}, we know that $\Phi$ is strictly convex and coercive,\footnote{Function $\Phi$ is coercive if $\Phi(\mathbf{t})\to\infty$ when $\|\mathbf{t}\|\to\infty$.} then $\Phi$ has a unique optimal solution $\mathbf{t}^\star$. Therefore, this point must satisfy $\nabla\Phi(\mathbf{t}^\star) = 0$. Let us compute $\nabla\Phi$: For each $a\in\A$, we get
\[
\frac{\partial \Phi}{\partial t_a} = \ell_a^{-1}(t_a) - \sum_{s\in\S}\sum_{d\in\D^s}\sum_{i\neq d}g_{i,d}^s\cdot\frac{\partial \tau_{i,d}^s}{\partial t_a}(\mathbf{t}).
\]
Due to Lemma~\ref{lemma:implicit_derivative} in Appendix~\ref{app:existence_uniqueness}, we know that for all $s\in\S$, $d\in\D^s$ and $a\in\A$ we have that
\[
\sum_{ i\neq d}g_{i,d}^s\cdot\frac{\partial \tau_{i,d}^s}{\partial t_a}(\mathbf{t}) = v_{a,d}^s(\mathbf{t}).
\]
Thus, we obtain
\[
\frac{\partial \Phi}{\partial t_a} = \ell_a^{-1}(t_a) - \sum_{s\in\S}\sum_{d\in\D^s}\sum_{i\neq d}g_{i,d}^s\cdot\frac{\partial \tau_{i,d}^s}{\partial t_a}(\mathbf{t}) =  \ell_a^{-1}(t_a) - \sum_{s\in\S}\sum_{d\in\D^s}v_{a,d}^s(\mathbf{t}) = 0,
\]
which means that the unique optimal solution $\mathbf{t}^\star$ must satisfy $\ell_a^{-1}(t_a) =\sum_{s\in\S}\sum_{d\in\D^s}v_{a,d}^s(\mathbf{t})$, i.e., $\mathbf{t}^\star$ defines a MTE flow vector $\mathbf{f}$ via $f_a = \ell_a^{-1}(t^\star_a)$ for all $a\in\A$.
\Halmos
\endproof
We remark that to obtain Theorem~\ref{thm:existence_uniqueness} we cannot immediately apply the result in~\citep{baillon2008markovian}. This is because the cost function of each arc is stratum-dependent and one could na\"ively try to define a flow equilibrium $\mathbf{f}$ with $\tau_{\cdot,d}^s$ solving \eqref{eq:fixed_point_opt2_general} for $c_a^s$ satisfying~\eqref{eq:expected_arc_cost_strata}, but inverting this as a function of $f_a$ is not possible. However, in our setting the crucial aspect is that the expected cost function of each arc is a separable function of the flow-dependent time function and the monetary cost function that only depends on the prices. This allows us to define a flow equilibrium with $\tau_{\cdot,d}^s$ solving \eqref{eq:fixed_point_opt2_general} for $t_a=\ell_a(f_a)$, i.e., only accounting for the flow-dependent time cost. It is an open question whether an MTE admits a variational characterization for the general model with costs that depend both on the stratum and the flow of the corresponding arc. 
We emphasize that Theorem~\ref{thm:existence_uniqueness} applies for any stratum-dependent monetary cost function that depends solely on the price of the arc. In our case study, we assume that users pay a price that depends only on basic characteristic of the arc.

Due to Theorem~\ref{thm:existence_uniqueness}, we can guarantee that, for a fix pricing scheme, an appropriate descent method will converge to the unique equilibrium. We formalize our method in Algorithm~\ref{alg:mte_main} that is adapted to include other modeling choices such as the possibility of users taking an outside option (e.g., public transit).

%% file: 3_Algorithm.tex

\section{Modeling Details and Algorithm}\label{sec:model_details_alg}
In this section, we present the specific users' choice models that we consider for our case study of Bogot\'a and, also, we introduce the outside option. Later, we present our main method to compute equilibria. We summarize the most relevant notation in Table~\ref{tab:notation2}.

\subsection{Choice Models}\label{sec:choice_models}
 In our setting, we assume that agents make decisions according to a logit choice model. Specifically, recall that for a stratum $s\in\S$, a destination $d\in\D^s$ and an arc $a\in\A$, the cost to-go value is $\tilde{z}_{a,d}^s = z_{a,d}^s+\tilde{\varepsilon}_{a,d}^s$. Given $i\in\N$, we assume that $\tilde{\varepsilon}_{a,d}^s$ for all $a\in \A_i^+$ are i.i.d. Gumbel random variables with \revised{scale parameter $\beta^{s}\geq 0$ which can be interpreted as the sensitivity to the total cost to-go (time plus monetary)}.
 This implies that the function $\varphi_{i,d}^s$ is
\begin{equation}\label{eq:choice_model}
\varphi_{i,d}^s(\mathbf{z}_d^s) = -\frac{1}{\beta^{s}}\cdot\log\sum_{a\in \A_i^+}\exp\left(-{\beta^{s}\cdot z_{a,d}^s}\right),
\end{equation}
which gives us the classic multinomial-logit transition probabilities, i.e., for $i\in\N$, $a=(i,j)\in\A_i^+$ 
\begin{equation}\label{eq:choice_model_transition_probabilities}
P_{i,j}^{s,d}=\frac{\partial\varphi_{i,d}^s}{\partial z_{a,d}^s}(\mathbf{z}_d^s) = \frac{\exp\left(-{\beta^{s}\cdot z_{a,d}^s}\right)}{\sum_{e\in \A_i^+}\exp\left(-{\beta^{s}\cdot z_{e,d}^s}\right)},
\end{equation}
where the expected cost to-go value is $z_{a,d}^s = t_a+\frac{1}{\theta^s}\cdot p_a^s+\tau_{j_a,d}^s$. 
From this expression, note that since $\beta^s$ and $\theta^s$ are non-negative, then a flow or price increase in arc $a$ will decrease the probability of choosing $a$, as defined in \eqref{eq:choice_model_transition_probabilities}.

\begin{table}[htpb]
 \centerline{\begin{minipage}[t]{1.0\textwidth}
\caption{Summary of the notation in Section~\ref{sec:model_details_alg}. Notation for the outside option uses a circle accent.}
\label{tab:notation2}
\centering
\begin{tabular}{ll}
\toprule
$\beta^{s}, \;\theta^s$ & Sensitivity to cost and willingness-to-pay of stratum $s$   \\
$\mathring{\beta}^{s}$  & Sensitivity to the cost of the outside option for stratum $s$\\
$\mathring{c}_{i,d}$ & Cost of the outside option for OD pair $(i,d)$   \\
$\mathring{P}_{i,d}^{s}$ & Probability of taking the outside option for stratum $s$ and OD pair $(i,d)$\\
\bottomrule 
\end{tabular}
\end{minipage}}
\end{table}

\subsection{Adding an Outside Option}\label{sec:outside_option}
In this work, we assume that the demand is elastic, i.e., a trip that is either too expensive or too time-consuming will be made through an alternative option. One way to model this is: For each OD pair, add an arc that connects those nodes directly and add the corresponding cost function. However, with this approach, agents would be allowed to interrupt their routes at any moment, which is unrealistic because travelers rarely switch their trip means in the middle, unless this is planned in advance.

Our goal is to model the case when an agent decides not to take a trip by car because is costly and prefers to take an outside option, which in the following means using public transit.\footnote{Our framework allows to also consider the case in which a trip is completely canceled or the case of both canceling and public transit at the same time. However, for simplicity here we restrict to the use of public transit.} For this, we assume that the agent decides---at their starting node---whether to initiate the trip by comparing the cost to-go of the available routes versus the cost of the outside option. In other words, the demand of each OD pair $g_{i,d}^s$ is adjusted by the fraction of users of stratum $s$ that are willing to drive given the current observed flows.

\begin{figure}[htpb]
\centering
\caption{Meta-network with two layers. Layer in red indicates the road network (cars) and the layer in blue indicates the outside option network (e.g., public transit). The demand $g_{i,d}^s$ splits according to probabilities $1-\mathring{P}_{i,d}^{s}$ and $\mathring{P}_{i,d}^{s}$.}
\label{fig:meta_network}
\tdplotsetmaincoords{100}{50}

\scalebox{1.4}{\begin{tikzpicture}[tdplot_main_coords]
\coordinate (A1) at (0,0,0);
\coordinate (B1) at (4,0,0);
\coordinate (C1) at (4,3,0);
\coordinate (D1) at (0,3,0);

\coordinate (A2) at (0,0,2);
\coordinate (B2) at (4,0,2);
\coordinate (C2) at (4,3,2);
\coordinate (D2) at (0,3,2);

\filldraw[fill=black!10, draw=blue, opacity=0.5] (A1) -- (B1) -- (C1) -- (D1) -- cycle;

\filldraw[fill=black!10, draw=red, opacity=0.5] (A2) -- (B2) -- (C2) -- (D2) -- cycle;

\draw[dashed] (A1) -- (A2);
\draw[dashed] (B1) -- (B2);
\draw[dashed] (C1) -- (C2);
\draw[dashed] (D1) -- (D2);

\begin{scope}[shift={(0,0,0)}]
    \coordinate (P1a) at (0.5, 0.5);
    \coordinate (P2a) at (3.5, 2.5);
    
    \fill[black] (P1a) circle (1pt) node[anchor=north east,above] {\tiny $i_2$};
    \fill[black] (P2a) circle (1pt) node[anchor=south west,above] {\tiny $d$};
    
    \draw[thin, black] (P1a) .. controls (1.5, 2) and (2.5, 1) .. (P2a);
\end{scope}

\begin{scope}[shift={(0,0,2)}]
    \coordinate (P1b) at (0.5, 0.5);
    \coordinate (P2b) at (3.5, 2.5);
    
    \fill[black] (P1b) circle (1pt) node[anchor=north east,above] {\tiny $i_1$};
    \fill[black] (P2b) circle (1pt) node[anchor=south west,above] {\tiny $d$};
    
    \draw[thin, black] (P1b) .. controls (1.5, 2) and (2.5, 1) .. (P2b);
\end{scope}

\coordinate (F) at (0.5,-1.5,1);
\fill[black] (F) circle (1pt) node[anchor=south] {\tiny $i$};

\coordinate (F1) at (0.4,-2.2,0.92);

\draw[->, thin, black] (F) node[inner sep=0pt, below right=10pt and 0pt of F] {\tiny $\mathring{P}_{i,d}^{s}$} -- (0.45,0.45,0) ;
\draw[->, thin, black] (F) node[inner sep=0pt, above right=10pt and -6pt of F] {\tiny $1-\mathring{P}_{i,d}^{s}$} -- (0.45,0.45,2) ;
\draw[->, thin, black] (F1) node[left=0pt and -2pt of F1] {\tiny $g^s_{i,d}$} -- (0.41,-1.5,1);
\end{tikzpicture}}
\end{figure}
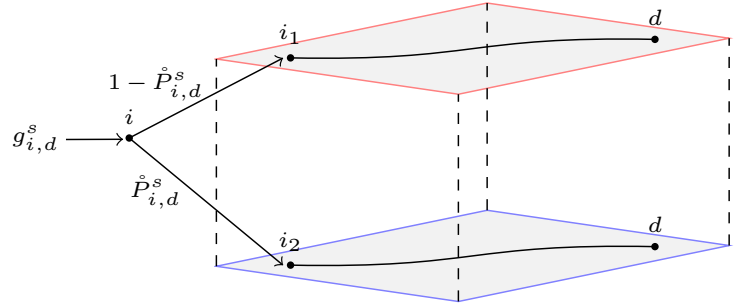

Formally, for each OD pair $(i,d)$ and stratum $s\in\S$, \revised{we denote the deterministic cost of the outside option as $\mathring{c}_{i,d}\geq 0$ and the sensitivity to it as $\mathring{\beta}^s\geq 0$. For simplicity, we assume that $\mathring{c}_{i,d}$ is a known deterministic value that does not depend on the strata and can be computed in advance, for instance, by considering the expected travel time and ticket cost of public transit. Note that it is also possible to model this cost as a flow-dependent function, however, our goal in this work is to understand how pricing schemes affect driving behavior and not choices among different transportation modes.}
Given this, the probability of an agent of stratum $s\in\S$ of choosing the outside option instead of starting the trip between the OD pair $(i,d)$ is 
\begin{equation}\label{eq:outside_option_probability}
\mathring{P}_{i,d}^{s} = \frac{\exp\Big(-\mathring{\beta}^{s}\cdot \mathring{c}_{i,d}\Big)}{\exp\left(-\mathring{\beta}^{s}\cdot \mathring{c}_{i,d}\right) + \sum_{a\in\A_i^+}\exp\left(-\beta^{s}\cdot z_{a,d}^s\right)}.
\end{equation}
Note that this probability depends on the current network flows because of the expected costs to-go values $z_{a,d}^s$. Finally, the demand of stratum $s$ for the OD pair $(i,d)$ is scaled by a factor $1-\mathring{P}_{i,d}^{s}$ since this is the fraction of users willing to drive. Therefore, the flow conservation constraints can be written as
\begin{subequations}\label{eq:functional_flow_conservation2}
\begin{align}
    v^s_{a,d} &= x^s_{i,d}\cdot P^s_{i,j}, \hspace{9em} \text{for all} \ i\in\N, \; a=(i,j)\in\A_i^+,\;  s\in\S, \; d\in\D^s,\\
  x^s_{i,d} &= g^s_{i,d}\cdot(1-\mathring{P}_{i,d}^{s}) + \sum_{e\in \A_i^-}v^s_{e,d}, \qquad \text{for all} \ s\in\S, \; i\in\N, \; d\in\D^s.
\end{align} 
\end{subequations}
We note that Theorem~\ref{thm:existence_uniqueness} still holds in this context. This is because one can model the inclusion of an outside option as a \emph{meta-network} of two layers. The first layer accounts for the car road network and the second layer corresponds to the network of the outside option (public transit). Then, for each stratum $s\in\S$ and OD pair $(i,d)$ with $d\in\D^s$, we add a dummy node $i'$ which we relabel as $i$ that gets demand $g_{i',d}^s=g_{i,d}^s$. We also add an arc $(i,i_1)$ where $i_1$ is in the first layer and an arc $(i,i_2)$ where $i_2$ is in the second network layer. The cost of the arc $(i,i_2)$ corresponds to the cost of the outside option. Assuming that users follow a logit choice model, then we get transition probabilities $P_{i,i_2}^s = \mathring{P}_{i,d}^{s}$ and $P_{i,i_1}^s = 1-\mathring{P}_{i,d}^{s}$. Therefore, Theorem~\ref{thm:existence_uniqueness} holds for the congestion game that occurs over this meta-network.  We provide an example of this representation in Figure~\ref{fig:meta_network}. We note that we are only interested in the flow conservation equations of the first network, so we penalize the demand by a factor $1-\mathring{P}_{i,d}^{s}$ as we did in Equations~\eqref{eq:functional_flow_conservation2}. One could imagine a more complex model with multiple layers modeling different modes of transportation, however, this is out of the scope of this work.


\subsection{Finding the Equilibrium Flow Given Prices}\label{sec:main_algorithm}

Given the guaranteed existence and uniqueness shown in Theorem~\ref{thm:existence_uniqueness} and the introduction of an outside option in the previous section, we can now present our main method to compute a Markovian traffic equilibrium.
\begin{algorithm}[htpb]
 \caption{Method to compute a Markovian traffic equilibrium}\label{alg:mte_main}
 \begin{algorithmic}[1]
 \Require Graph $\G=(\N,\A)$, destinations $\D$, demands $g_{i,d}^s$, tolerance $\delta>0$, prices $p_{a}^s$
 \Ensure Traffic equilibrium flow $\mathbf{f} = (f_a)_{a\in\A}$.
 \State Let $\mathbf{f},\mathbf{f}'$ be flow vectors such that $\mathbf{f}=\mathbf{0}$ and $\mathbf{f}'=\infty$.
 \While{$\|\mathbf{f}-\mathbf{f}'\| > \delta$}
 \State Compute costs $c_a^s = \ell_a(f_a)+\frac{1}{\theta^s}\cdot p_a^s$ for all $s\in\S, \ a\in\A$.
 \For{$s\in\S,\; d\in\D$} 
 \State Solve $\tau_{i,d}^s = \varphi^s_{i,d}\left(\big(c_a^s+\tau_{j_a,d}^s\big)_{a\in \A_i^+}\right)$ via fixed-point iteration, with $\varphi_{i,d}^s$ defined in \eqref{eq:choice_model}.  
 \State Compute matrices ${\mathbf{P}}^{s,d}$, ${\mathbf{Q}}^{s,d}$ using $\tau_{i,d}^s$ values.
  \State Build vector $\mathbf{y}$ with components $y_i = g_{i,d}^{s}\cdot(1-\mathring{P}_{i,d}^{s})$ for every $i\in\N, \ i\neq d$.
  \State Solve flow conservation constraints ${\mathbf{x}}^{s}_d = [\mathbf{I} - ({\mathbf{P}}^{s,d})^\top ]^{-1}\mathbf{y}$.
 \State Compute $\mathbf{v}_d^s = {\mathbf{Q}}^{s,d}\cdot{\mathbf{x}}^{s}_d$.
 \EndFor
 \State Compute $\mathbf{f}' = \sum_{s\in\S}\sum_{d\in\D}\mathbf{v}_d^s$.
\State $\mathbf{f}\leftarrow\mathbf{f} - \alpha_k\cdot(\mathbf{f} - \mathbf{f}')$, where $\alpha_k$ is the step size at the $k$-th iteration.
 \EndWhile
 \end{algorithmic}
 \end{algorithm}
First, let us recall the vectorial notation that we used in previous sections. For a given stratum $s\in\S$ and destination $d\in\D^s$, we denote by ${\mathbf{x}}^{s}_d = (x_{i,d}^s)_{i\neq d}$ the vector where each component is the flow that enters node $i\neq d$ towards destination $d$ and ${\mathbf{v}}^{s}_d = (v_{a,d}^s)_{a\in\A}$ the vector whose components are the expected flow that traverses arc $a$ towards destination $d$. Similarly, let ${\mathbf{g}}^{s}_d = (g_{i,d}^s)_{i\neq d}$ be the demand vector with components indexed by $i\neq d$, ${\mathbf{P}}^{s,d} = (P_{i,j}^{s,d})_{i,j\neq d}$ be the transition matrix and, finally, ${\mathbf{Q}}^{s,d} = (Q_{i,a}^{s,d})_{i\neq d,a\in\A}$ be an auxiliary matrix indexed by node $i\neq d$ and arc $a$ such that $i_a=i$, specifically,
\[
Q_{i,a}^{s,d} = \left\{\begin{matrix}P_{i,j_a}^s & \text{if } i=i_a \\ 0 & \text{otherwise}\end{matrix}\right.
\]
with the transition probabilities defined in \eqref{eq:choice_model_transition_probabilities}.

We formalize our method in Algorithm~\ref{alg:mte_main}. This algorithm receives as an input a fixed set of roadway prices and proceeds to compute a flow equilibrium by using the following first-order descent algorithm (Problem~\eqref{eq:convex_program} can be viewed as an unrestricted convex problem with a unique optimum) with a fixed-point method as a subroutine: 
\begin{enumerate}
\item[(i)] In the inner loop, for each stratum $s\in\S$ and destination $d\in\D^s$, we find values $\tau_{i,d}^s$ via a fixed-point iteration (Step 5). With these values, the transition probabilities can be computed (Step 6). Then, a linear system is solved to obtain flows $x^s_{i,d}$ and, consequently, $v^s_{a,d}$ (Steps 8-9). The latter are aggregated to obtain the incumbent flow vector ${\mathbf{f}'}$ (Step 10). 
\item[(ii)] In the outer loop, we obtain the next flow vector $\mathbf{f}$ via a first-order update using ${\mathbf{f}'}$ (Step 11).
\end{enumerate}
The process above continues until the norm of the difference between $\mathbf{f}$ and $\mathbf{f}'$ is smaller than the desired tolerance. 
 Note that Algorithm~\ref{alg:mte_main} follows the same high-level idea as the method provided in \citep{baillon2008markovian}, i.e., a first-order descent method with a fixed-point iteration as an inner loop. The main differences between Algorithm~\ref{alg:mte_main} and the method proposed by \citet{baillon2008markovian} are the penalization of the demand due to the outside option (Step 7) and the for-loop over multiple strata (Step 4). We remark that Step 5 is solved by a standard fixed-point iteration method, i.e., we iterate over $n\in\NN$
 \begin{equation}\label{eq:fixed_point_iteration}
 \tau_{i,d}^{s,n+1} = \varphi^s_{i,d}\left(\big(c_a^s+\tau_{j_a,d}^{s,n}\big)_{a\in \A_i^+}\right)
 \end{equation}
 until the difference between the current and the incumbent points has a norm below the desired threshold. For our computational study, we implemented various acceleration tools to improve the running-time performance of Algorithm~\ref{alg:mte_main} in large-scale instances, as several steps are computationally expensive; we describe these accelerations in Section~\ref{sec:parameter_tested}.

%% file: 3-4_Price_Opt.tex

\section{Welfare and Revenue via Price Optimization}\label{sec:price_opt}
{\color{black} In this section, we formalize our notion of welfare and pricing schemes.
Our main goal is to study how congestion pricing changes users' behavior and, ultimately, optimize the balance between welfare (of a given stratum or the whole society) and the total revenue collected. 
Moreover, we investigate the effect of each pricing policy on other metrics such as the proportion of trips started, distance traveled, road efficiency, and road type usage. 
Our focus is to optimize over \emph{arc-based pricing schemes}, i.e., agents of stratum $s\in\S$ must pay $p_a^s$ units of money for using arc $a\in\A$. 
For this, Algorithm~\ref{alg:mte_main} is key as it finds the unique equilibrium flow \textit{for a given set of prices}. In the remainder of this section, we assume that we have access to this algorithm as an oracle. Formally, we write the oracle as $\alg(\mathbf{p})$ that, for a vector of prices $\mathbf{p}$, returns the unique flow equilibrium vector $\mathbf{f}(\mathbf{p})$. 
Observe that each choice of prices directly impacts the flow equilibrium in the network as per the definition of the expected cost of each arc in \eqref{eq:expected_arc_cost_strata} and transition probabilities~\eqref{eq:choice_model_transition_probabilities}. 

Let us now formalize our notion of welfare. We define the welfare of stratum $s$ as the expected time difference (if the agent either drives or takes the outside option) -- compared to the no price equilibrium -- minus the monetary cost of the trip. In other words, the welfare measures how much time (on average) the agents are gaining or losing for what they are paying for, compared to the corresponding setting without any prices. Formally, for a stratum $s\in\S$ and an OD pair $(i,d)$ with $d\in\D^s$, let $t_{i,d}^s(\mathbf{p}) = t_{i,d}^s(\mathbf{f}(\mathbf{p}))$ be the total expected travel time\footnote{Recall that the travel time is dependent of the stratum because their choice behavior is captured by their transition probabilities, which may thus differ based on sensitivity to pricing. In other words, groups more sensitive to pricing may opt for secondary roads, increasing travel time.} and
$\kappa_{i,d}^s(\mathbf{p}) = \kappa_{i,d}^s(\mathbf{f}(\mathbf{p}))$ be the total expected amount paid by stratum $s$ when traveling between $i$ and $d$ when the pricing scheme is $\mathbf{p}\in\RR_+^{\A\times\S}$. 
Note that both expressions actually depend on the flow equilibrium $\mathbf{f}(\mathbf{p})$ when imposing pricing scheme $\mathbf{p}$, but to ease the exposition, we avoid this extra notation unless otherwise stated.

\begin{definition}[Welfare]\label{def:welfare_def}
We define the {welfare} of the non-atomic agents in stratum $s\in\S$ as the function $W^s:\RR_+^{\A\times\S}\to\RR$ which, for a given pricing scheme $\mathbf{p}$, outputs
\begin{align*}
W^s(\mathbf{p}) = \frac{1}{|G^s|}\cdot\sum_{(i,d)\in G^s}\Bigg[\left(t^s_{i,d}(\mathbf{0})-t^s_{i,d}(\mathbf{p})-\frac{1}{\theta^s}\cdot \kappa^s_{i,d}(\mathbf{p})\right)\cdot (1-\mathring{P}_{i,d}^{s}(\mathbf{p})) + \left(t^s_{i,d}(\mathbf{0})- \mathring{c}_{i,d}\right)\cdot \mathring{P}_{i,d}^{s}(\mathbf{p})\Bigg],
\end{align*}
where $G^s = \left\{(i,d): \ i\in\N,\; d\in\D^s \text{ such that } g_{i,d}^s>0\right\}$ is the set of OD pairs that have positive demand by stratum $s$ and $t^s_{i,d}(\mathbf{0})$ corresponds to the total expected travel time when there is no pricing on the roads.
\end{definition}
\vspace{0.5em}
Observe that we minimize notation in Definition~\ref{def:welfare_def} as we do not write the dependence of the welfare value on the flow equilibrium, i.e., $W^s(\mathbf{p}) = W^s(\mathbf{f}(\mathbf{p}))$. Also, recall that the outside option probabilities also depend on the equilibrium flows that result from the chosen pricing schemes.
The first term of the welfare represents when the agent chooses to start the trip with probability $(1-\mathring{P}_{i,d}^{s})$. In that case, the expected travel time for the OD pair without congestion pricing $t^s_{i,d}(\mathbf{0})$ is compared to the cost of the trip with congestion pricing that accounts for time and  monetary costs: $t^s_{i,d}(\mathbf{p}) + \frac{1}{\theta^s}\cdot \kappa^s_{i,d}(\mathbf{p})$. We can interprate this metric as follows: If the difference is positive, then it means that agents have a surplus in time (trips are shorter) that is higher than the welfare loss due to the monetary cost paid. If the difference is negative, then needing to pay worsens the agents' welfare (in terms of time and money).
The second term in Definition~\ref{def:welfare_def} reflects the case of the user choosing to take the outside option. In that case, the expected travel time without pricing $t^s_{i,d}(\mathbf{0})$ is compared to the cost of the outside option 
$\mathring{c}_{i,d}$. In summary, the welfare value represents the change in each traveler's welfare when compared to no pricing, averaged over all the OD pairs with positive demand. Finally, note that for $\mathbf{p}=\mathbf{0}$, $W^s(\mathbf{0})$ only quantifies the welfare when choosing the outside option. In general, it may occur that $W^s(\mathbf{0})>0$ due to $\mathring{P}_{i,d}^{s}(\mathbf{0})>0$ for some OD pair $(i,d)$, however, in our case study we did not observe such cases and, instead, we observe $W^s(\mathbf{0})=0$.

Using the above notation, the \emph{total welfare} is then defined as the sum over all stratum-dependent welfare values, namely
\begin{equation*}
W(\mathbf{p}) = \sum_{s\in\S}W^s(\mathbf{p}).
\end{equation*}
Our first objective is to understand how a change in price may affect each stratum-dependent welfare, relative to the total welfare. For this, we study the Pareto frontier between the welfare of each stratum $s$ and the total welfare. Specifically, for each $\lambda\in[0,1]$ and $s\in\S$, we solve
\begin{subequations}\label{eq:optimization_problem}
\begin{align}
    \max_{\mathbf{p}} & \quad \lambda\cdot W^s(\mathbf{f}(\mathbf{p})) + (1-\lambda)\cdot  W(\mathbf{f}(\mathbf{p}))\\
    s.t. & \quad \mathbf{f}(\mathbf{p}) = \alg(\mathbf{p}), \  \mathbf{p}\in \P,
\end{align} 
\end{subequations}
where $\P$ is the set of feasible pricing schemes and $\alg(\mathbf{p})$ is the function that runs Algorithm~\ref{alg:mte_main} with price input $\mathbf{p}$ and outputs the flow equilibrium vector $\mathbf{f}(\mathbf{p})$. 

Our second goal is to study the tradeoffs between the welfare of each stratum and the total revenue. Therefore, we define the \emph{expected revenue} collected from a given stratum $s$ when the pricing scheme is $\mathbf{p}$ as 
\begin{equation*}
R^s(\mathbf{p})=R^s(\mathbf{f}(\mathbf{p})) = \sum_{a\in\A}f_a^s\cdot p_a^s.
\end{equation*}
Then, the \emph{total expected revenue} collected by the decision-maker is given by
\[
R(\mathbf{p})  = \sum_{s\in\S}R^s(\mathbf{p}).
\]
As in \eqref{eq:optimization_problem}, we study the Pareto frontier between the welfare of stratum $s$ and the total revenue. Namely, for each $\lambda\in[0,1]$ and $s\in\S$, we solve
\begin{subequations}\label{eq:optimization_problem_revenue}
\begin{align}
    \max_{\mathbf{p}} & \quad \lambda\cdot W^s(\mathbf{f}(\mathbf{p})) + (1-\lambda)\cdot  R(\mathbf{f}(\mathbf{p}))\\
    s.t. & \quad \mathbf{f}(\mathbf{p}) = \alg(\mathbf{p}), \  \mathbf{p}\in \P.
\end{align} 
\end{subequations}

\paragraph{Pricing Schemes.} Due to the highly non-linear structure of the objective functions in \eqref{eq:optimization_problem} and \eqref{eq:optimization_problem_revenue}, it is unclear whether or not there exists a provably efficient algorithm that jointly finds an equilibrium flow and an optimal pricing policy. Consequently, we focus our analysis on low-dimension pricing regions $\mathcal{P}$ so we can empirically solve those problems via a simple grid search. 
%
For each pricing strategy, we first assume that the price $p_a^s$ is weighted by the distance traveled in that road, namely $p_a^s = r_a^s\cdot l_a$ where $l_a$ is the length of arc $a$ and $r_a^s$ is the rate paid per distance unit. Second, we differentiate the rate $r_a^s$ as follows:
\begin{enumerate}
\item \emph{Uniform pricing}: $r_a^s \in\{0,r\}$ for all $s\in\S, a\in\A$ with $r\geq0$.
\item \emph{Personalized pricing}: $r_a^s \in\{0,r^s\}$ for all $s\in\S, a\in\A$ with $r^s\geq0$. 
\item \emph{Area pricing}: $r_a^s \in\{0,r_a\}$ for all $s\in\S, a\in\A$ with $r_a\geq0$ depending on the geographical location of $a$ in the network $\G$. 
\end{enumerate}
Note that both personalized and area pricing policies are (different) generalizations of uniform pricing. Namely, one can set the same rate for every stratum in the case of personalized pricing or the same price for every area in area pricing (we simulate this setting in Appendix \ref{sec:synthetic}). However, whether they are strictly better (for a given objective) than uniform pricing depends on the distribution of OD pairs. For instance, consider a network with the same uniform OD pairs distribution for each stratum, then intuitively area pricing may not be strictly better than uniform pricing. 
Additionally, a comparison between personalized pricing and area pricing is not direct. Neither generalizes the other and which is better on any given metric depends on the topology of the network and the distribution of OD pairs of each stratum. For example, if the OD pairs of each stratum are in different regions of the networks, then we can recover personalized pricing via area pricing and vice versa. We further note that, for computational simplicity, we do not consider personalized area prices. In our data application, we will designate ``secondary'' (e.g., local roads) and ``primary'' roads (e.g., motorways, highways) -- prices for secondary roads will be 0, while all primary road prices may potentially vary per stratum or area. Finally, we also do not consider policies such as cordon-based pricing. Preliminary test showed its effectiveness in Bogot\'a is limited since our data shows that most drivers choose higher-capacity roads that run across the city, as opposed to the case of New York City~\citep{nyt} where there is a higher demand in local districts.
}

%% file: 4_Data.tex

\section{Case study: Data and Setup Description}\label{sec:data_description}
We now perform an empirical analysis in the city of Bogot\'a, one of the most congested cities in the world \citep{bogota_top,bogota_top2}. For this, we use OD pairs of demand data collected by our industry partner ClearRoad in early 2022. We now describe the experimental setup, in particular, the cost functions and sensitivity parameters, among others. Then, we will proceed to describe the road network and data.

\subsection{Time and Monetary Cost Functions}
One of the main aspects of our model is the expected cost function presented in \eqref{eq:expected_arc_cost_strata}, which is a separable function of time and price cost components. Let us recall this cost: For an arc $a\in\A$ and stratum $s\in\S$ is defined as
\[
c_a^s = \ell_a(f_a) + \frac{1}{\theta^s}\cdot p_a^s,
\]
where $\ell_a(f_a)$ is the time cost of the flow $f_a$ that traverses arc $a$ and $p_a^s$ is the price that stratum $s$ pays for using that arc. In our computational study, we use the strictly increasing time cost function
\begin{equation}\label{eq:time_cost_function}
\ell_a(f_a) = t_{a}^{0}\cdot\left(1+\gamma_a\cdot\Big(\frac{f_a}{b_a}\Big)^{\nu_a}\right),
\end{equation}
where $t_{a}^0=\ell_a(0)$ is the travel time
through $a$ with empty roads (i.e., without flow), $b_a$ is the capacity of the arc, and $\nu_a,\gamma_a$ are fitting parameters. The function in \eqref{eq:time_cost_function} is known as the Bureau of Public Roads power function \citep{roads1964traffic} which is widely used in the transportation literature to model how travel time increases with arc flow.

For the monetary cost, we consider two aspects of the road (arc): its length and whether is a primary or secondary road, where primary corresponds, roughly speaking, to major high-capacity or higher-speed roads. Formally, we use the strictly increasing price cost function 
\begin{equation}
\label{eq:price_cost_function}
p_a^s = \left\{\begin{matrix}r_a^s\cdot l_a & \quad \text{if arc} \; a \; \text{is primary}\\ 0 & \quad \text{otherwise}\end{matrix}\right.\;,
\end{equation}
where $l_a$ is the travelling distance of arc $a$. 
Note that in this price cost function, agents are paying for the distance traveled only when using a primary road. This function is also commonly used for tolling, where only a subset of roads are tolled and tolls are based on the length traveled. We note that our framework is flexible to other choices of \eqref{eq:price_cost_function}, such as cordon-based policies in which travels must pay a fixed cost to enter an area (in that case, one can charge a fixed price on each incoming arc to that area).

\subsection{Data Description}\label{sec:data_details}
\begin{figure}[htpb]
\begin{minipage}[t]{0.49\textwidth}
\centering
\captionof{table}{Instance Summary. Last 3 rows: The value in parentheses indicate the number of OD pairs.}
\label{tab:instance}
 \vspace{2em}
\begin{tabular}{lc}
\toprule
Center Coordinates & 4.67172, -74.11290  \\
City Radius & 10 km.  \\
Nodes & 663 \\
Roads (Arcs) &  1486 \\
Primary Roads & 306 \\
High-income Demand & 1495 trips (1064)\\
Mid-income Demand & 5224 trips (4073)\\
Low-income Demand & 2810 trips (2266)\\
\bottomrule
\end{tabular}
\end{minipage}
\begin{minipage}[t]{0.49\textwidth}
\centering
\captionof{figure}{Road Network: Nodes are in red and dashed lines indicate area splits.}\label{fig:network}
\includegraphics[width=0.8\textwidth,bb=0 0 455 457]{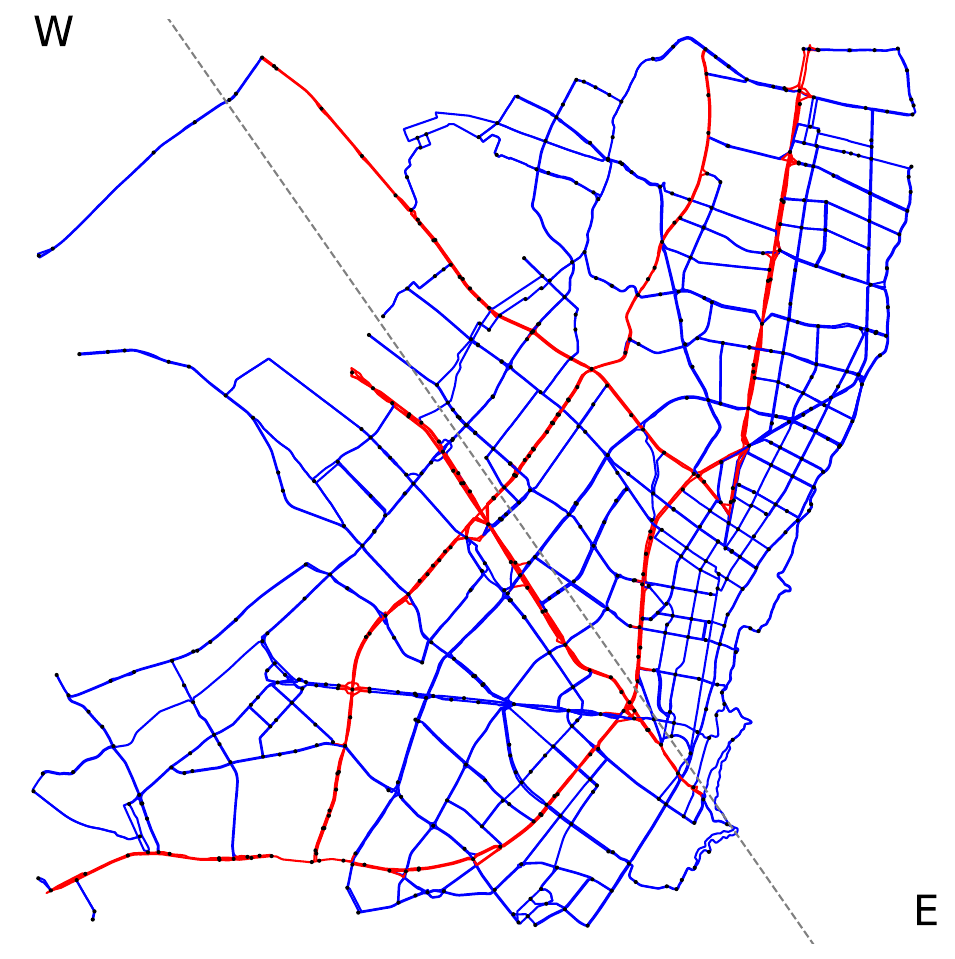}
\end{minipage}
\end{figure}

We now specify our data, summarized in Table~\ref{tab:instance}.
\paragraph{Road Network.} We extract the road network using the Open Street Maps Networkx (osmnx) \citep{Boeing25} library in Python. We chose as the city center the point $(4.67172, -74.11290)$ and a radius of 10 kilometers. The types of roads chosen were: motorway, motorway link, trunk, trunk link, primary, primary link, and secondary. To make the problem tractable, we consolidated intersections within a 70 meters radius.\footnote{This can create duplicated edges, but we consider only the main arc and not its copies.} Finally, we pre-processed the graph as follows: We got the largest strongly connected component and then we removed any node with 0 outgoing arcs.\footnote{This means that every node could be a trip origin. Also note that because of strong connectivity, every node is reachable.} A map of the resulting network is depicted in Figure~\ref{fig:network}, where we color major roads that we price in red (these were selected by a preliminary analysis of our GPS data) and the rest in blue. We note that the main road in the middle (that goes from NW to SE) corresponds to the primary road that connects the airport with the city. Historic center (most touristic area) is on SE corner. The wealthiest neighborhoods of Bogot\'a are on the E area. On the other hand, the poorest neighborhoods of Bogot\'a are on the SW area and bottom of the network. For more details we refer to the Strata Data detailed below and Bogot\'a stratification maps in Figure~\ref{fig:stratification} in the Appendix.

\paragraph{Strata Data.} Since the 1990s, neighborhoods in each district (\emph{localidad} in Spanish) of Bogot\'a are socioeconomically classified from 1 to 6, where 1 is the poorest and 6 is the wealthiest; for more details (in Spanish) we refer to \citep{secretaria} or \citep{the_guardian}, and for a map of the stratification we refer to Figure~\ref{fig:stratification} in the Appendix. 
The goal of this stratification is the subsidy of utilities and services such as electricity, telephone bills, and trash collection. In other words, the high-income population pays more to subsidize the lower-income population -- notably, then, personalized congestion pricing schemes have precedence in the city. In our model, we aggregate the strata in the following way: We say that agents are \emph{low-income} if they belong to strata 1-2, \emph{mid-income} if they belong to stratum 3, and \emph{high-income} if they belong to strata 4-6. This means that our strata set is $\S =\{\text{high-income},\text{mid-income},\text{low-income}\}$ which we also denote as $\S=\{\text{h},\text{m},\text{l}\}$. 
As apparent in \Cref{tab:instance}, our data is primarily composed of low- and mid-income trips.\footnote{We choose not to calibrate the data according to demographic data on where each stratum lives -- it is unclear how to calibrate driving demand, given the population distribution.}

\begin{figure}
\caption{Instance Demand Data}
\label{fig:instance_demand_data}
\begin{subfigure}[t]{0.49\textwidth}
\centering
\caption{\footnotesize{\% of Trips Area to Area (Rel. to Stratum Total)}}\label{fig:percentagetrips_areatoarea}
\includegraphics[width=\textwidth,bb=0 0 413 330]{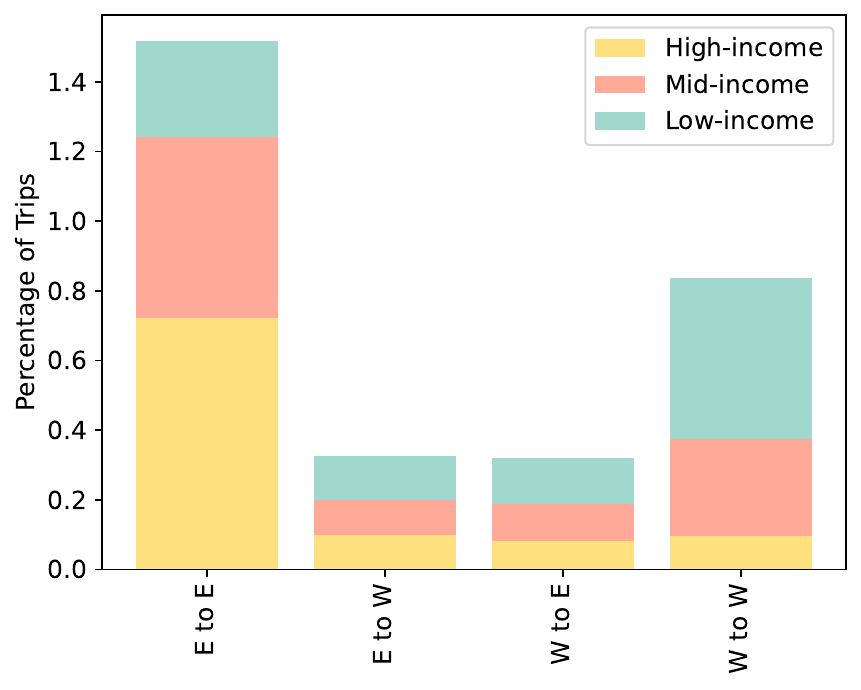}
\end{subfigure}
\begin{subfigure}[t]{0.49\textwidth}
\centering
\caption{\footnotesize{\% of Trips From/To Area (Rel. to Area Total)}}\label{fig:percentagetrips_fromtoarea}
\includegraphics[width=\textwidth,bb=0 0 413 299]{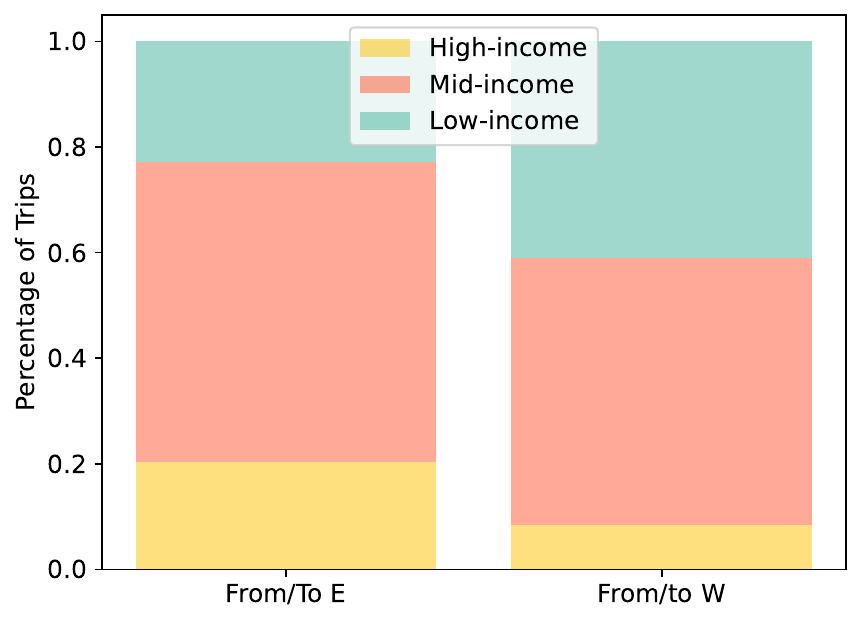}
\end{subfigure}
\end{figure}

\paragraph{Demand Data.} The anonymized data was provided to us by ClearRoad, which piloted a mobile application-centric approach to charge congestion prices. 
The data is organized into trips, each conducted by one of the pilot participants. Each trip is composed of a sequence of GPS points where each GPS point comes with the following information: Latitude/Longitude value, a time-stamp, and an estimate by ClearRoad of whether the user is currently in a private vehicle (as opposed to walking, cycling, or public transportation). In this work, we use data collected over 139 days on 204 users, starting from February 24, 2022, to July 13, 2022. 
Because we are analyzing the road network, we focus on the trips that contain vehicle rides.  Out of 64287 trips, there are only 38553 with vehicles. However, some of these trips with vehicles also contain a part where the user is not in a vehicle. Thus, we extract the tracked points during which users are estimated to be in a private vehicle.

\revised{We used the trips' GPS data to construct the demand matrix of OD pairs as follows: For each trip that crosses the network area in Figure~\ref{fig:network}, we trimmed the trip to those GPS points that lie in the area. We consider the (updated) starting and ending GPS points as the origin and destination, respectively. Since these may not perfectly correspond to nodes in our graph (Figure~\ref{fig:network}), we project them to the nearest nodes in the network. This resulted in 1495 trips of high-income demand, 5224 trips of mid-income demand, and 2810 trips of low-income demand.}

\revised{To implement our \emph{area pricing} scheme, we partition the network area with a simple two-sided grid, where the nodes on the left belong to W and the nodes on the right belong to E.
Given this partition, we are able to analyze the demand more closely as shown in Figure~\ref{fig:instance_demand_data}. First, in Figure~\ref{fig:percentagetrips_areatoarea}, we show the percentage of trips between given areas relative to the total number of trips of a stratum (i.e., the sum of the bars of the same color is 1). We observe that most of the demand is concentrated on trips inside E. For example, E receives about 20\% of the total low-income demand,~40\% of the total mid-income demand, and~70\% of the total high-income demand. 
On the other hand, trips inside W receive more than 40\% of the total low-income demand,~25\% of the total mid-income demand and less than 10\% of the total high-income demand. 
We observe that these demand trips are correlated with the stratification of the city (Figure~\ref{fig:stratification}) 
since low-income strata lives in W and high-income population lives in the E area.
Notably, trips that start or end in W receive almost no high-income demand and most of the demand comes from low-income strata which, again, may be due to the segregation in the city. 
The rest of the demand is minor compared to the ones mentioned before. Second, in Figure~\ref{fig:percentagetrips_fromtoarea}, we show the percentage of trips that enter/exit a given area, relative to the total of that area (i.e., each bar sums to 1). We clearly see the differences between the demand of low-income and high-income in E and W. We also observe that most of the demand comes from mid-income as shown in the details in Table~\ref{tab:instance}. We remark that further dividing the city in more sub-regions was ineffective in terms of pricing since most of the demand concentrates in the east. Moreover, we prioritize lower-dimensional pricing policies for tractability purposes.}

\subsection{Parameters Tested}\label{sec:parameter_tested}
In this section, we briefly present each of the parameters' and inputs' values that we tested.

\paragraph{Parameters in the Time Cost Function.} {\color{black} Parameters $\nu_a$ and $\gamma_a$ in Equation~\eqref{eq:time_cost_function} significantly affect the convergence of our method, so we consider the values that led to the most stable results: $\gamma_a = 0.02$\footnote{We stress that this value is not small enough to make travel times constant.} and travel time scaling exponent $\nu_a = 2$ (used also in \citet{baillon2008markovian}, preliminary results also showed no differences when using $\nu_a =4$). Also, we computed the travel time of an empty arc $t_{a}^{0}$ by using the maximum speed and the length of the arc (both pieces of information can be obtained using the osmnx library). Finally, since our data spans multiple days and not specific rush hours, we chose to set the arcs' capacities $b_a$ as follows: we multiply the number of lanes by the length (distance in kilometers) of the arc and divide by the average size of a car. 
We remark that we selected the parameters above for convergence purposes since the granularity of our dataset does not allow us to estimate each of them, in contrast of, e.g., \citet{maheshwari2024congestion} who consider a significantly smaller network with aggregated trips over different larger areas.

\paragraph{Cost Sensitivity and Willingness-to-pay.}
We assume that all strata are equally sensitive to time; specifically, we set $\beta^{s}=1$ for all $s\in\S$. However, we assume that strata differ in their willingness-to-pay -- that high-income strata are less sensitive to higher costs. We set these parameters to $\theta^{\text{h}}=2$ for high-income, $\theta^{\text{m}}=10/7$ for mid-income and $\theta^{\text{l}}=1$ for low-income. In other words, this means that the higher-income population is willing to pay twice more for using an arc than lower-income agents. Our framework is flexible to this choice. 

\paragraph{Outside Option Parameters.} We computed the cost of the outside option $\mathring{c}_{i,d}$ as follows: For every OD pair $(i,d)$ we compute the lowest total travel time between $i$ and $d$ when the roads are empty (i.e., no flow) multiplied by a factor of 3 and we set the monetary cost of the outside option as $500$ (e.g., the cost of the public transit ticket is $500$).\footnote{These values were chosen to make the outside option attractive relative to driving.} We then transform this monetary cost in time units using a  rate of 1 time/money, i.e., $500$ units of money corresponds to $500$ units of time.\footnote{We chose the same rate for all strata as our goal is to study the impact of congestion pricing rather than the interactions between choosing the outside option or driving. The framework is flexible to this choice, and results are qualitatively robust: alternative choices would simply scale the value of the outside option for each stratum.}
Finally, we set the sensitivity to cost of the outside option as follows: $\mathring{\beta}^{s}=1.2$ for $s=\text{high-income}$, $\mathring{\beta}^{s}=1.1$ for $s=\text{mid-income}$ and $\mathring{\beta}^{s}=1$ for $s=\text{low-income}$. Note that the closer to zero this parameter is, the less likely that stratum is to start a trip by car, according to probabilities in Equation~\eqref{eq:outside_option_probability}; thus, overall, the low-income group is assumed to be more likely to take the outside option (public transit). 

\begin{remark}
We performed different robustness analyses (Appendix~\ref{app:robustness_analysis}) to test the impact of: (i) willingness to pay of the strata, (ii) outside option parameters, (iii) Bogot\'a at different scales and (iv) demand perturbations. Perhaps surprisingly, the only parameters that matter are the differences in willingness to pay. We also confirm that our equilibrium flows lead to ``realistic'' path choices, namely that they do not result into a random walk-like behavior (Appendix~\ref{app:equilibrium_analysis}).
\end{remark} }

\subsection{Method Overview}\label{sec:method_overview}
In the following, we provide details on how we implemented our Algorithm~\ref{alg:mte_main}, the acceleration techniques that we used and, finally, how we run our simulations using the outputs of our method.
\paragraph{Algorithm~\ref{alg:mte_main} Setup.} Let us briefly summarize how the experiments were performed. We run Algorithm~\ref{alg:mte_main} with step size $\alpha_k = \max\{0.125,1/(k+1)\}$ as in \citet{baillon2008markovian} where $k$ is the iteration number of the outer-loop. The fixed-point iteration method was implemented with a tolerance of $10^{-1}$ or for a maximum of 1000 iterations, whichever is first. We set the maximum number of iterations of the descent method to 10 and a tolerance of $10^1$, whichever comes first. Robustness tests showed no significant change in the flow vector or transition and starting probabilities, when the tolerance was decreased or the maximum number of iterations increased. \revised{For uniform pricing, we tested a rate of $r\in[0,2500]$ with steps of 100, which gives us 26 options. For personalized pricing, we consider low-income rate $r_{\textsf{l}}$, mid-income rate $r_{\textsf{m}}$, high-income rate $r_{\textsf{h}}\in[0,2500]$ with steps of 100 and a restriction of $r_{\textsf{l}}\leq r_{\textsf{m}}\leq r_{\textsf{h}}$, which gives us roughly 3250 possible combinations. Note that we do not test other ordering of rates as they would result in worse welfare for lower-income strata; in practice, it would be impractical and unfair to charge a higher price to lower-income strata.
Finally, for area pricing, we consider $r_{\textsf{W}},r_{\textsf{E}}\in[0,2500]$  with steps of 100, resulting in about 650 price combinations.}

\paragraph{Acceleration Tools.} There are several parts in Algorithm~\ref{alg:mte_main} that are computationally expensive. First, the logit choice model is known to be numerically unstable, in particular, when one of the arguments $z_{a,d}^s$ is much higher or lower than the rest. For this, we use the \texttt{logsumexp} function in the scipy library which successfully deals with numerical unstability. This function was implemented in such a way that, for a given $s\in\S$ and $d\in\D^s$, we can efficiently compute $\varphi_{i,d}^s$ (in Step 5) and $\mathbf{P}^{s,d}$ (in Step 6) for all $i\neq d$. Second, note that the for-loop in Step 4 only needs to go over those $s\in\S$ and $d\in\D^s$ such that $g_{i,d}^s>0$. Third, there are two main computationally expensive parts in the algorithm: (i) the fixed-point iteration in Step 5 and (ii) solving the linear system in Step 8. For (i), we significantly reduce the number of iterations in \eqref{eq:fixed_point_iteration} by setting an appropriate initial point $\tau_{i,d}^{s,0}$: Using the arc-flows $f_a$'s obtained in the previous outer-loop iteration, we set $\tau_{i,d}^{s,0}$ as the lowest travel cost between $i$ and $d$ where each arc $a$ has a cost $\ell_a(f_a)+(1/\theta^s)\cdot p_a^s$. One can na\"ively consider 0 as an initial point, however, this requires  more iterations to converge. For (ii), we take advantage of the sparsity of the matrices, thus we use the corresponding scipy functions to solve sparse linear systems. Fourth, observe that the computations needed for each $s\in\S$ and $d\in\D^s$ in the for-loop (Step 4) are independent, therefore we can parallelize this section of the algorithm using multiple cores in a single machine. Considering all the above, for a given set of prices, we run Algorithm~\ref{alg:mte_main} using an average of 20 CPUs with each CPU having 8GB of RAM. Depending on the machine and on the pricing scheme, calculating the equilibrium flows for a given set of prices takes between 30 min and 3 hours. We ran all the combinations of prices in several machines on a cluster. For example, running all of the 3250 price combinations for personalized pricing took about 3-4 days.\footnote{To note the challenge of personalized pricing and of choosing a finer grid, for each price combination, several matrices (such as average travel time per OD pair) are stored. Then, the outputs of all of the price combinations used around 300-400GB of storage.} 

\paragraph{Simulation Setup to Calculate Metrics.} After equilibria flows were obtained from Algorithm~\ref{alg:mte_main}, we then ran simulations to mimic individual-level routing decisions. For each pricing scheme (and its corresponding flow equilibrium vector, transition probabilities, and starting probabilities) we run 10 simulations for each unit of demand in each OD pair to compute our \emph{evaluation metrics} such as the proportion of trips started, proportion of the flow using primary roads, revenue raised, travel time, and welfare. \revised{Note that some of these metrics can be computed directly using the flow vector without the need of simulations (e.g., total revenue collected).}

%% file: 5_Results.tex

\section{Case study: Results for Congestion pricing in Bogota}\label{sec:results}

We now apply our algorithmic framework to the design of congestion pricing in Bogot\'a, using the data and functional parameters as specified in the previous section.

\paragraph{Summary of the results.}
Our results can be summarized as follows: (a) uniform pricing is highly inequitable and inefficient (raising the least revenue); (b) personalized pricing can be highly equitable while raising the most revenue; (c) area pricing interpolates: it can raise a similar (though lesser) amount of revenue than personalized pricing \textit{and} can be far more equitable than uniform pricing (though not as equitable as personalized pricing), with relatively similar total welfare than personalized pricing. 

\begin{figure}[tbh]
\caption{The effects of Uniform Pricing, with each metric calculated under the equilibrium for the associated pricing. Overall, uniform pricing has highly inequitable effects, disproportionately harming the low-income group that is most price-sensitive. This group starts fewer trips, uses the primary roads less, and has a lower average speed. Uniform pricing also severely impacts the low-income and mid-income groups in terms of welfare. We remark that we do not plot the welfare at price 0 in (e) because we obtained a value of 0.}\label{fig:metrics_uniform_pricing}
\begin{subfigure}[t]{.51\textwidth}
    \centering
\caption{\footnotesize Contribution to and Total Revenue}\label{fig:revenue_uniform_pricing}
\includegraphics[scale=0.49,bb=0 0 447 314]{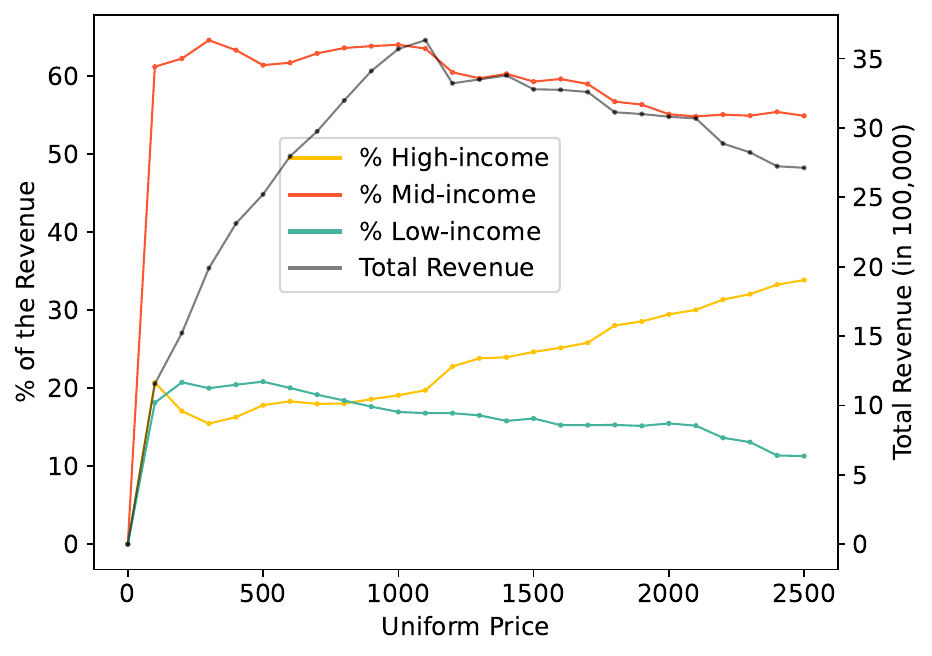}
\end{subfigure}%
\begin{subfigure}[t]{.48\textwidth}
    \centering
\caption{\footnotesize Proportion of Trips Started}\label{fig:proptripstarted_uniform_pricing}
\includegraphics[scale=0.49,bb=0 0 417 314]{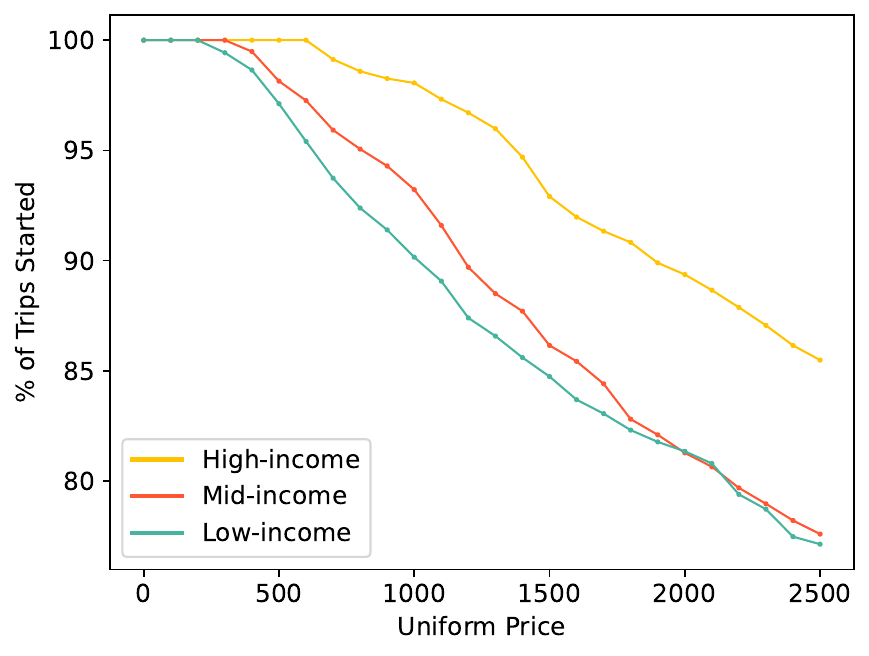}
\end{subfigure}
\begin{subfigure}[t]{.51\textwidth}
    \centering
\caption{\footnotesize Percentage of Primary Flow }\label{fig:percflowtypes_uniform_pricing}
\includegraphics[scale=0.49,bb=0 0 447 318]{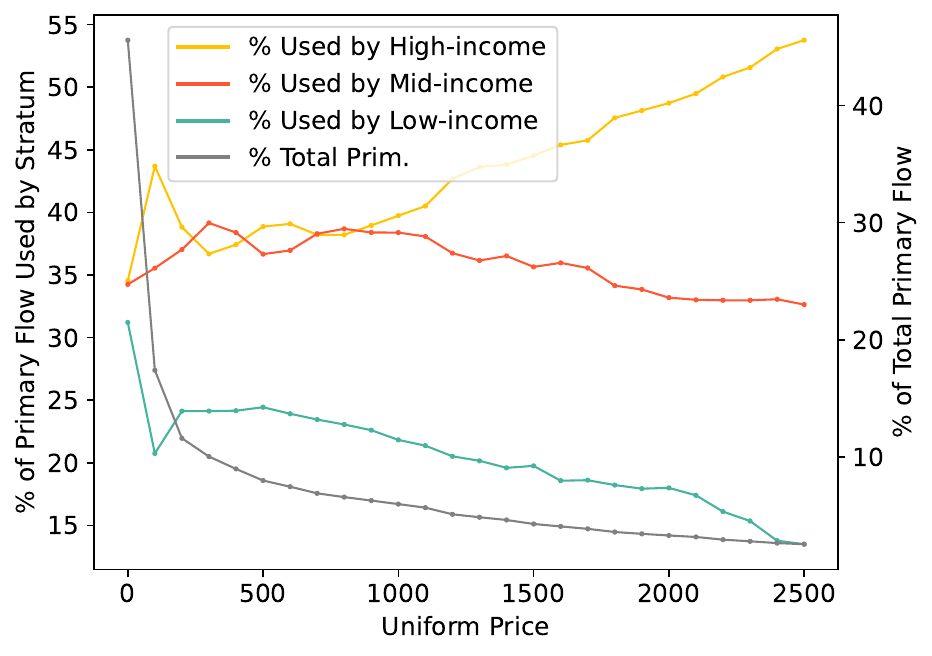}
\end{subfigure}
\begin{subfigure}[t]{.48\textwidth}
\centering
\caption{\footnotesize Average Speed (across trips started)}
\label{fig:speedvariation_uniformpricing}
\includegraphics[scale=0.49,bb=0 0 409 314]{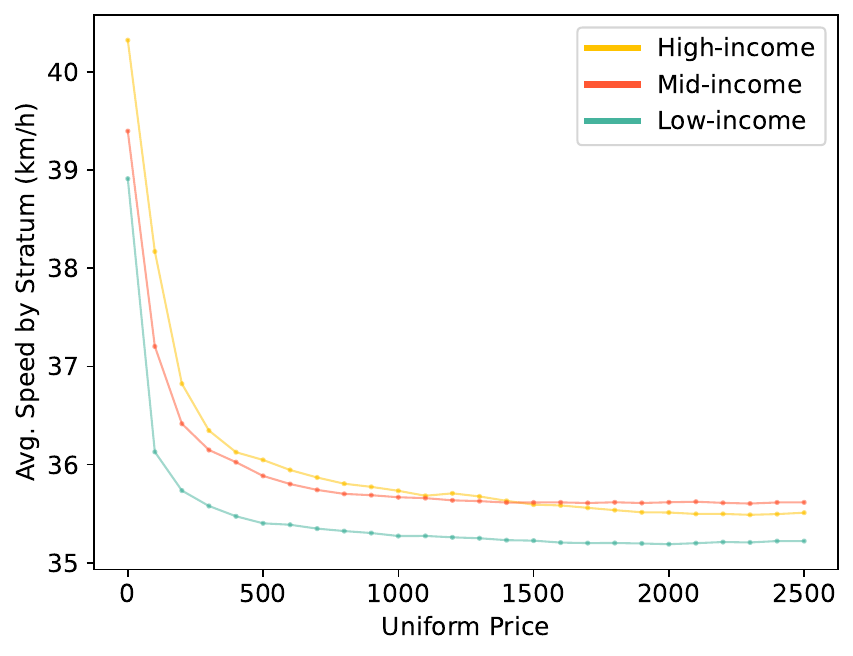}
\end{subfigure}
\centerline{\begin{subfigure}[t]{.48\textwidth}
\centering
\caption{\footnotesize Average Welfare}
\label{fig:welfare_uniformpricing}
\includegraphics[scale=0.49,bb=0 0 427 314]{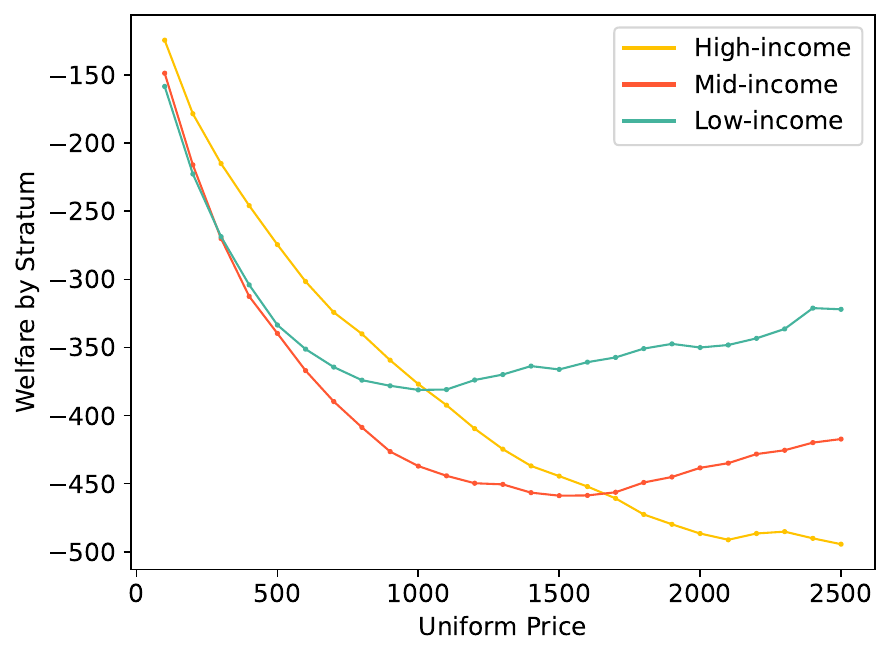}
\end{subfigure}}
\end{figure}

\subsection{The Inequitable Effects of Uniform Pricing}
We first study uniform pricing, with various metrics shown in \Cref{fig:metrics_uniform_pricing}, split by group.  
As shown in \Cref{fig:metrics_uniform_pricing}, for \textit{every} choice of the uniform price that we evaluate (in the range [0, 2500]), the outcomes are highly inequitable: the low-income group starts a smaller fraction of trips (instead choosing their outside option), a higher fraction of their flows use secondary instead of primary roads, and their average welfare and speed are lower. 

Several other facts are apparent from \Cref{fig:metrics_uniform_pricing}. At the revenue optimal price (around 1100), most trips still occur (approximately 90\% of the trips) -- thus, the revenue optimal price is ineffective at reducing congestion. This choice further leads to approximately the \textit{lowest} average speed: prices are high enough that commuters (especially low-income) opt to use secondary roads instead of the tolled primary roads, but not so high that they opt to instead use their outside option. At lower prices, more commuters use primary roads; at higher prices, fewer use primary roads but the proportion of trips started also decreases, and so the average speed on both primary and secondary stabilizes. Thus, welfare is also low at revenue-optimal uniform pricing, with few congestion benefits. The welfare of the high-income group decreases substantially at higher prices, not due to travel times but rather higher monetary costs. The welfare of the other strata recovers for higher prices, because they are using secondary roads.

More broadly, the uniform pricing results reflect that our algorithmic approach produces \textit{plausible} outputs, and can capture non-trivial phenomena -- such as the fact that traffic flows, average speeds, and revenue (an equilibria object that depends on prices and the decisions of other agents) are highly non-monotonic. 

\begin{table}[htpb]
\centering
\caption{Revenue Optimal Prices and their effects for each pricing scheme.}
\label{tab:revenue_opt_prices}
\begin{tabular}{lccc}
\toprule
Pricing Scheme & Revenue Opt. Price & Revenue &Low-income Welfare  \\
\midrule
Uniform $p$ &  1100 & 3,632,207 & -380\\
Personalized $(p_{\textsf{h}},p_{\textsf{m}},p_{\textsf{l}})$ & (2100,1100,1100) & 3,836,924 & -380\\
Area $(p_{\textsf{E}},p_{\textsf{W}})$ &(1000,1300) &3,667,911 &-377\\
\bottomrule
\end{tabular}
\end{table}

\subsection{Comparing Uniform to Personalized and Area Pricing}

\begin{figure}[tb]
\caption{Pareto frontiers for Welfare for a given stratum vs Total Welfare, for different pricing schemes. Shapes denote the stratum for which welfare is being plotted on the Y axis (low, medium, or high), colors denote the pricing scheme, and numbers indicate the price vector. We plot the Pareto curves for each scheme and stratum. For example, there is only one Pareto optimal uniform price ($p = 200$) for each pair of Total Welfare and Welfare for mid-income and high-income strata; for low-income stratum the Pareto optimal points are $p=1600$ and $p=200$. At $p=200$, welfare for the high-income group is higher than the welfare for the other groups. As another example, the optimal per-stratum pricing scheme (for each stratum), denoted in green, can yield more Total Welfare and per-stratum Welfare than the uniform pricing. The welfare achieved by area pricing is comparable to that of personalized pricing. Note that we omit uniformly 0 prices as those would generally dominate all other pricing (giving 0 welfare for each group), while providing no revenue.}\label{fig:w0_wt_maintext}
    \centering
\includegraphics[width=.6\textwidth,bb=0 0 430 381]{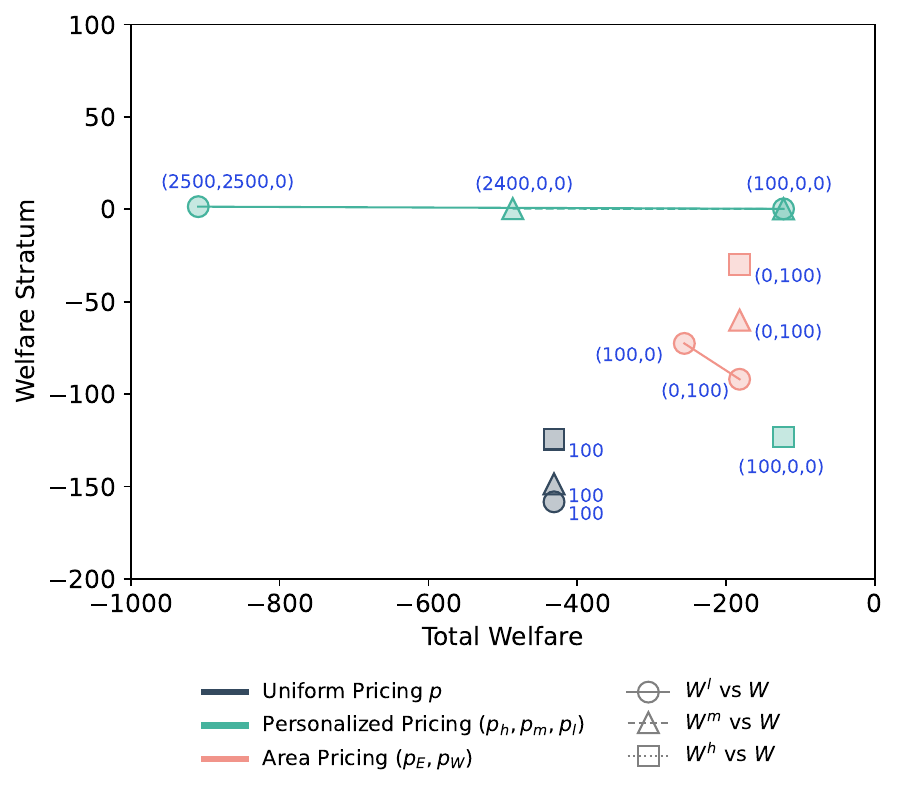}
\end{figure}


\begin{table}[htpb]
\centering
\caption{Metrics under the prices that are optimal for Low-Income Stratum's Welfare, for each pricing scheme.}
\label{tab:welfarelowincome_opt_prprices}
\begin{tabular}{lccc}
\toprule
Pricing Scheme & Low-income Opt. Price & Revenue & Low-income Welfare \\
\midrule
Uniform $p$ &  100 & 1,155,838 & -158 \\
Personalized $(p_{\textsf{h}},p_{\textsf{m}},p_{\textsf{l}})$ & (2500,2500, 0) & 2,407,866 & 1 \\
Area $(p_{\textsf{E}},p_{\textsf{W}})$ & (100,0) & 709,237 & -72 \\
\bottomrule
\end{tabular}
\end{table}
%

\begin{figure}[tb]
\caption{Pareto frontiers for Welfare for a given stratum vs Total Revenue, for different pricing schemes. Colors denote the pricing scheme. We plot the Pareto curves for each scheme and stratum. For example, for the low-income stratum, personalized pricing (in green) Pareto dominates area pricing (in red). We remark that in (c), the blue curve is over the green one, which is technically not possible, however, this is due to the price grid for personalized pricing that we considered, and linear interpolation between the analyzed points.}\label{fig:welfare_vs_revenue}
\begin{subfigure}[t]{.51\textwidth}
    \centering
\caption{\footnotesize Low-income Stratum}\label{fig:lowincome_welfare_revenue}
\includegraphics[scale=0.49,bb=0 0 427 314]{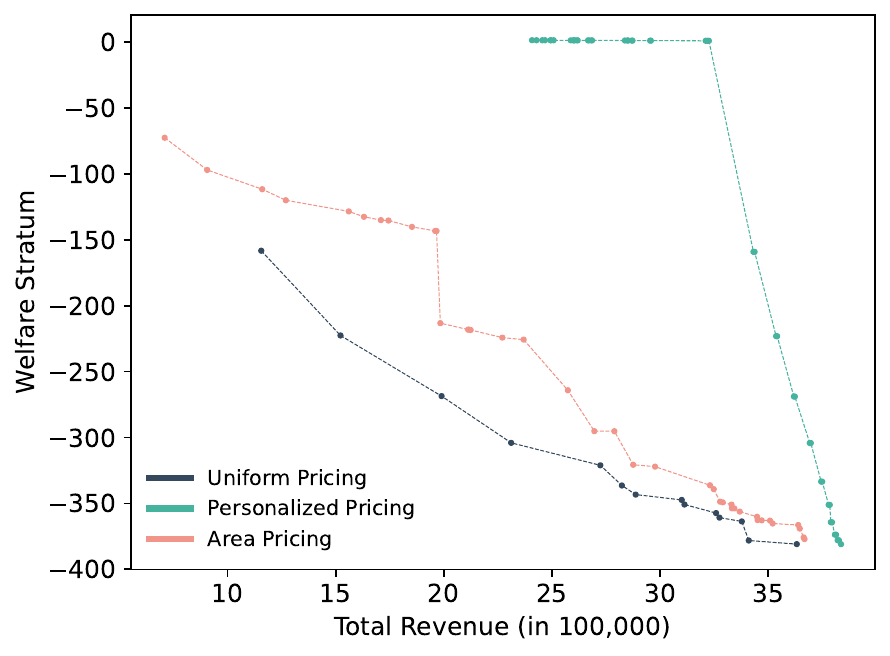}
\end{subfigure}%
\begin{subfigure}[t]{.48\textwidth}
    \centering
\caption{\footnotesize Mid-income Stratum}\label{fig:midincome_welfare_revenue}
\includegraphics[scale=0.49,bb=0 0 427 314]{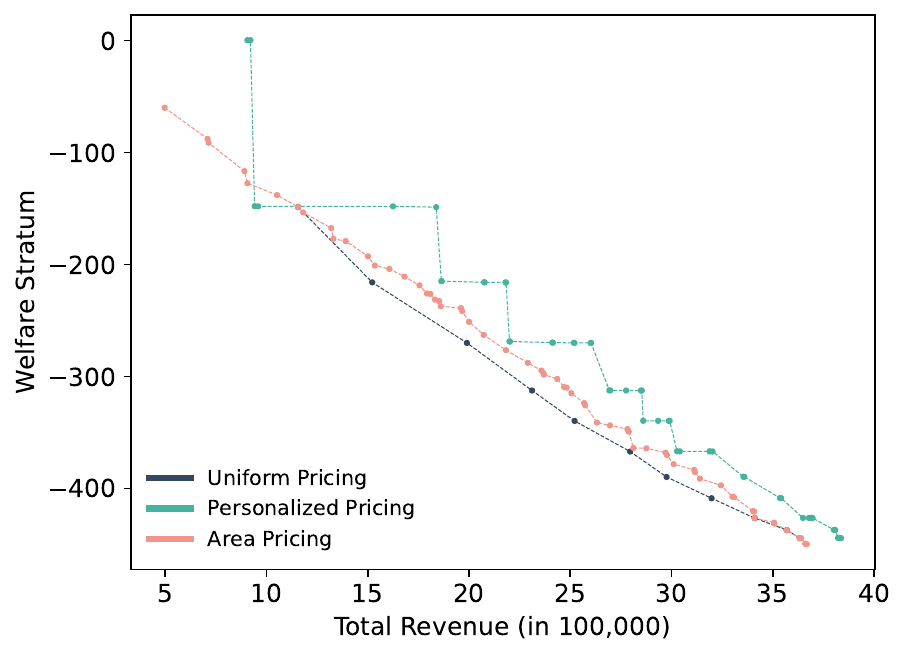}
\end{subfigure}
\centerline{\begin{subfigure}[t]{.51\textwidth}
    \centering
\caption{\footnotesize High-income Stratum }\label{fig:highincome_welfare_revenue}
\includegraphics[scale=0.49,bb=0 0 427 314]{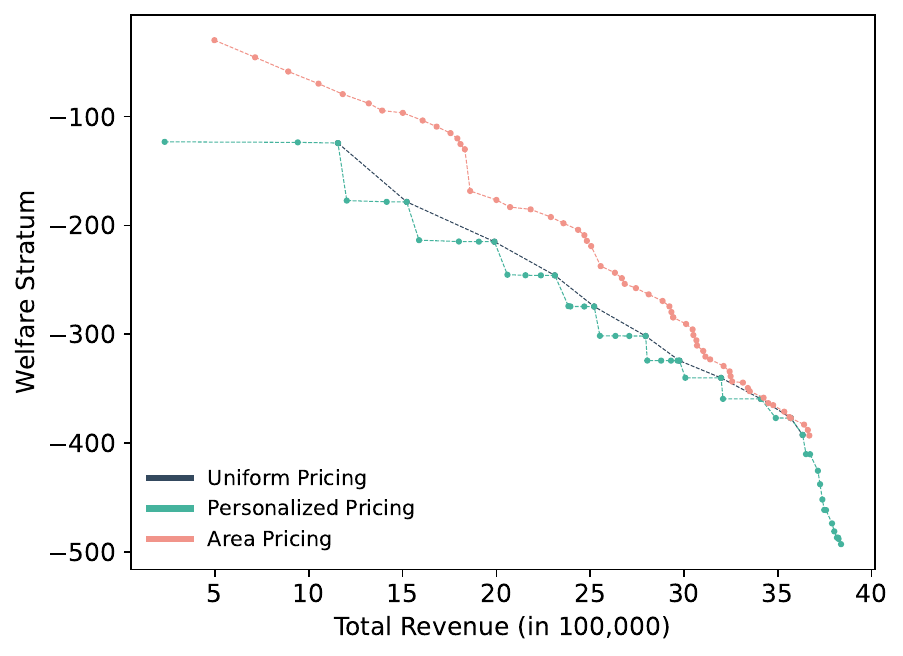}
\end{subfigure}}
\end{figure}

Next, we compare uniform pricing to personalized and area pricing. As these prices are multi-dimensional (one price per stratum or area), we cannot plot curves as in \Cref{fig:metrics_uniform_pricing}. Instead, we compare them by plotting Pareto curves  -- for each stratum, what is the Pareto curve 
in terms of a single stratum's welfare and total welfare achievable by each pricing scheme (\Cref{fig:w0_wt_maintext}). \revised{We remark that, in these curves, we did not include the welfare for price $\mathbf{0}$ since it leads to zero welfare -- which would Pareto-dominate most other points in terms of welfare, while providing no revenue.}
We also compare these schemes by plotting Pareto curves for stratum's welfare versus total revenue collected (Figure~\ref{fig:welfare_vs_revenue}); we further compare the schemes by the highest revenue that they can achieve (\Cref{tab:revenue_opt_prices}) and the welfare they provide to the low-income group (\Cref{tab:welfarelowincome_opt_prprices}). We now discuss these results in turn. 

\Cref{fig:w0_wt_maintext} illustrates the Pareto curves achievable by each scheme, in terms of welfare for a given stratum versus total welfare. \revised{The x-axis corresponds to the total welfare and the y-axis shows the stratum-dependent welfare. Each color represents different pricing schemes and each shape represents different stratum-dependent welfare versus total welfare comparisons. 
Each point in Figure~\ref{fig:w0_wt_maintext} is a Pareto point within their corresponding legend, i.e., it Pareto-dominates other points in that comparison (which we did not plot).
For example, the green circle curve is the Pareto curve of welfare for the low-income group versus total welfare, with the highest welfare point for the low-income stratum being at the price point (2500, 2500, 0), i.e., the other strata are charged maximal prices. In other words, this green circle dominates other points resulting from different personalized pricing combinations.  Several key insights emerge from these Pareto curves:
\begin{itemize}
    \item As expected, personalized pricing can achieve the optimal per-strata welfare. In particular, with personalized pricing, we can reverse the effects of uniform pricing: we can achieve higher welfare for the low- and mid-income groups, at considerable welfare loss for the high-income group. (Compare the green Pareto curves to the black shapes; notably, only one Uniform price ($r = 100$) is Pareto optimal among Uniform prices; recall that we did not plot the results for $r=0$).
    \item Perhaps surprisingly, area pricing schemes yield a relatively high total welfare -- though they cannot achieve as high of per-stratum welfare as an appropriate personalized pricing can. (Compare the red curves to the green curves, and note that the latter are above and to the right of the red curves).
    \item The area pricing scheme Pareto dominates uniform pricing: Note that the schemes $(0,100)$ and $(100,0)$, indicating just pricing for roads in the W area or E area, have higher total welfare and per-stratum welfare for each stratum than does the Pareto optimal uniform price $r = 100$. (Note that the red curves are above and to the right of the corresponding black curve of the same shape).
    \item Neither area pricing nor personalized pricing dominate each other. Personalized pricing can achieve higher welfare for low- and mid-income strata, at the cost of (much) lower total and high-income welfare. (Neither green or red set of curves is above and to the right of the other set).
    \item The Pareto optimal schemes for area pricing are given by the set of prices $(100,0)$ and $(0,100)$ which means that either E or W area get priced at the lowest rate. For the low-income welfare, point $(100,0)$ leads to slightly improved per-stratum welfare.
\end{itemize}}

\revised{To further compare our pricing schemes, Figure~\ref{fig:welfare_vs_revenue} contains  Pareto curves for stratum welfare versus total revenue, with analogous insights as before. For the low-income stratum, in Figure~\ref{fig:welfare_vs_revenue} (a), personalized pricing (in green) Pareto dominates uniform pricing (in blue) and area pricing (in red). One interesting result here is that the green curve (personalized pricing) suddenly jumps to zero welfare since those points represent the case when $r^s=0$ for $s=$low-income, which might be less realistic from a policy standpoint. Instead, area pricing has a wider range of options which is reflected by how the Pareto curve (in red) changes more smoothly for different pricing points. This larger set of pricing options may be more attractive to policymakers. Note as well, in Figure~\ref{fig:welfare_vs_revenue} (a), that area pricing leads to higher welfare and revenue than uniform pricing. Surprisingly, for the mid-income group in Figure~\ref{fig:welfare_vs_revenue} (b), there is no dominance between area pricing and personalized pricing, and in fact personalized pricing is closer to uniform pricing. Finally, for the high-income stratum in Figure~\ref{fig:welfare_vs_revenue} (c), there is no difference between personalized pricing and uniform pricing, while area pricing almost dominates them.}


Next, we consider the prices that optimize revenue (\Cref{tab:revenue_opt_prices}) and the prices that optimize low-income welfare under each pricing scheme (\Cref{tab:welfarelowincome_opt_prprices}). These results tell a similar story as the welfare and revenue Pareto curves: Uniform pricing is dominated by what the other schemes \textit{can} achieve. However, each given choice of personalized or area prices chooses a different point of the tradeoff curve. In the Appendix, Figures~\ref{fig:metrics_pricing_per_stratum} and \ref{fig:metrics_pricing_per_area}, we show more evidence of this tradeoff. In particular, we plot the percentage of trips started versus revenue and primary flow usage for personalized and area pricing. Overall, we observe an analogous pattern than uniform pricing, however, these pricing schemes provide more flexibility to balance revenue, primary flow usage and percentage of trips started.

\vspace{0.5em}
\noindent Putting things together, our results suggest that either personalized or area pricing -- depending on relative implementation feasibility -- are promising strategies for Bogot\'a. Area pricing (in particular for the E area) can achieve higher per-stratum (for each stratum) welfare, total welfare, and revenue than uniform pricing. 
Personalized pricing can achieve higher welfare for low- and mid-income groups, at a substantial cost to the high-income group. In Appendix~\ref{sec:synthetic}, we study through synthetic experiments how the topology of the network and the segregation in the demand affects the benefits of personalized and area pricing. 

%% file: 7_Conclusions.tex

\section{Conclusions}\label{sec:conclusions}
In this work, we study equitable congestion pricing schemes under the Markovian traffic equilibrium model. An essential aspect of this model is its ability to capture the variability in route costs perceptions. We extended this model in several directions: (i) We allow for multiple users' socioeconomic types, (ii) we introduce a generalized cost function that accounts for travel time and pricing costs, and (iii) we include an outside option choice at the start of each trip. In this setting, we show that there exists a unique equilibrium. We test our methodology with a dataset from Bogot\'a provided by our industry partner. Our empirical results have practical insights. First, uniform pricing negatively impacts every single metric, except revenue, especially inequitably. Second, personalized pricing can achieve higher low-income and mid-income welfare, at the expense of high-income welfare. Finally, area pricing appropriately interpolates both of the previous pricing schemes. 

Several research directions remain to be explored. First, it would be interesting to analyze other pricing cost functions, for example, not dependent on distance. Second, other pricing schemes that could potentially interpolate between personalized pricing and uniform pricing are paying per destination. 
Third, as we observed in our empirical results, congestion pricing impacts everyone's welfare, potentially inequitably, further suggesting the need for implementing revenue refunding schemes or using the funding to improve, for example, public transportation. Finally, from a theoretical standpoint, the question of how to globally solve the price optimization Problems \eqref{eq:optimization_problem} and \eqref{eq:optimization_problem_revenue} is an interesting direction for future research.


%% file: 999_Appendix_A.tex
\section{Appendix to Section~\ref{sec:existence_uniqueness}}\label{app:existence_uniqueness}
\paragraph{Notation.} In the following, for each stratum $s$ and destination $d\in\D^s$, we consider the following notation:
\begin{itemize}
\item Transition matrix $\mathbf{P}^{s,d}(\mathbf{z}^s_d) =(P_{i,j}^{s,d})_{i,j\neq d}$.
\item Auxiliary matrix $\mathbf{Q}^{s,d}(\mathbf{z}^s_d)= (Q_{i,a}^{s,d})_{i\in\N,a\in\A}$, where $Q_{i,a}^{s,d} = P_{i,j_a}^{s,d}$ if $i = i_a$ and zero otherwise.
\end{itemize} 
To ease the exposition, in the remainder we avoid the explicit dependency on $\mathbf{z}^s_d$ of the matrices above.

We now provide the technical lemmas to prove Theorem~\ref{thm:existence_uniqueness}.
First, note that $\mathcal{C}$ as defined in \eqref{eq:good_set} is open, convex (because functions $\varphi_{i,d}^s$ are concave) and upward closed, i.e., for each $\mathbf{t}\in\mathcal{C}$ then for any $\mathbf{t}'\geq\mathbf{t}$ component-wise we have $\mathbf{t}'\in\mathcal{C}$ (because functions $\varphi_{i,d}^s$ are component-wise nondecreasing). The first technical lemma we need is 
\begin{lemma}\label{lemma:technical}
Fix $s\in\S$ and $d\in\D^s$. Assume that $\tau_{\cdot,d}^s$ solves \eqref{eq:fixed_point_opt2_general} for a given $\mathbf{t}\in\mathcal{C}$ and let $z_{a,d}^s = t_a+\frac{\beta^{s,p}}{\beta^{s,t}}\cdot\kappa_a(p_a^s)+\tau^s_{j_a,d}$. Then,
\begin{enumerate}
\item For each $i\neq d$ there exists $j\in\N$ with $P_{i,j}^{s,d}>0$ and $\delta_{j,d}^s <\delta_{i,d}^s$, where $\delta_{i,d}^s =\tau_{i,d}^s - \hat{\tau}_{i,d}^s$ with  $\hat{\tau}_{\cdot,d}^s$ such that for all $i\neq d$
\[
\hat{\tau}_{i,d}^s < \varphi_{i,d}^s\left(\left(t_a+\frac{\beta^{s,p}}{\beta^{s,t}}\cdot\kappa_a(p_a^s)+\tau^s_{j_a,d}\right)_{a\in\A_i^+}\right).
\]
\item The matrix $\mathbf{I}-\mathbf{P}^{s,d}$ is invertible, with $\mathbf{I}$ being the identity matrix.
\item Equations~\eqref{eq:functional_flow_conservation_general} have a unique solution $\mathbf{v}^s_d = (\mathbf{Q}^{s,d})^{\top}\mathbf{x}^s_d$ with $\mathbf{x}_d^s = [\mathbf{I}-(\mathbf{P}^{s,d})^{\top}]^{-1}\cdot \mathbf{g}^s_{d}$.
\end{enumerate}
\end{lemma}
\proof{Proof.}
The proof is analogous to Lemma 2.1 and Proposition 2.2 (pages 225-226) in \citep{cominetti2012modern}, so we just make a few remarks on where there are differences.
\begin{enumerate}
\item The proof follows similarly by defining $\hat{z}^s_{a,d} = t_a +\frac{\beta^{s,p}}{\beta^{s,t}}\cdot\kappa_a(p_a^s)+\hat{\tau}^s_{j_a,d}$ and ${z}^s_{a,d} = t_a +\frac{\beta^{s,p}}{\beta^{s,t}}\cdot\kappa_a(p_a^s)+{\tau}^s_{j_a,d}$. Then, we use the concavity and smoothness of function $\varphi_{i,d}^s$ and note that $\hat{z}^s_{a,d} -{z}^s_{a,d} = \hat{\tau}^s_{j_a,d}-{\tau}^s_{j_a,d}$.
\end{enumerate}
The extra term $\frac{\beta^{s,p}}{\beta^{s,t}}\cdot\kappa_a(p_a^s)$ in $z_{a,d}^s$ does not affect the proofs of 2. and 3.\Halmos
\endproof

\begin{proposition}\label{prop:technical}
If $\mathbf{t}\in\mathcal{C}$ then for each stratum $s\in\S$ and destination $d\in\D^s$ the Equations~\eqref{eq:fixed_point_opt2_general} have unique solutions $\tau_{i,d}^s = \tau_{i,d}^s(\mathbf{t})$. Moreover, the maps $\mathbf{t}\mapsto\tau_{i,d}^s(\mathbf{t})$ are smooth, concave and component-wise nondecreasing.
\end{proposition}
Since the monetary costs do not depend on the flow of the arc, then the extra term $\frac{\beta^{s,p}}{\beta^{s,t}}\cdot\kappa_a(p_a^s)$ in $z_{a,d}^s$ does not affect the proof in Proposition 2.3 (page 226) in \citep{cominetti2012modern}, so we conclude Proposition~\ref{prop:technical} by using the results in Lemma~\ref{lemma:technical}. We finally remark that $\tau_{i,d}^s(\mathbf{t})\leq \bar{\tau}_{i,d}^s(\mathbf{t})$ where $\bar{\tau}_{i,d}^s(\mathbf{t})$ are the shortest travel times that satisfy 
\begin{equation}\label{eq:shortest_travel_time}
\bar{\tau}_{i,d}^s(\mathbf{t}) = \min\left\{t_a +\bar{\tau}_{j_a,d}^s(\mathbf{t}): \ a\in\A_i^+\right\},
\end{equation}
which can be easily computed with a shortest path algorithm where each arc's weight is $t_a$.

The main consequence of Proposition~\ref{prop:technical} is that Equations~\eqref{eq:fixed_point_opt2_general} and~\eqref{eq:functional_flow_conservation_general} define unique implicit functions $x_{i,d}^s = x_{i,d}^s(\mathbf{t})$, $v_{a,d}^s = v_{a,d}^s(\mathbf{t})$ and $\tau_{i,d}^s = \tau_{i,d}^s(\mathbf{t})$ which is crucial to Theorem~\ref{thm:existence_uniqueness}. We finish this section by presenting missing proofs.

\begin{lemma}\label{lemma:convex_coercive}
Function $\Phi$ defined in \eqref{eq:convex_program} is strictly convex and coercive in region $\{\mathbf{t}\geq \mathbf{t}^0\}\subseteq\mathcal{C}$.
\end{lemma}
\proof{Proof.}
The proof is analogous to the first part of the proof in Theorem 2.5 (page 228) in \citep{cominetti2012modern}.
\Halmos
\endproof

\begin{lemma}\label{lemma:implicit_derivative}
For all $s\in\S$, $d\in\D^s$ and $a\in\A$ we have that
\[
\sum_{ i\neq d}g_{i,d}^s\cdot\frac{\partial \tau_{i,d}^s}{\partial t_a}(\mathbf{t}) = v_{a,d}^s(\mathbf{t}).
\]
\end{lemma}
\proof{Proof.}
Now recall that $\tau_{i,d}^s$ satisfies Equation~\eqref{eq:fixed_point_opt2_general}, which can be represented as a function of $\mathbf{t}$ as follows:
\[
\tau^s_{i,d}(\mathbf{t}) = \varphi^s_{i,d}\left(\big(z_{a,d}^s(\mathbf{t})\big)_{a\in \A_i^+}\right),
\]
where $z_{a,d}^s(\mathbf{t}) = t_a+\frac{\beta^{s,p}}{\beta^{s,t}}\cdot\kappa_a(p_a^s)+\tau^s_{j_a,d}(\mathbf{t})$. In the following, we omit the notation for the explicit dependency on $\mathbf{t}$. Then, by taking the implicit derivative with respect to $t_a$, we obtain with the chain rule
\begin{align}\label{eq:system_eqs}
\frac{\partial \tau_{i,d}^s}{\partial t_a} &=\sum_{e\in \A_i^+} \frac{\partial \tau_{i,d}^s}{\partial z_{e,d}^s}\cdot \frac{\partial z_{e,d}^s}{\partial t_a}=\frac{\partial \tau_{i,d}^s}{\partial z_{a,d}^s}\cdot\left(1+ \frac{\partial \tau^s_{j_a,d}}{\partial t_a}\right) + \sum_{\substack{e\in \A_i^+,\\e\neq a}} \frac{\partial \tau_{i,d}^s}{\partial z_{e,d}^s}\cdot \frac{\partial \tau^s_{j_e,d}}{\partial t_a} \notag\\
&= \frac{\partial \tau_{i,d}^s}{\partial z_{a,d}^s} + \sum_{e\in \A_i^+} \frac{\partial \tau_{i,d}^s}{\partial z_{e,d}^s}\cdot \frac{\partial \tau^s_{j_e,d}}{\partial t_a} \notag\\
&= P^s_{i,j_a} + \sum_{e\in \A_i^+} P^s_{i,j_e}\cdot \frac{\partial \tau^s_{j_e,d}}{\partial t_a},
\end{align}
where in the second equality we used that $\frac{\partial z_{e,d}^s}{\partial t_a} = 1+\frac{\partial \tau^s_{j_a,d}}{\partial t_a}$ for $e=a$ and $\frac{\partial z_{e,d}^s}{\partial t_a} = \frac{\partial \tau^s_{j_a,d}}{\partial t_a}$ for all $e\in\A_i^+$ with $e\neq a$. In the last equality, we used that
  that the transition probability satisfies $P^s_{i,j_e}=\frac{\partial \tau_{i,d}^s}{\partial z_{e,d}^s}$ with $e = (i,j_e)$ for all $e\in\A_i^+$. Therefore, we can write~\eqref{eq:system_eqs} in vector form for $i\in\N$ such that $i\neq d$ as follows
\[
\frac{\partial \tau_{\cdot,d}^s}{\partial t_a} = \mathbf{Q}_{\cdot,a}^{s,d} + \mathbf{P}^{s,d}\cdot\frac{\partial \tau_{\cdot,d}^s}{\partial t_a},
\]
where $\mathbf{Q}_{\cdot,a}^{s,d} = (P_{i,j_a})_{i\neq d}$ and $\mathbf{P}^{s,d} = (P_{i,j})_{i,j\neq d}$. So, since $\mathbf{I}-\mathbf{P}^{s,d}$ is invertible (due to Lemma~\ref{lemma:technical}) with $\mathbf{I}$ being the identity matrix, then
\begin{equation}\label{eq:aux_proof}
\frac{\partial \tau_{\cdot,d}^s}{\partial t_a} = [\mathbf{I}-\mathbf{P}^{s,d}]^{-1}\cdot\mathbf{Q}_{\cdot,a}^{s,d}
\end{equation}
Using~\eqref{eq:aux_proof}, we can compute the following for a given destination $d\in\D^s$
\[
\sum_{i\neq d}\frac{\partial \tau_{i,d}^s}{\partial t_a}g_{i,d}^s = \left(\frac{\partial \tau_{\cdot,d}^s}{\partial t_a}\right)^{\top}\mathbf{g}_{\cdot,d}^s = \left(\mathbf{Q}_{\cdot,a}^{s,d}\right)^{\top}\cdot[\mathbf{I}-\left(\mathbf{P}^{s,d}\right)^{\top}]^{-1}\mathbf{g}_{\cdot,d}^s = \left(\mathbf{Q}_{\cdot,a}^{s,d}\right)^{\top}\mathbf{x}_{\cdot,d}^s = v_{a,d}^s,
\]
where the last two inequalities follow from the flow conservation constraints~\eqref{eq:functional_flow_conservation_general}.\Halmos
\endproof

%% file: 999_Appendix_B.tex

\section{Appendix to Section~\ref{sec:data_details}}\label{app:data_description}


\begin{figure}[htpb]
\captionof{figure}{Bogot\'a stratification, 2019. Map and legend taken from~\citep{bogota_stratification}.}
\label{fig:stratification}
\begin{minipage}[t]{.4\textwidth}
\centerline{
\includegraphics[scale=0.6,bb=0 0 134 243]{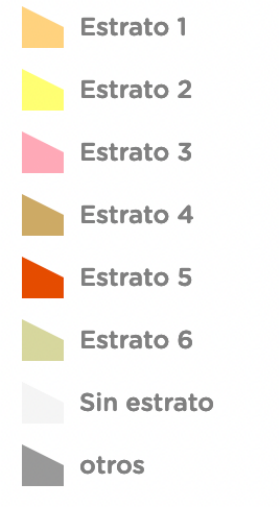}
}
\end{minipage}
\begin{minipage}[t]{.6\textwidth}
\centerline{
\includegraphics[scale=0.35,bb=0 0 514 728]{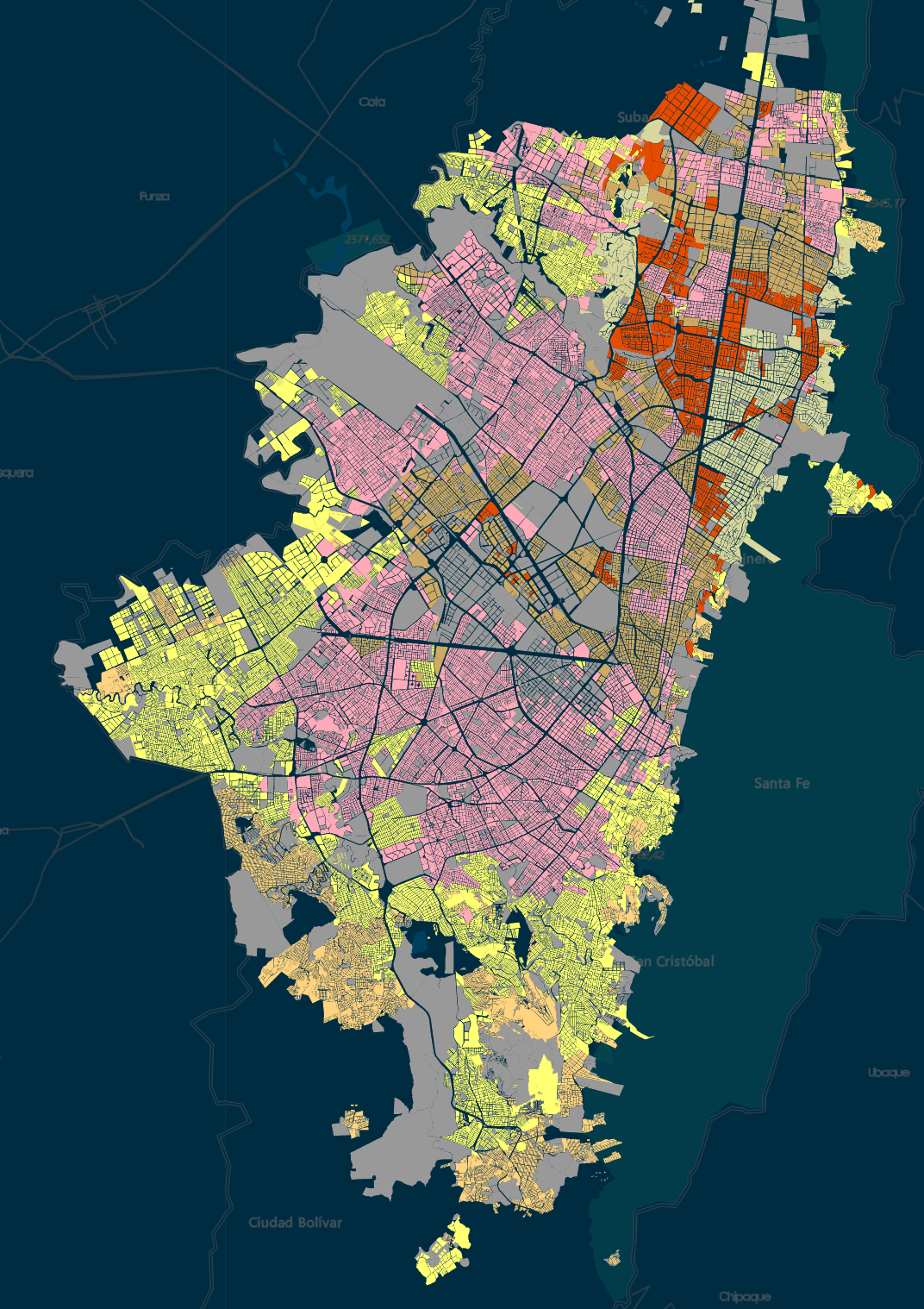}
}
\end{minipage}
\end{figure}

\clearpage

%% file: 999_Appendix_C.tex

\subsection{Additional Figures}

\begin{figure}[htpb]
\centering
\caption{In (a), we show revenue versus percentage of trips started (each point corresponds to a set of prices). In (b), we present the percentage of primary flow versus percentage of trips started. Overall we observe a similar pattern than for uniform pricing: For the low-income stratum, as the proportion of started trips decreases then the revenue contribution decreases since they use mostly secondary roads. The opposite pattern applies to the high-income stratum whose contribution to revenue increases. For primary flow, we observe that the percentage used by high-income stratum increases, while the rest decreases (as the percentage of trips started decreases).}\label{fig:metrics_pricing_per_stratum}
\begin{subfigure}[t]{.43\textwidth}
    \centering
\caption{\footnotesize Revenue vs. Prop. Trips Started}\label{fig:revenue_vs_proptripstarted_pricing_per_stratum}
\includegraphics[width=\textwidth,bb=0 0 417 314]{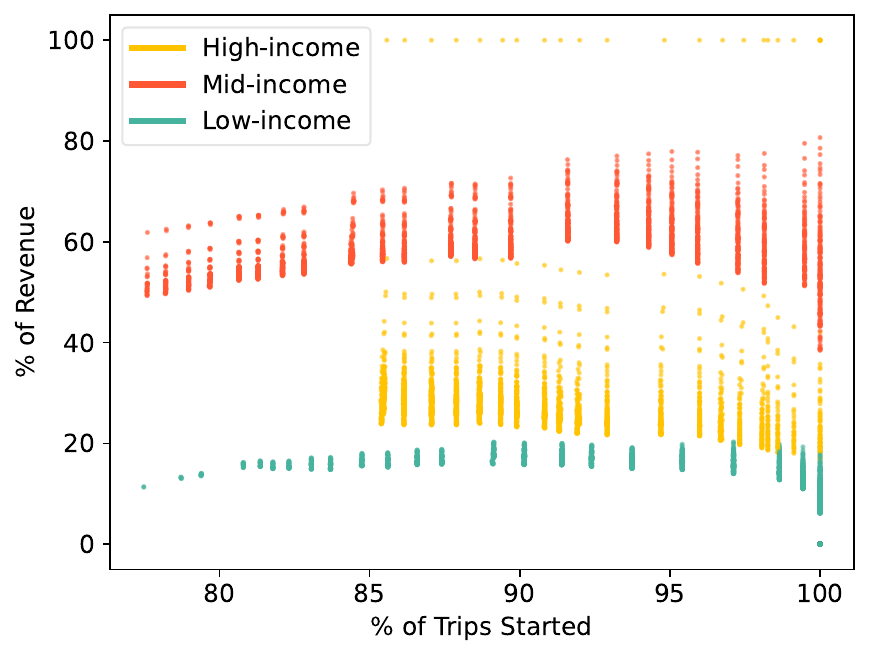}
\end{subfigure}%
\begin{subfigure}[t]{.43\textwidth}
    \centering
\caption{\footnotesize Prim. Flow vs. Prop. Trips Started}\label{fig:primflow_vs_proptripstarted_pricing_per_stratum}
\includegraphics[width=\textwidth,bb=0 0 409 314]{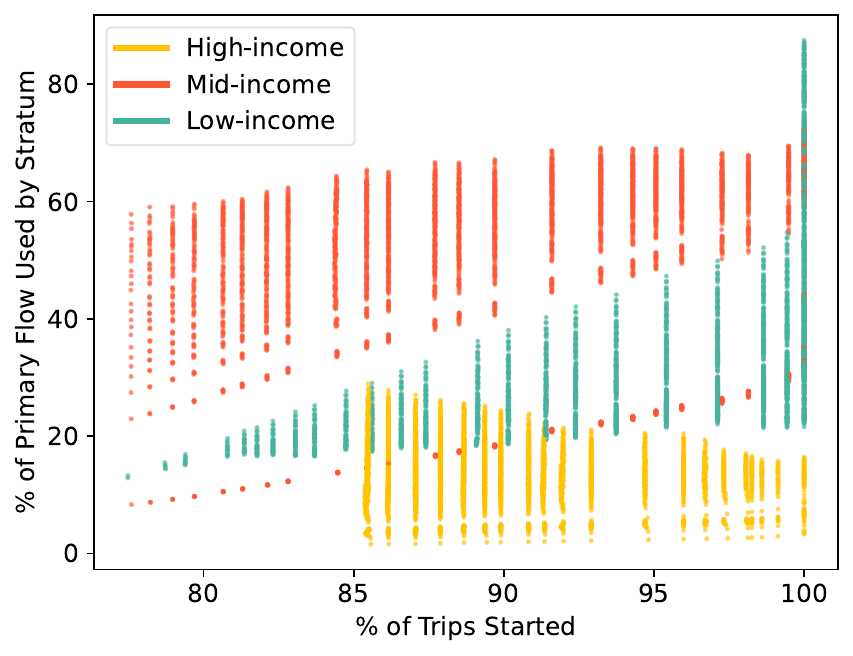}
\end{subfigure}
\end{figure}

\begin{figure}[htpb]
\centering
\caption{In (a), we show revenue versus percentage of trips started (each point corresponds to a set of prices). In (b), we present the percentage of primary flow versus percentage of trips started. The pattern in this plot is less apparent, but we can observe that most of the revenue collected comes from the mid-income stratum (regardless of the percentage of trips started). In terms of primary flow, we observe again that the high-income stratum tends to benefits.}\label{fig:metrics_pricing_per_area}
\begin{subfigure}[t]{.43\textwidth}
    \centering
\caption{\footnotesize Revenue vs. Prop. Trips Started}\label{fig:revenue_vs_proptripstarted_pricing_per_area}
\includegraphics[width=\textwidth,bb=0 0 409 314]{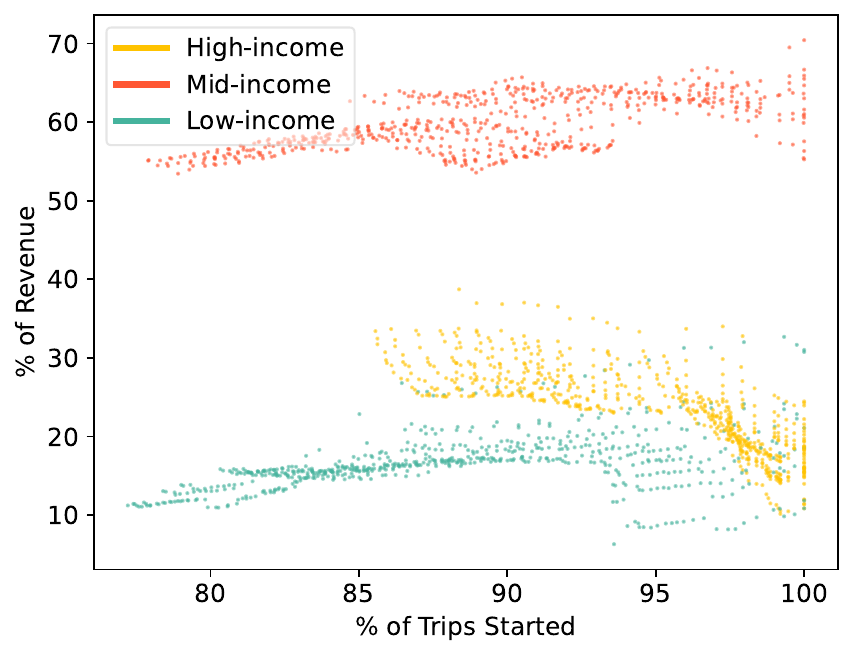}
\end{subfigure}%
\begin{subfigure}[t]{.43\textwidth}
    \centering
\caption{\footnotesize Prim. Flow vs. Prop. Trips Started}\label{fig:primflow_vs_proptripstarted_pricing_per_area}
\includegraphics[width=\textwidth,bb=0 0 409 314]{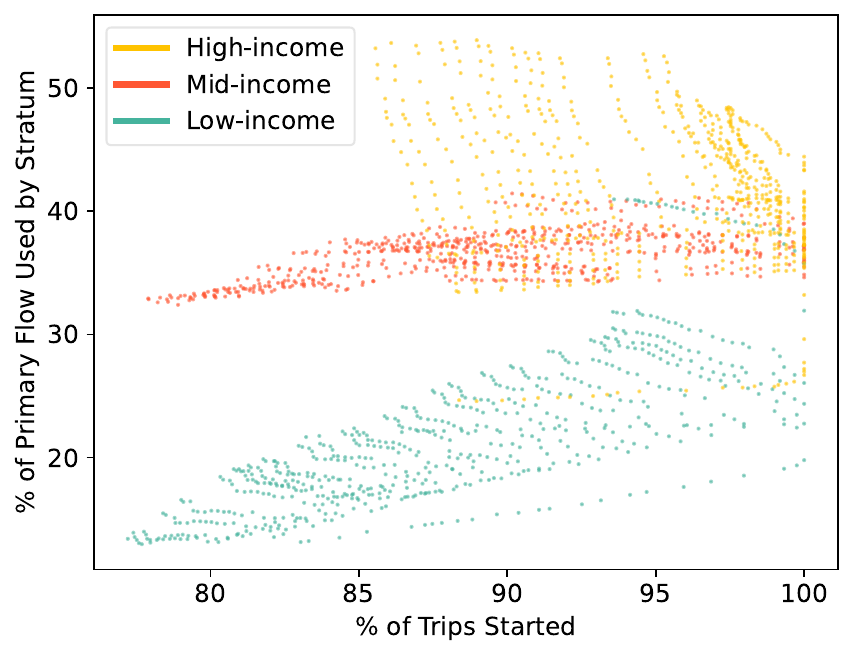}
\end{subfigure}
\end{figure}

%% file: 999_Appendix_D.tex
\section{Results on Synthetic Data}\label{sec:synthetic}
In this section, we briefly discuss our experimental study on synthetic instances. We show how the topology of the network and the segregation in the demand affects the benefits of personalized and area pricing. We find, in particular, that the benefit of per-area pricing in the above analysis depends crucially on the network and demand structure. 
\begin{figure}[htpb]
\begin{minipage}[t]{0.48\textwidth}
\captionof{figure}{Single OD Network.}
\label{fig:single_OD}
\vspace{2em}
\centerline{
\scalebox{1.2}{\begin{tikzpicture}
  \node[draw, circle] (A) at (0,0) {\footnotesize $0$};
  \node[draw, circle] (B) at (3,0) {\footnotesize $3$};
  \node[draw, circle] (C) at (1.5,1.5) {\footnotesize $1$};
  \node[draw, circle] (D) at (1.5,-1.5) {\footnotesize $2$};

  \draw[->, bend left=30,thick,color=red] (A) to (B);
  \draw[->,thick,color=blue] (A) to (C);
  \draw[->,thick,color=blue] (C) to (B);
  \draw[->, bend left=30,thick,color=red] (B) to (A);
  \draw[->,thick,color=blue] (B) to (D);
  \draw[->,thick,color=blue] (D) to (A);
\end{tikzpicture}}}
\end{minipage}
\begin{minipage}[t]{0.48\textwidth}
\captionof{figure}{10 by 10 grid network.}
\label{fig:grid_graph}
\centerline{
\includegraphics[scale=0.4,bb=0 0 456 457]{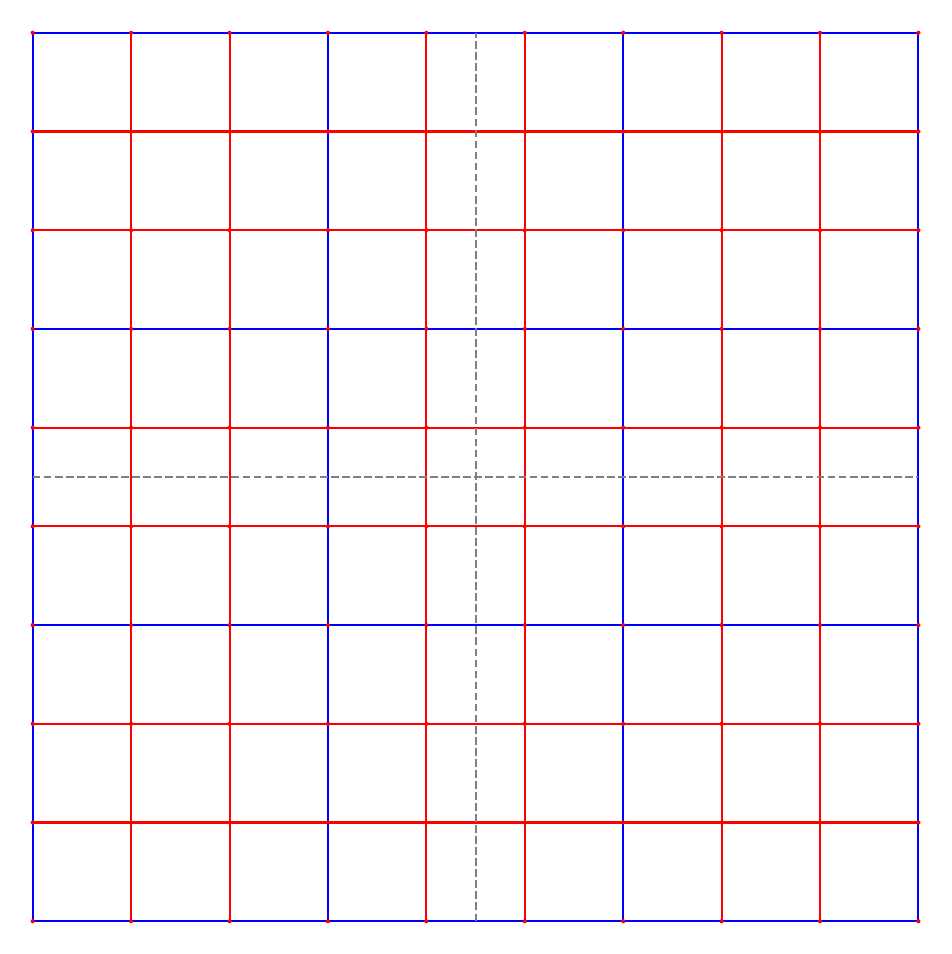}
}
\end{minipage}
\end{figure}

\subsection{Single OD Pair: Area Pricing is not Strictly Better than Uniform Pricing}\label{sec:single_OD}
We consider a simple network composed of 4 nodes and 6 arcs (2 primary),\footnote{Some of the arcs are dummy arcs to make the network strongly connected, which do not have any flow.} depicted in Figure~\ref{fig:single_OD}: Blue arcs denote secondary roads and red arcs indicate primary roads. Secondary arcs length is 3km and speed is 40km/h, while primary arcs (which have 3 lanes) length is 5km and speed is 80km/h. We computed the capacity of each arc as in the Bogot\'a case study. We considered 3 strata: high, mid, and low. Each stratum has the same OD pair 0 to 3, and the same demand equal to 500 trips. Most of the parameters that we chose for this instance are the same as in our experiments for Bogot\'a (we do not change the sensitivity parameters), except for the outside option costs for which we consider: travel time is 1.2 times the time between 0 and 3 when roads are empty and a ticket cost of 300.\footnote{We changed these values so the outside option is attractive for users.} 
In this setting, we tested a price grid in $[0,200]$ with a step of 25 for uniform and personalized pricing. For area pricing, we consider a 2 by 2 grid resulting in zones: NW, NE, SW, SE; the price options for each area are $\{0,200\}$. The only primary road that matters goes from 0 to 3, which belongs to the NW area, therefore the pricing scheme $(200,0,0,0)$ is the only relevant.  

\begin{figure}[htpb]
\captionof{figure}{Results for a Single OD Pair.}
\begin{subfigure}[t]{0.49\textwidth}
    \centering
    \caption{\footnotesize Percentage of Primary Flow}
    \label{fig:prop_prim_flow_line}
    \includegraphics[scale=0.48,bb=0 0 462 314]{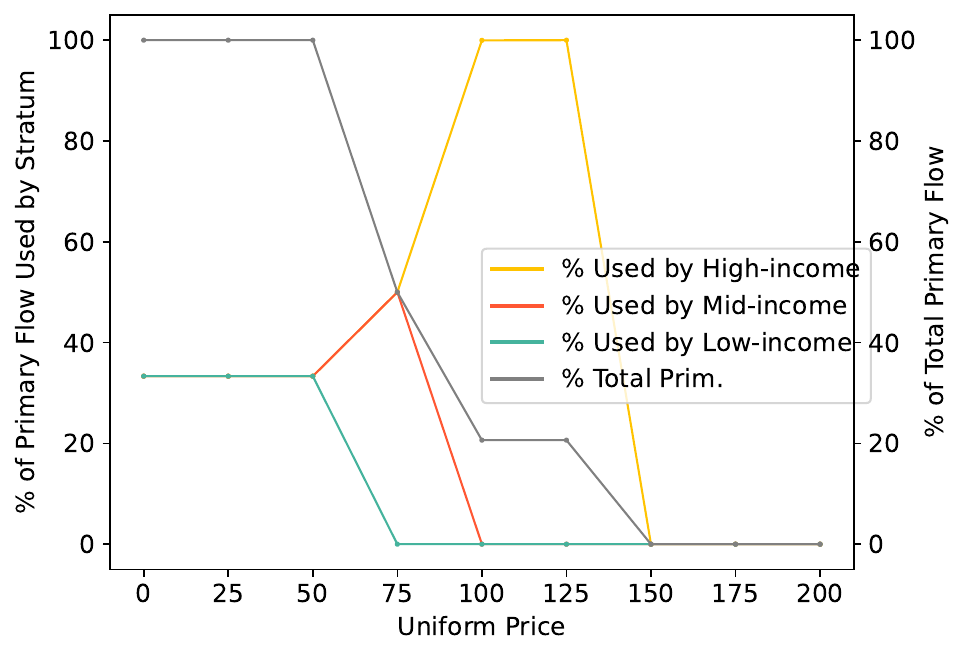}
\end{subfigure}
\begin{subfigure}[t]{0.49\textwidth}
    \centering
    \caption{\footnotesize Stratum-Welfare vs Total Welfare}
    \label{fig:pareto_welfare_line}
    \includegraphics[scale=0.45,bb=0 0 427 377]{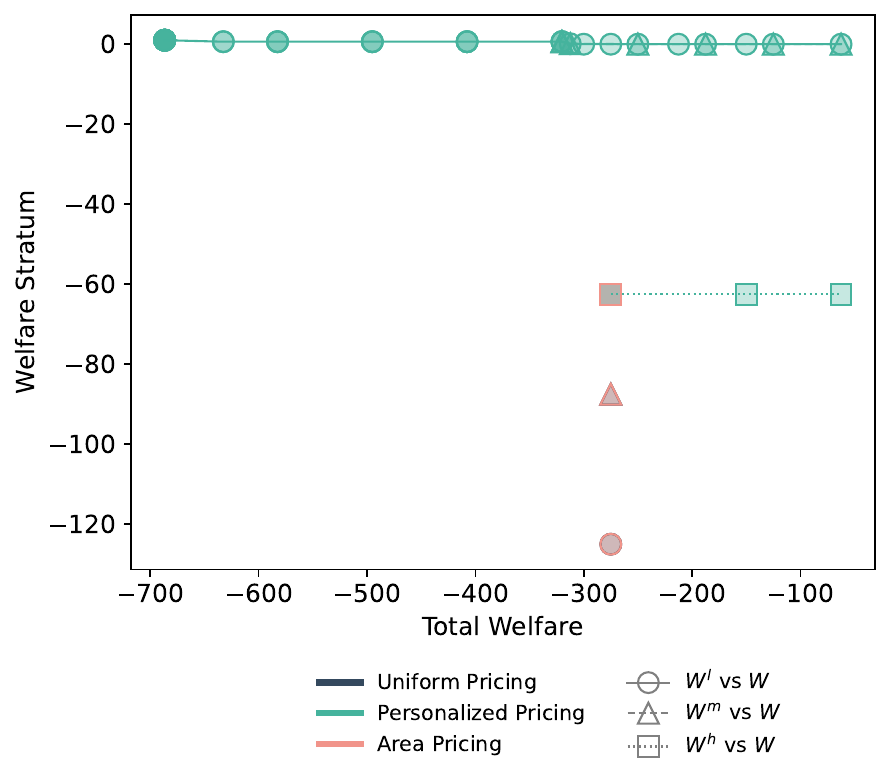}
\end{subfigure}
\centerline{\begin{subfigure}[t]{0.48\textwidth}
\caption{\footnotesize Low-Income Welfare vs Total Revenue.}
\label{fig:welfare_totalrev_line}
\includegraphics[scale=0.52,bb=0 0 428 314]{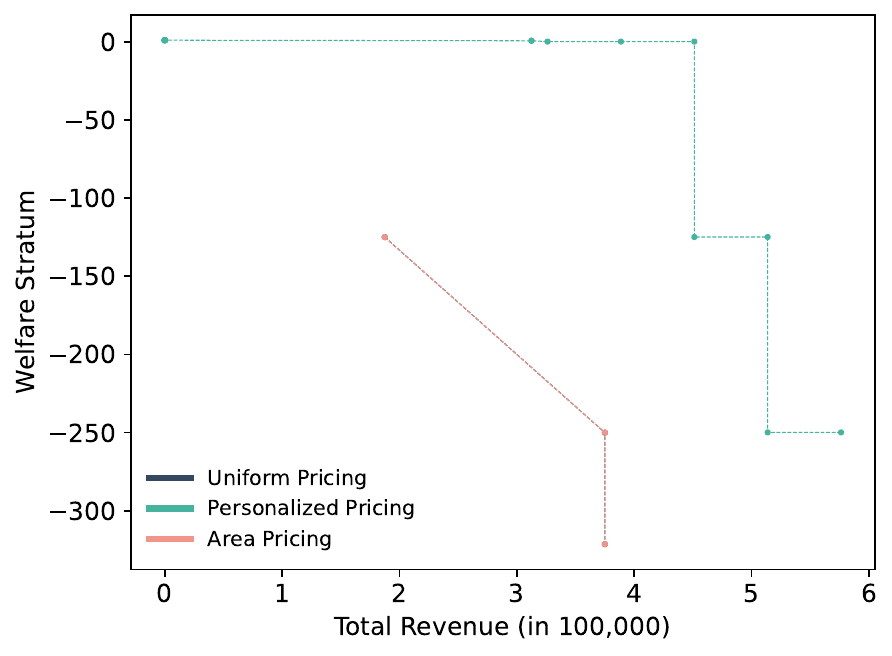}
\end{subfigure}}
\end{figure}

In Figures~\ref{fig:prop_prim_flow_line},~\ref{fig:pareto_welfare_line} and~\ref{fig:welfare_totalrev_line}, we present the results for this set of experiments.\footnote{We obtain similar results than for Bogot\'a in terms of trips started, average speed, revenue generated and welfare, so to avoid redundancy we do not present those.} Specifically, in (a) we show the primary road usage for different uniform prices, in (b) we present the stratum-welfare vs total welfare Pareto curves, in (c) we show the stratum-welfare versus total revenue Pareto curves. Analogous to the Bogot\'a case study, uniform pricing is highly inequitable for the low-income population in every metric. In particular, the usage of primary roads by the low-income stratum goes to zero after the optimal revenue price which is 50. Similarly, the mid-income stratum is negatively affected by the uniform price after $p=75$. In (b), we further observe the benefit of personalized pricing for any strata. However, we note that the Pareto curves for uniform and area pricing are the same (single points in red and in blue which cover each other). This empirically shows that area pricing is not strictly better than uniform pricing. Analogous conclusions can be observed in the stratum-welfare versus total revenue Pareto curves in Figure~\ref{fig:welfare_totalrev_line} (note that the area pricing curve in red covers the uniform pricing curve in blue). Technically, this occurs because there is a single OD pair and, for area pricing, we price the arc according to the geographical position of the entry node, i.e., in this setting the location of 0. We note that no matter how we split the network, the only affected arc is $(0,3)$, so whatever we price the area in which 0 is, then that it coincides with uniform pricing. 
\begin{remark}
One could also similarly show that personalized pricing is not strictly better than uniform pricing. For this, consider three copies of the singled OD network in Figure~\ref{fig:single_OD} which are connected arbitrarily to keep the network strongly connected. Then, each stratum has a single OD pair demand in only one of the copies. We can observe that in this setting, setting a price $p$ for a specific stratum would lead to the same equilibrium in the corresponding copy than pricing everyone at $p$.
\end{remark}

\begin{figure}[htpb]
\captionof{figure}{Results on a Grid Network.}
\begin{subfigure}[t]{0.49\textwidth}
    \centering
    \caption{\footnotesize Percentage of Primary Flow}
    \label{fig:prop_prim_flow_10by10}
    \includegraphics[scale=0.48,bb=0 0 447 314]{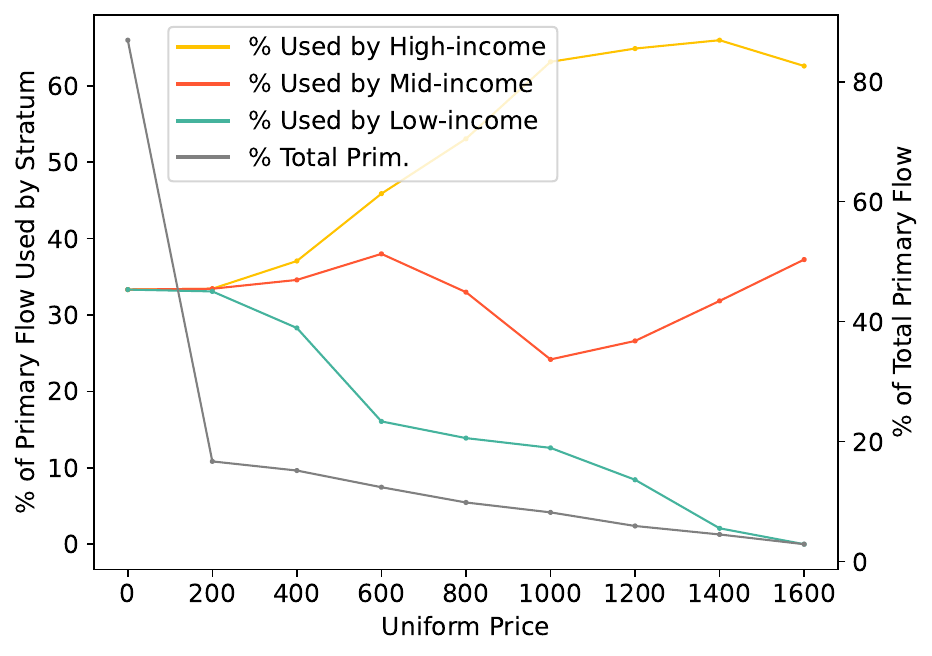}
\end{subfigure}
\begin{subfigure}[t]{0.49\textwidth}
    \centering
    \caption{\footnotesize Stratum-Welfare vs Total Welfare}
    \label{fig:pareto_welfare_10by10}
    \includegraphics[scale=0.45,bb=0 0 427 377]{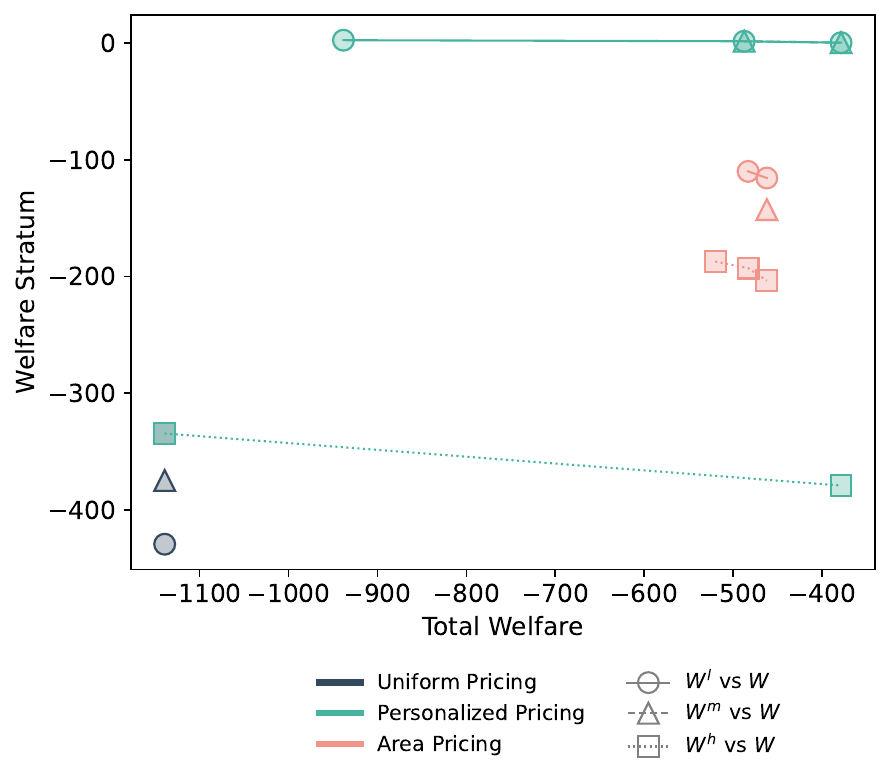}
\end{subfigure}
\centerline{\begin{subfigure}[t]{0.48\textwidth}
\caption{figure}{\footnotesize Low-Income Welfare vs Total Revenue.}
\label{fig:welfare_totalrev_10by10}
\includegraphics[scale=0.52,bb=0 0 427 314]{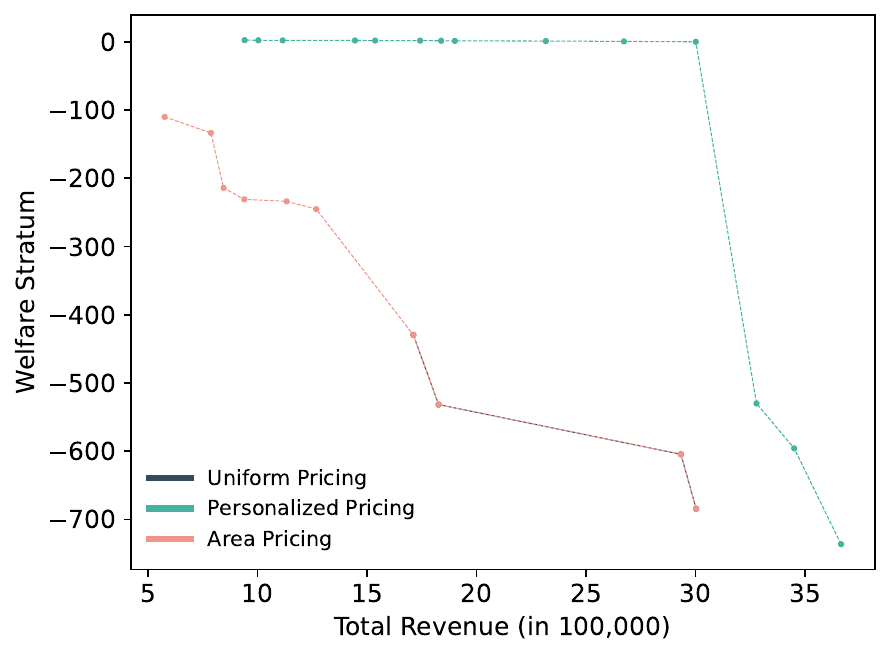}
\end{subfigure}}
\end{figure}

\subsection{Grid Network with Uniform Demand}
We now test a more complex network than the single OD pair studied above. We consider a 10 by 10 grid network composed of 100 nodes and 252 arcs (108 are primary), which is depicted in Figure~\ref{fig:grid_graph}: Blue arcs denote secondary roads (in both directions per pair of nodes), red arcs indicate primary roads (one-way direction per pair of nodes). Primary roads (which have 3 lanes) length is 1.2km and speed is 80km/h, while secondary roads length is 0.6km and speed is 30km/h. Capacities were computed as before. Dashed lines in gray indicate the area split. We note that the area split is a 2 by 2 grid resulting in areas: NW, NE, SW, SE.  

We consider the same set of strata and the demand is constructed as follows: Among all possible pairs of nodes, we consider only those that have a minimum traveling distance (shortest path) of at least 5km. Then, we group them according to area-to-area pairs, e.g., NW to NE. For each group, we sampled 10 OD pairs for which we set a demand (equally for all strata) of 10 trips---thus, demand may vary spatially (through randomness of the 10 pairs) but is identical across strata. The rest of the parameters are set as in the Bogot\'a study, except for the ticket cost of the outside option which we set to 400 to make it more attractive. The price grid is taken from the interval $[0,1600]$ with step of 200 and for area pricing we only consider combinations with prices $0$ and $1600$.\footnote{We do not test other combinations due to the results in the Bogot\'a study.}

We present our results in Figure~\ref{fig:prop_prim_flow_10by10}, Figure~\ref{fig:pareto_welfare_10by10}, and  Figure~\ref{fig:welfare_totalrev_10by10}. We obtain the same results as in the previous experiments: uniform pricing highly affects the low-income population, which can be observed in the percentage of primary roads used (Figure~\ref{fig:prop_prim_flow_10by10}). The Pareto curves in Figure~\ref{fig:welfare_totalrev_10by10} show that personalized pricing can significantly benefit the low- and mid-income groups at the cost of the high-income stratum. We also note that area pricing is strictly dominated by personalized pricing, due to the uniform distribution of the demand. In terms of total welfare area pricing dominates uniform pricing, however, if one aims to maximize revenue (Figure~\ref{fig:welfare_totalrev_10by10}), then area pricing achieves the same levels of low-income welfare as uniform pricing (bottom right), due to the uniform demand (the red curve covers the blue curve). 

The results above suggest that the benefits of area pricing may be inherently correlated with the segregation of the demand across the city. We expect the effectiveness of area pricing to generalize in cities with geographic population heterogeneity. We further expect similar results in terms of other demographic and socioeconomic heterogeneity, since cities have geographic segregation across such dimensions.


%% file: 999_Appendix_E.tex
\section{Robustness Analysis for Uniform Pricing}\label{app:robustness_analysis}
Our goal in this section is to test different parameters in the model to find which are the most relevant for the insights in our work. For this, we test Bogot\'a at different scales and demand, sensitivity to willingness to pay, sensitivity parameters of the outside option, demand perturbations {and equilibrium solutions}. Our main takeaways from these results are: (i) the only parameters that drive uniform pricing inequity are the willingness to pay differences; (ii) the equilibrium solution lead to realistic sample paths. Uniform pricing inequity is monotonic in differences and does not change when the higher strata relatively prefer the outside option, the lower stratum has higher overall demand, the lower stratum is more likely to use primary roads without pricing (as in the 10km scale), etc. Similarly, per-strata beats per-area for low-income welfare, but per-area beats per-strata for overall welfare consistently as the sensitivity to willingness to pay varies. In the following, to avoid redundancy, we just present the results for the usage flow of primary roads (we obtained similar results for other metrics and pricing schemes).

\subsection{Bogot\'a at Different Scales}
\begin{figure}[htpb]
\centering
\caption{Bogot\'a at different scales.}\label{fig:bogota_different_scales}
\begin{subfigure}[t]{.45\textwidth}
    \centering
\caption{\footnotesize 5km radius}\label{fig:primflow_2k}
\includegraphics[width=\textwidth,bb=0 0 447 317]{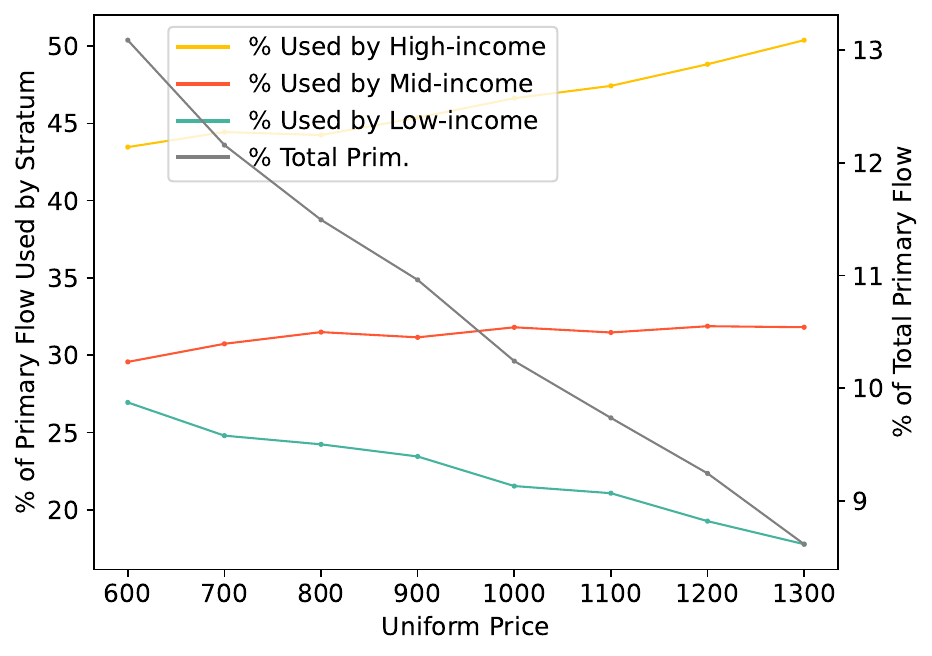}
\end{subfigure}%
\begin{subfigure}[t]{.45\textwidth}
    \centering
\caption{\footnotesize 7km radius}\label{fig:primflow_5k}
\includegraphics[width=\textwidth,bb=0 0 447 314]{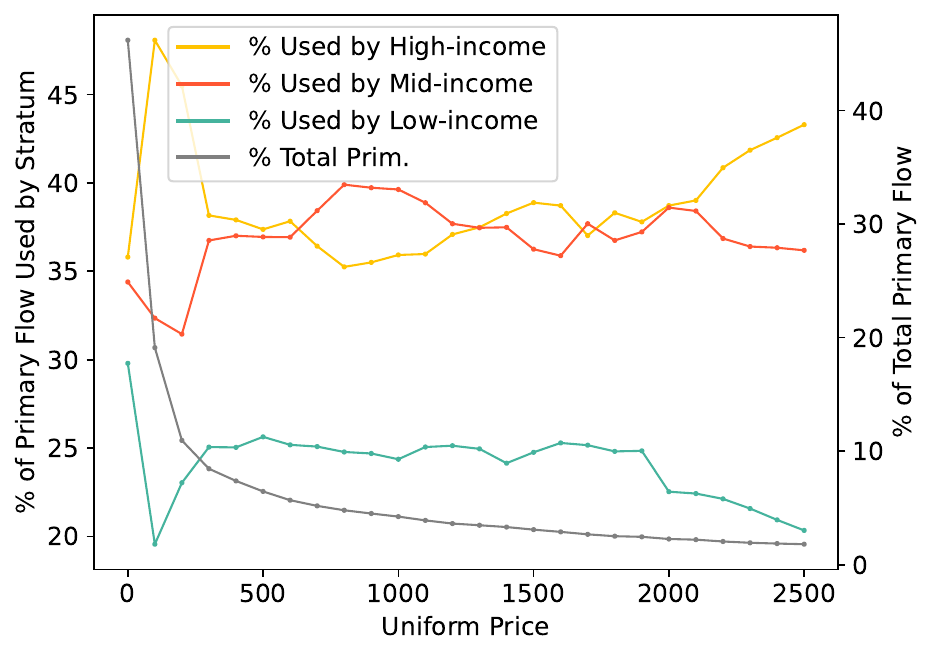}
\end{subfigure}
\end{figure}

In this set of experiments, we consider the same setting as in Section~\ref{sec:data_description} with the same parameters, however, we change the radius of the city. We consider a radius of 5km and 7km. This affects the demand matrix and the proportion of trips across strata. For the 5km network (480 edges and 241 nodes), the demand is as follows: 552 trips for high-income, 2913 trips for mid-income, and 1569 trips for low-income. On the other hand, for the 7km network (886 edges and 407 nodes), the demand is 1029 trips for high-income, 4046 trips for mid-income, and 2160 trips for low-income. As we observed in Figure~\ref{fig:bogota_different_scales}, no matter which scale of the city we consider, uniform pricing negatively impacts the low-income stratum. The only difference is the impact on the mid-income group, however, as the price increases the high-income group's usage of primary roads always increases.

\subsection{Different Sensitivity to Willingness to Pay}
\begin{figure}[htpb]
\centering
\caption{Different willingness-to-pay ratios.}\label{fig:different_wtp}
\begin{subfigure}[t]{.45\textwidth}
    \centering
\caption{\footnotesize $\theta^{\textsf{l}}<\theta^{\textsf{m}}<\theta^{\textsf{h}}$}\label{fig:wtp_increasing}
\includegraphics[width=\textwidth,bb=0 0 409 314]{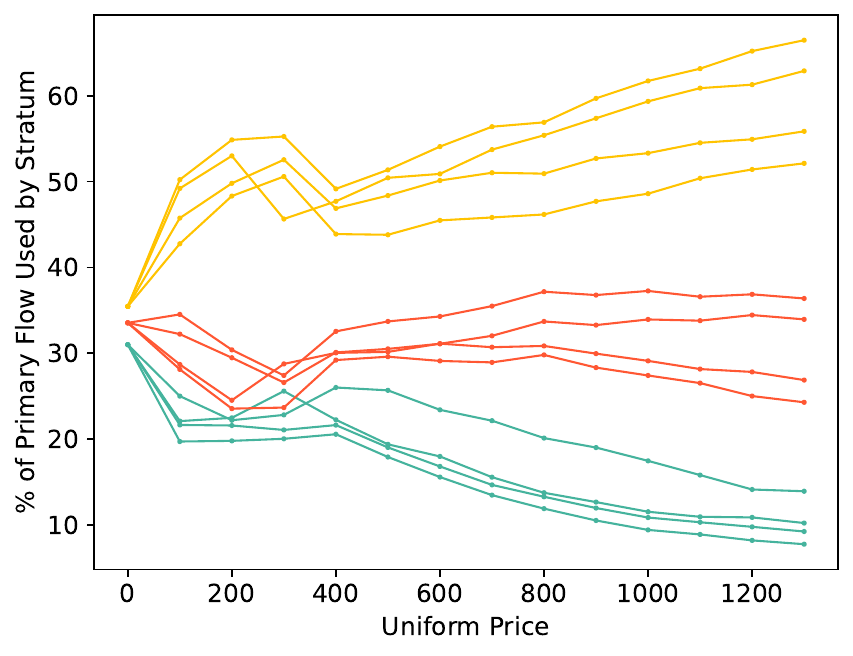}
\end{subfigure}%
\begin{subfigure}[t]{.45\textwidth}
    \centering
\caption{\footnotesize $\theta^{\textsf{l}}>\theta^{\textsf{m}}>\theta^{\textsf{h}}$}\label{fig:wtp_decreasing}
\includegraphics[width=\textwidth,bb=0 0 409 314]{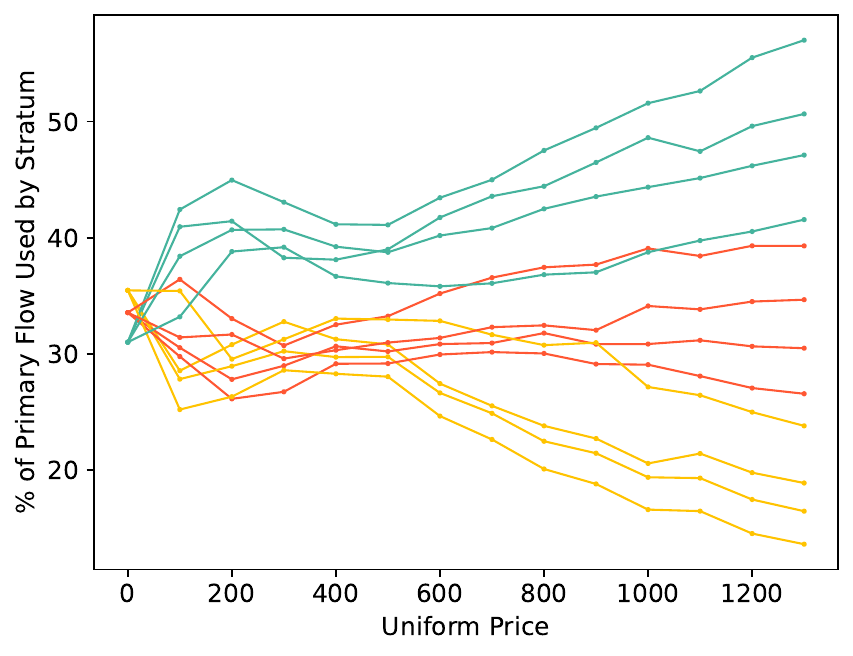}
\end{subfigure}
\begin{subfigure}[t]{.45\textwidth}
    \centering
\caption{\footnotesize $\theta^{\textsf{m}}<\theta^{\textsf{l}}<\theta^{\textsf{h}}$}\label{fig:wtp_mid_low_high}
\includegraphics[width=\textwidth,bb=0 0 409 314]{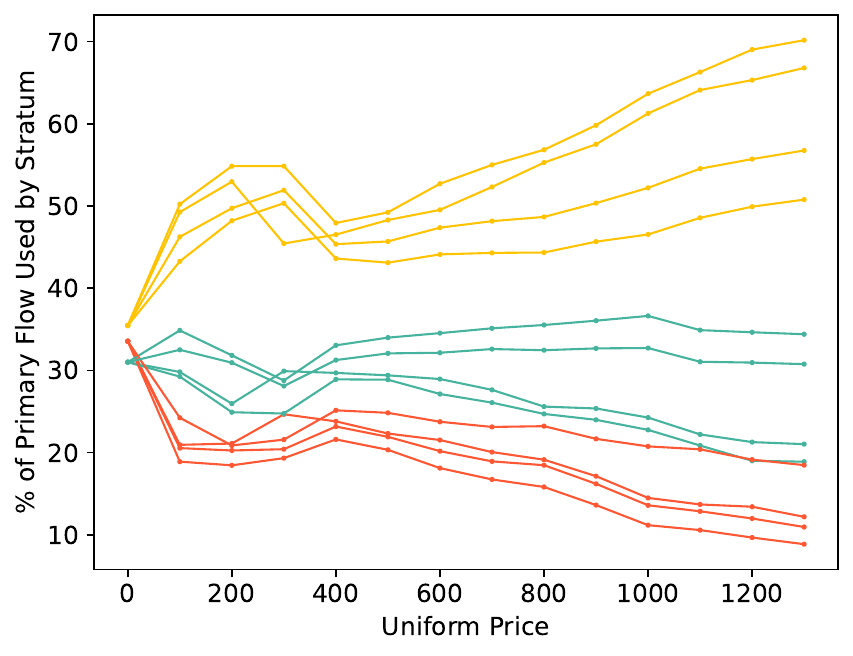}
\end{subfigure}
\begin{subfigure}[t]{.45\textwidth}
    \centering
\caption{\footnotesize $\theta^{\textsf{h}}<\theta^{\textsf{l}}<\theta^{\textsf{m}}$}\label{fig:wtp_high_low_mid}
\includegraphics[width=\textwidth,bb=0 0 409 314]{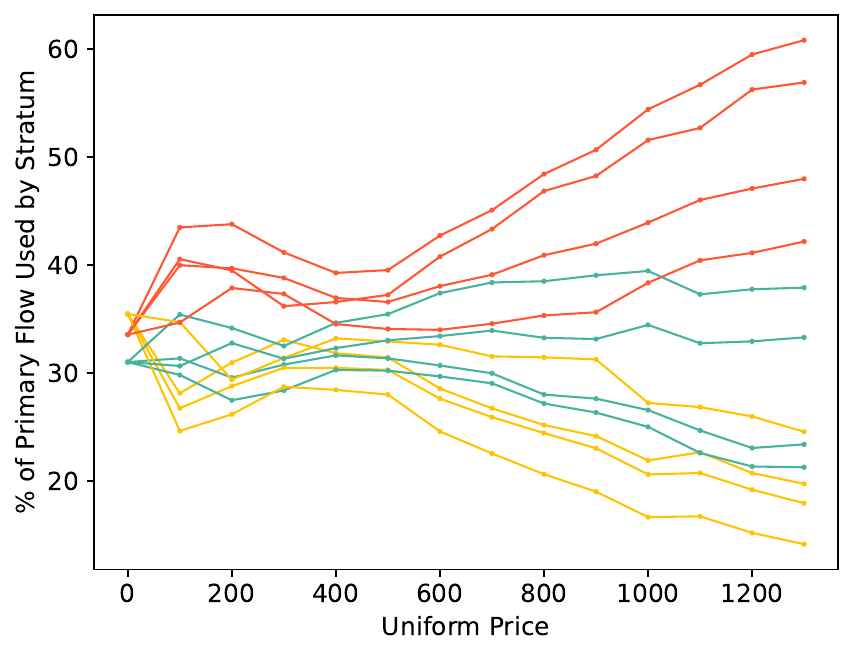}
\end{subfigure}
\end{figure}

For these experiments, we consider the instance of Bogot\'a with a radius of 5km. To ease the exposition, we only consider a range of prices of $[0,1300]$ with a step of 100. 
We consider $\theta^s\in\{0.5,0.8,1.5,2.0\}$ and $\beta^s=1$ for all $s\in\{\textsf{l},\textsf{m},\textsf{h}\}$. We present the results in Figure~\ref{fig:different_wtp}, where the color legends are the same as the previous figures, i.e., green for low-income, orange for mid-income, and yellow for high-income. Each curve corresponds to a different WTP ratio for a given stratum. In each subplot, we present different orderings of $\theta^s$.
\revised{The most intuitive ordering is shown in Figure~\ref{fig:wtp_increasing}, where the ratios are increasing with income. We note that, in any combination of ratios, 
the stratum with the highest WTP ratio is the one that benefits the most. Figures~\ref{fig:wtp_mid_low_high} and~\ref{fig:wtp_high_low_mid} confirm this trend: in both plots, we observe that uniform pricing is inequitable for the strata with the lowest WTP ratio (the magnitude of the inequity is determined by each sequence of values). }

\subsection{Different Sensitivity Parameters for The Outside Option}
\begin{figure}[htpb]
\centering
\caption{Different WTO ratios.}\label{fig:different_wto}
\includegraphics[width=.45\textwidth,bb=0 0 409 314]{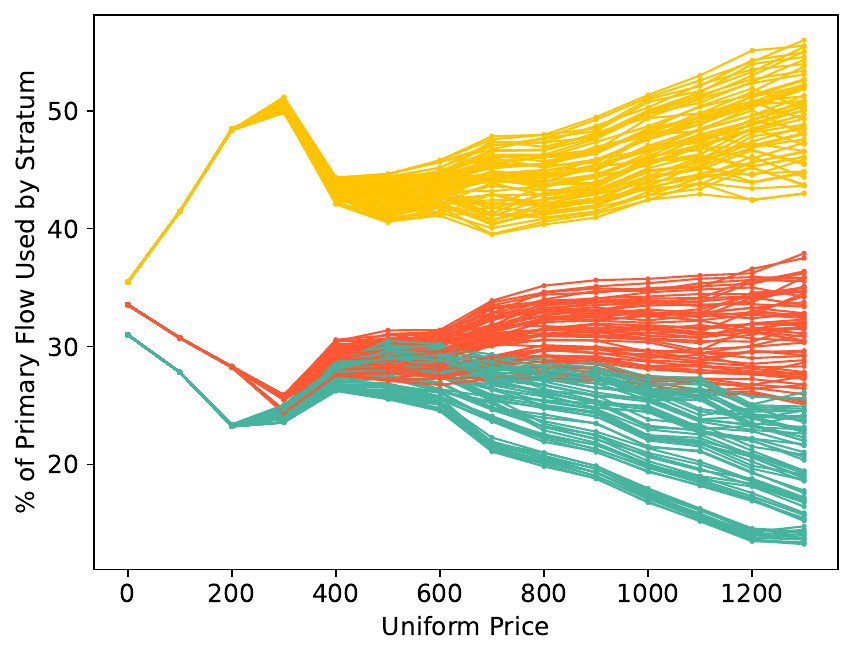}
\end{figure}

In this section, we also consider the instance of Bogot\'a with a radius of 5km and a range of prices of $[0,1300]$ with a step of 100. The rest of the parameters are as in the 10km instance, except for the sensitivity parameters of the outside option. 
We consider $\mathring{\beta}^s\in\{0.8,0.9,1.0,1.1,1.2\}$ for all $s\in\{\text{\sf l},\text{\sf m},\text{\sf h}\}$, where the highest this value is the least willing the stratum is to take the outside option. We present the results in Figure~\ref{fig:different_wto}, where the color legends are: green for low-income, orange for mid-income, and yellow for high-income. For a fix stratum, each curve corresponds to a different sensitivity value. We observe that there is no significant change in trend, uniform pricing negatively impacts the low-income stratum. The only difference is when the green curves start to decrease which depends on the relation between the revenue optimal uniform price and the outside option cost.
\subsection{Demand Matrix Perturbations}
\begin{figure}[htpb]
\centering
\caption{Perturbations of the Demand Matrix.}\label{fig:demandpert}
\includegraphics[width=.45\textwidth,bb=0 0 409 314]{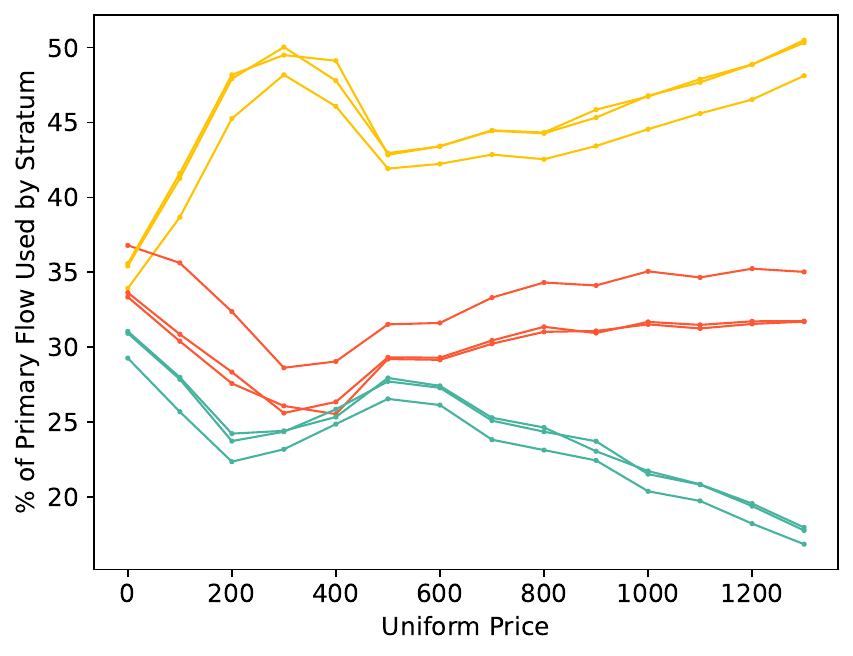}
\end{figure}

Finally, we test different perturbations of the demand, with uniform pricing. The setting is as follows: Bogot\'a with a 5km radius, the same parameters (including WTP and sensitivity to the outside option cost) as in the 10km instance, and a price range of $[0,1300]$ with a step of 100. We consider the same OD pairs as in the original instance and we amplify the demand of each stratum in each OD pair uniformly at random by a certain factor. 
These factors were chosen arbitrarily to make the perturbed matrices significantly different, in terms of numbers of trips, relative to the original one. 
We obtain three demand matrices: (i) 2791 trips for high-income, 951 trips for mid-income, 3182 trips for low-income; (ii) 2246 trips for high-income, 2922 trips for mid-income, 3661 trips for low-income; (iii) 2336 trips for high-income, 3121 trips for mid-income, 1786 trips for low-income. We present the results in Figure~\ref{fig:demandpert} from which we observe no significant change in our previous takeaways: Even if the demand of the low-income population is higher than the rest, they are negatively affected by uniform pricing.

\newpage

{\color{black}
\subsection{Equilibrium Solution Analysis}\label{app:equilibrium_analysis}
In Figure~\ref{fig:equilibrium_samples}, we show various sample paths (in orange) for different OD pairs where the equilibrium flow is obtained with uniform pricing ($r\in\{0,1100\}$). In green we present the original GPS path taken by the driver, which was projected over our network.
\begin{figure}[htpb]
\centering
\caption{Sample Routes. Route in green denotes the GPS data and in yellow the sample path resulting from the equilibrium flow at the given uniform price. Note that both routes may overlap when $r=0$.}\label{fig:equilibrium_samples}
\begin{subfigure}[t]{.4\textwidth}
    \centering
\caption{\footnotesize OD pair $(158,120)$, $r=0$}\label{fig:od_pair_1_p0}
\includegraphics[width=\textwidth]{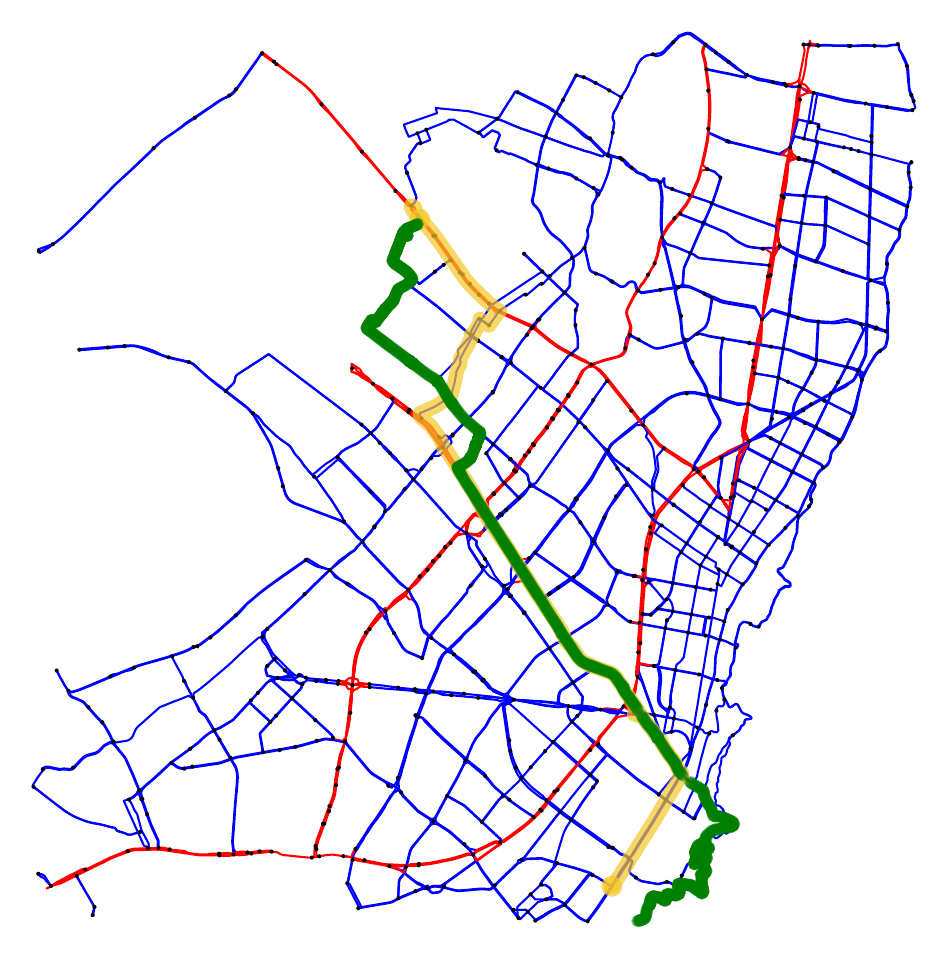}
\end{subfigure}%
\begin{subfigure}[t]{.4\textwidth}
    \centering
\caption{\footnotesize OD pair $(158,120)$, $r=1100$}\label{fig:od_pair_1_p1100}
\includegraphics[width=\textwidth]{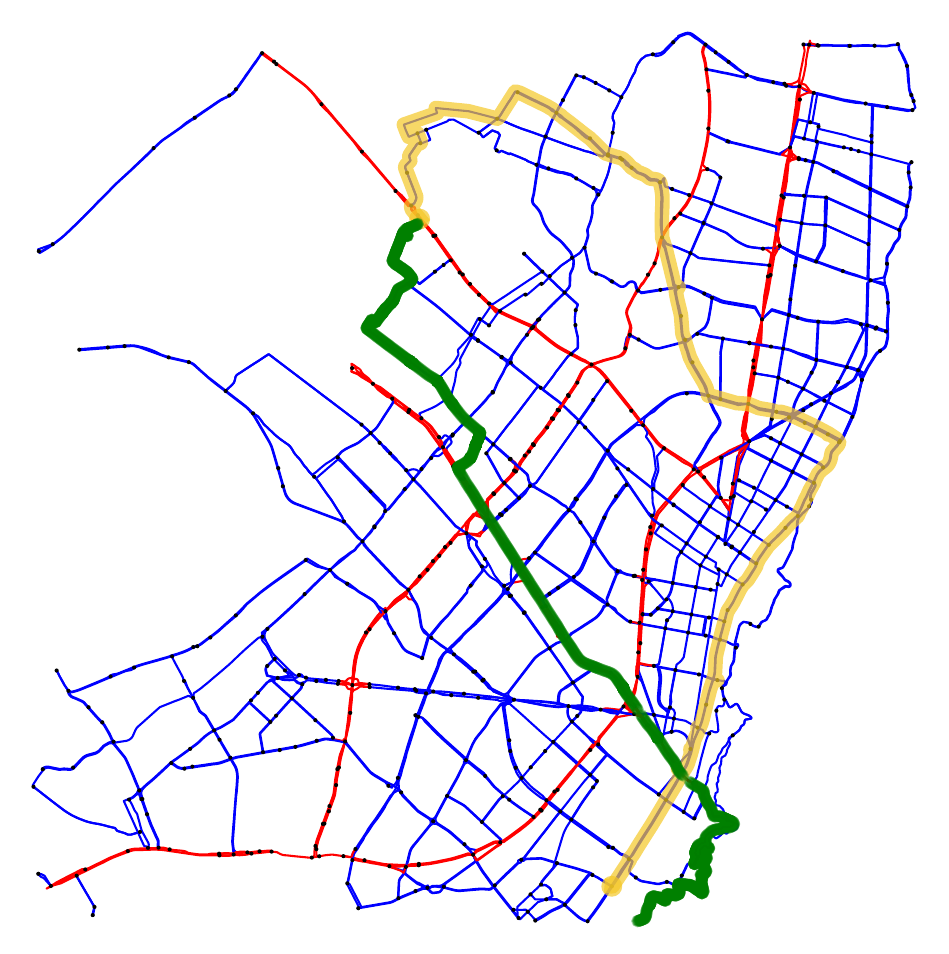}
\end{subfigure}%
\\
\begin{subfigure}[t]{.4\textwidth}
    \centering
\caption{\footnotesize OD pair $(660,513)$, $r=0$}\label{fig:od_pair_2_p0}
\includegraphics[width=\textwidth]{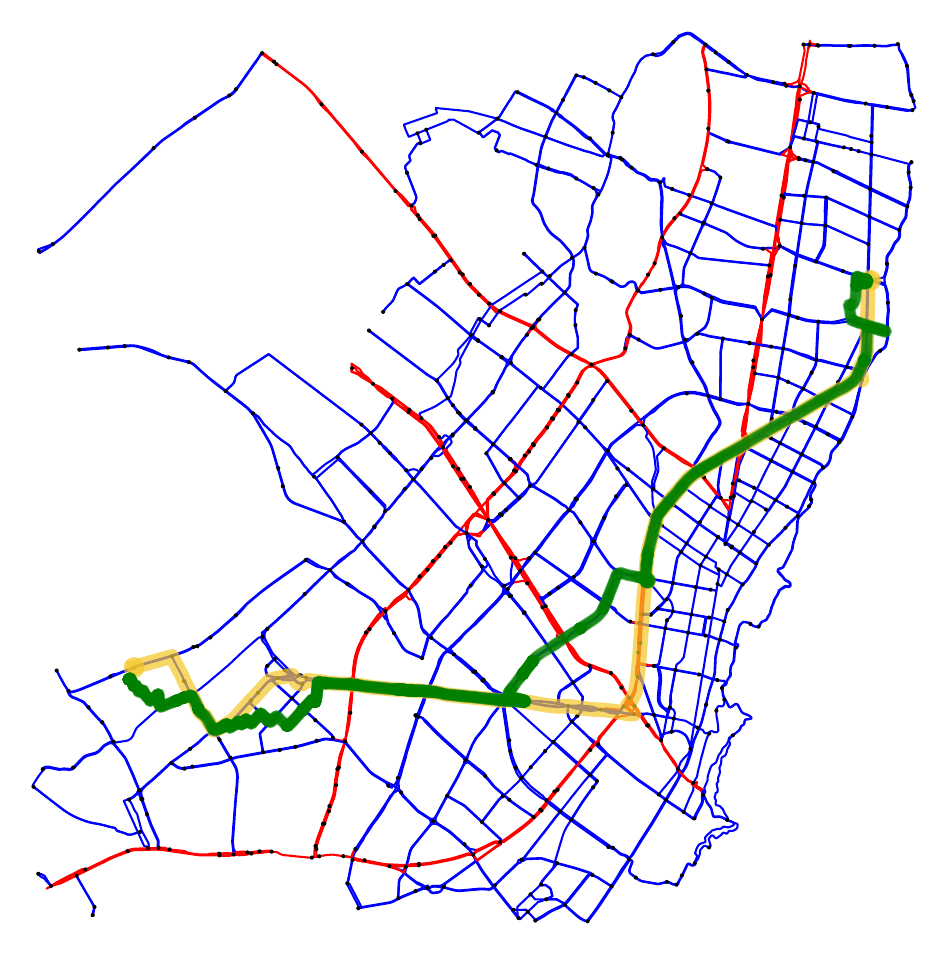}
\end{subfigure}%
\begin{subfigure}[t]{.4\textwidth}
    \centering
\caption{\footnotesize OD pair $(660,513)$, $r=1100$}\label{fig:od_pair_2_p1100}
\includegraphics[width=\textwidth]{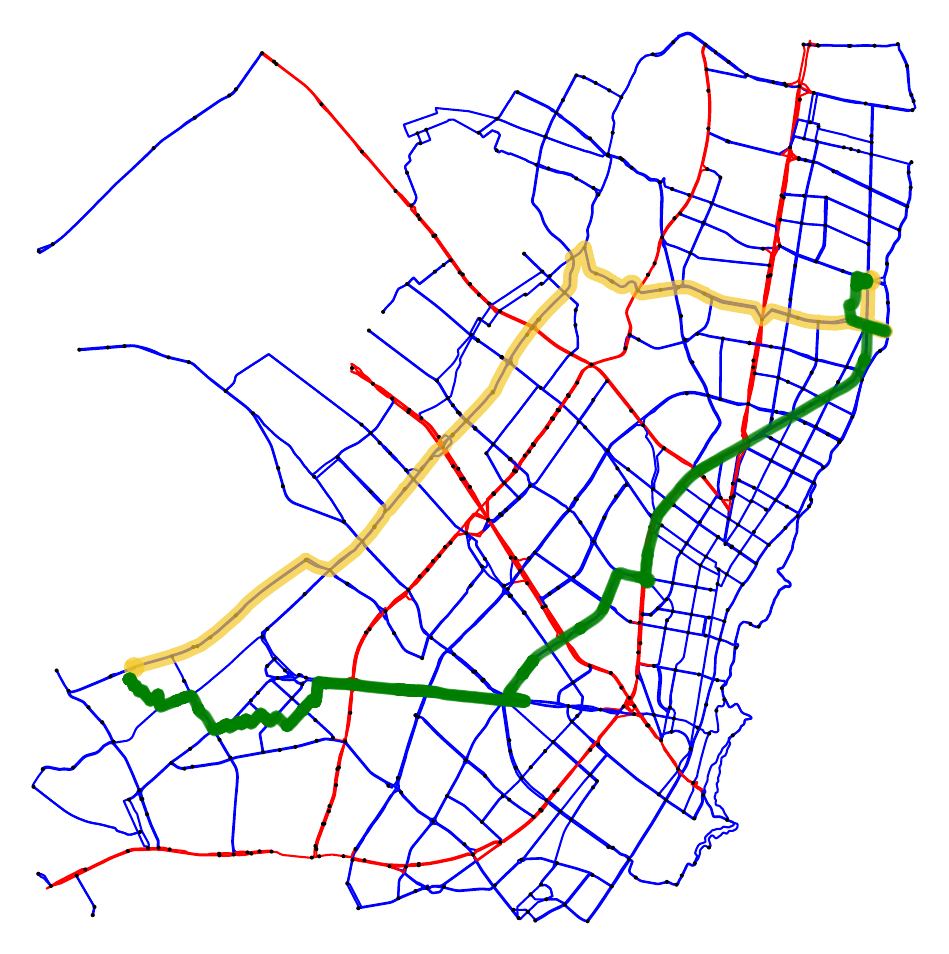}
\end{subfigure}%
\end{figure}
}